%% file: main.tex
\documentclass[fleqn,usenatbib]{mnras}

\usepackage{newtxtext,newtxmath}

\usepackage[T1]{fontenc}

\DeclareRobustCommand{\VAN}[3]{#2}
\let\VANthebibliography\thebibliography
\def\thebibliography{\DeclareRobustCommand{\VAN}[3]{##3}\VANthebibliography}

\usepackage{graphicx}	
\usepackage{amsmath}	
\usepackage{amssymb}	
\usepackage{times} 
\usepackage{latexsym}
\usepackage{amsfonts}
\usepackage{rotfloat} 
\usepackage{rotating} 
\usepackage{float}
\usepackage{natbib}
\usepackage{supertabular}
\usepackage{longtable}
\usepackage{url}
\usepackage{dcolumn}
\usepackage{placeins}
\usepackage{multirow} 
\usepackage{textcomp}
\usepackage{multicol}
\usepackage{bm}
\usepackage{pdflscape}
\usepackage[utf8]{inputenc}
\usepackage{ae,aecompl}
\usepackage[caption=false]{subfig}

\newcommand{\mgii}{\mbox{Mg\,{\sc ii}}}

\newcommand{\feii}{\mbox{Fe\,{\sc ii}}}

\newcommand{\civ}{\mbox{C\,{\sc iv}}}
\newcommand{\ovi}{\mbox{O\,{\sc vi}}}

\newcommand{\sifour}{\mbox{Si\,{\sc iv}}}
\newcommand{\ha}{H\,$\alpha$}

\newcommand{\oii}{[O\,{\sc ii}]}
\newcommand{\oiii}{[O\,{\sc iii}]}
\newcommand{\nii}{[N\,{\sc ii}]}

\newcommand{\lya}{\ensuremath{{\rm Ly}\alpha}}

\newcommand{\zqso}{$z_{\rm QSO}$}

\newcommand{\wmgii}{$\rm W_{\rm r}^{\mgii}$}
\newcommand{\wciv}{$\rm W_{\rm r}^{\civ}$}

\newcommand{\mstar}{$\rm M_*$}
\newcommand{\mhalo}{$\rm M_{\rm h}$}

\newcommand{\fc}{$\rm f_{\rm c}$}
\newcommand{\kms}{km\,s$^{-1}$}
\newcommand{\cms}{cm$^{-2}$}

\newcommand{\ergscm}{$\rm erg\,s^{-1}\,cm^{-2}$}

\newcommand{\msun}{$\rm M_\odot$}
\newcommand{\msunyr}{$\rm M_\odot yr^{-1}$}
\title[Metals across galaxy overdensities]{Metal-enriched halo gas across galaxy overdensities over the last 10 billion years}

\author[R. Dutta et al.]{Rajeshwari Dutta$^{1,2}$\thanks{E-mail: rajeshwari.dutta@unimib.it}, 
Michele Fumagalli$^{1,3}$,
Matteo Fossati$^{1,2}$,
Richard M. Bielby$^{4,5}$,
John P. Stott$^{6}$,
\newauthor{Emma K. Lofthouse$^{1,2}$,
Sebastiano Cantalupo$^{1,7}$,
Fergus Cullen$^{8}$,
Robert A. Crain$^{9}$,
Todd M. Tripp$^{10}$,}
\newauthor{J. Xavier Prochaska$^{11}$,
Fabrizio Arrigoni Battaia$^{12}$,
Joseph N. Burchett$^{13}$,
Johan P. U. Fynbo$^{14}$,}
\newauthor{Michael T. Murphy$^{15}$,
Joop Schaye$^{16}$,
Nicolas Tejos$^{17}$,
Tom Theuns$^{18}$}
\\
$^{1}$Dipartimento di Fisica G. Occhialini, Universit\`a degli Studi di Milano Bicocca, Piazza della Scienza 3, 20126 Milano, Italy \\
$^{2}$INAF - Osservatorio Astronomico di Brera, via Bianchi 46, 23087 Merate (LC), Italy \\
$^{3}$INAF - Osservatorio Astronomico di Trieste, via G. B. Tiepolo 11, 34143 Trieste, Italy \\
$^{4}$Centre for Extragalactic Astronomy, Department of Physics, Durham University, South Road, Durham DH1 3LE, UK \\
$^{5}$Data Insights and Statistics Division, Department for Education, Bishopsgate House, Feethams, Darlington, DL1 5QE, UK \\
$^{6}$Department of Physics, Lancaster University, Lancaster LA1 4YB, UK\\
$^{7}$Department of Physics, ETH Zurich, Wolfgang-Pauli-Strasse 27, CH-8093 Zurich, Switzerland \\
$^{8}$Institute for Astronomy, University of Edinburgh, Royal Observatory, Edinburgh EH9 3HJ, UK \\
$^{9}$Astrophysics Research Institute, Liverpool John Moores University, 146 Brownlow Hill, Liverpool, L3 5RF, UK \\
$^{10}$Department of Astronomy, University of Massachusetts-Amherst, 710 North Pleasant Street, Amherst, MA 01003-9305, USA \\
$^{11}$UCO/Lick Observatory, University of California, Santa Cruz, CA, USA \\
$^{12}$Max-Planck-Institut f\"ur Astrophysik, Karl-Schwarzschild-Str 1, D-85748
Garching bei M\"unchen, Germany \\
$^{13}$Department of Astronomy, New Mexico State University, Las Cruces, NM 88003, USA \\
$^{14}$Niels Bohr Institute, University of Copenhagen, Lyngbyvej 2, DK-2100 Copenhagen, Denmark \\
$^{15}$Centre for Astrophysics and Supercomputing, Swinburne University of Technology, Hawthorn, Victoria 3122, Australia \\
$^{16}$Leiden Observatory, Leiden University, PO Box 9513, 2300 RA Leiden, the Netherlands \\
$^{17}$Instituto de F\'isica, Pontificia Universidad Cat\'olica de Valpara\'iso, Casilla 4059, Valpara\'iso, Chile\\
$^{18}$Institute for Computational Cosmology, Department of Physics, Durham University, South Road, Durham, DH1 3LE, UK \\
}

\date{Accepted XXX. Received YYY; in original form ZZZ}

\pubyear{2021}

\begin{document}
\label{firstpage}
\pagerange{\pageref{firstpage}--\pageref{lastpage}}
\maketitle

\begin{abstract}
We present a study of metal-enriched halo gas traced by \mgii\ and \civ\ absorption at $z<2$ in the MUSE Analysis of Gas around Galaxies survey and the Quasar Sightline and Galaxy Evolution survey. Using these large and complete galaxy surveys in quasar fields, we study the dependence of the metal distribution on galaxy properties and overdensities, out to physical projected separations of 750\,kpc. We find that the cool, low-ionization gas is significantly affected by the environment across the full redshift range probed, with $\approx2-3$ times more prevalent and stronger \mgii\ absorption in higher overdensity group environments and in regions with greater overall stellar mass and star formation rates. Complementary to these results, we have further investigated the more highly ionized gas as traced by \civ\ absorption, and found that it is likely to be more extended than the \mgii\ gas, with $\approx$2 times higher covering fraction at a given distance. We find that the strength and covering fraction of \civ\ absorption show less significant dependence on galaxy properties and environment than the \mgii\ absorption, but more massive and star-forming galaxies nevertheless also show $\approx$2 times higher incidence of \civ\ absorption. The incidence of \mgii\ and \civ\ absorption within the virial radius shows a tentative increase with redshift, being higher by a factor of $\approx$1.5 and $\approx$4, respectively, at $z>1$. It is clear from our results that environmental processes have a significant impact on the distribution of metals around galaxies and need to be fully accounted for when analyzing correlations between gaseous haloes and galaxy properties.
\end{abstract}

\begin{keywords}
galaxies: groups: general – galaxies: haloes – quasars: absorption lines.
\end{keywords}

%
\section{Introduction}
\label{sec_introduction}

The formation and evolution of galaxies are shaped by both internal and external processes. The extent to which the properties of galaxies are determined by these processes can vary with time and population, and it is not always possible to trivially disentangle the effects of secular evolution and environment. In addition to internal processes of feedback from stars and active galactic nuclei which can suppress star formation and eject metals \citep[e.g.][]{dekel1986,silk1998}, external processes of gravitational interactions between group or cluster members like tidal stripping, and hydrodynamic interactions between galaxies and the hot intra-group/cluster medium like ram-pressure stripping can significantly affect the gas and star formation in galaxies \citep[e.g.][]{gunn1972,merritt1983,boselli2006}.

Several studies have investigated the statistical effects of the environment on galaxy properties using large samples of galaxies and various techniques to describe the galaxy environment \citep[see][for a comparison of methods]{muldrew2012}, including $N$th nearest neighbour methods \citep{dressler1980,baldry2006}, number of galaxies in fixed or adaptive apertures \citep{hogg2003,croton2005}, adaptive smoothing \citep{scoville2007} and Voronoi tessellation \citep{ebeling1993}. Such studies in general have found that on top of internal processes, environmental effects regulate the star formation activity and morphology of galaxies, and that the galaxy population progresses from less massive, bluer and star forming spirals to more massive, redder and passive ellipticals with increasing galaxy density \citep[e.g.][]{dressler1980,balogh2004,kauffmann2004,baldry2006,peng2010,wetzel2012,hirschmann2014,balogh2016,fossati2017,lemaux2019,old2020}.

Most studies to date have focused on how the environmental processes affect the properties of the galaxy rather than its interplay with the ambient gaseous halo or the circumgalactic medium (CGM), which is now established to be a crucial component of the galaxy ecosystem \citep{tumlinson2017}. The CGM mediates the gas flows and the baryon and metal cycles in and around galaxies \citep[see][for a review]{peroux2020}. Observations of the CGM can place strong constraints on how gas accretes onto and is ejected from galaxies in models of galaxy formation and evolution \citep[e.g.][]{fumagalli2011,suresh2015,turner2017,oppenheimer2018,stern2018}.

Absorption lines detected in the spectrum of a bright background source like a quasar have been the most relied upon tool in the literature to probe the diffuse gas in the CGM. Different ionic transitions have been used to trace different phases of the gas. For example, the \mgii\ doublet lines have long been used to trace the cool ($T\sim10^4$\,K) and low-ionization (ionization potential, IP = 15.0 eV) gas \citep[e.g.][]{lanzetta1990,steidel1992,churchill2000,chen2010,nielsen2013,zhu2013,dutta2020}, while the \civ\ doublet lines on the other hand have been used extensively to trace the potentially warmer ($T\sim10^5$\,K) and high-ionization (IP = 64.5 eV) gas \citep[e.g.][]{sargent1988,chen2001,schaye2003,cooksey2013,bordoloi2014,burchett2016}.

Using samples of absorber-galaxy pairs, the distribution of the multiphase gas around galaxies and its dependence on galaxy properties have been studied at $z\lesssim0.5$ \citep[e.g.][]{prochaska2011,stocke2013,tumlinson2013,werk2014,liang2014,burchett2016,prochaska2017,chen2018,berg2019,burchett2019,wilde2021} and at $z\gtrsim2$ \citep[e.g.][]{rudie2012,turner2014,turner2015,fumagalli2015,bielby2017b,rudie2019}. Such studies have greatly increased our knowledge of the CGM and found that both the neutral hydrogen and metal absorbing gas show strong statistical correlation with distance from galaxies, that the gas becomes progressively more ionized with increasing distance from galaxies, and that the distribution and extent of the different gas phases can depend on mass and star formation activity of the galaxies. However, most of the above studies have been based upon imaging to pre-select galaxies at close separation from quasars for spectroscopic follow-up. Consequently, these studies were preferentially biased towards the more luminous (more massive) galaxies, especially at high redshifts, and lacked a complete census of their local galaxy environment. Furthermore, there have been relatively few studies of the galaxy-halo connection in the intermediate redshift range ($z\approx0.5-2$) due to observational constraints. In order to obtain a holistic understanding of the co-evolution of gas and galaxies from the cosmic noon ($z\approx1-2$) to the present, the galaxy-halo connection needs to be interpreted in the context of the environment in which galaxies reside. Therefore, it is imperative to bridge the gap between studies of galaxy environment and gaseous haloes.

There have been few studies that have looked at the CGM in the context of the local environment, particularly focusing on the warm-hot gas phase in small samples probing cluster/group environments \citep{tejos2016,stocke2014,pointon2017,burchett2016,burchett2018}. Further, there have been studies of the cool gas traced by \mgii\ absorption in individual groups and clusters \citep{whiting2006,lopez2008,kacprzak2010,gauthier2013}. Recently, with the advent of wide-field optical integral field unit (IFU) spectrographs like the Multi Unit Spectroscopic Explorer \citep[MUSE;][]{bacon2010} on the Very Large Telescope (VLT), there have been several identifications of groups associated with cool gas as traced by \mgii\ absorption \citep{bielby2017a,peroux2017,fumagalli2017,peroux2019,chen2019,fossati2019b,hamanowicz2020,dutta2020}. With a more complete identification of the small-scale galaxy environment and the galaxy population down to a fixed flux limit across the entire field of view, such studies suggest that environmental interactions play a role in enhancing the cross-section of \mgii\ absorbing gas around galaxies in groups \citep{fossati2019b,dutta2020}. 

Hubble Space Telescope (HST) near-infrared (NIR) grism observations can complement optical IFU observations in constructing unbiased (without pre-selection) flux-limited samples of galaxies over a large area. For example, \citet{bielby2019} studied the association of \ovi\ absorbing gas with galaxies at $z<1.4$ using such a grism survey. Initial results from complete and deep galaxy surveys using IFU and grism observations around quasars with high quality spectra have shown that neutral hydrogen and metal absorption can arise in a variety of galaxy environments \citep{bielby2019,fossati2019b,lofthouse2020,chen2020a,bielby2020,lundgren2021,muzahid2021,beckett2021}.

In this work, we study gaseous haloes up to $z\approx2$ with a systematic focus on the galaxy environment, taking advantage of the large and complete galaxy surveys that are becoming available now in quasar fields. In particular, we use data from two large, independent surveys - the MUSE Analysis of Gas around Galaxies \citep[MAGG;][]{lofthouse2020,dutta2020,fossati2021} survey and the Quasar Sightline and Galaxy Evolution \citep[QSAGE;][]{bielby2019,stott2020} survey. The structure of this paper is as follows. First in Section~\ref{sec_data}, we describe the galaxy and absorption line samples from MAGG and QSAGE used in this work. Then, in Section~\ref{sec_galaxies_groups}, we study the distribution of \mgii\ and \civ\ absorption around galaxies and groups in MAGG and QSAGE. Next, we provide a more statistical definition of the galaxy environment and investigate the dependence of \mgii\ and \civ\ absorption on galaxy overdensities in these large surveys in Section~\ref{sec_overdensity}. Finally, we discuss and summarize our results in Section~\ref{sec_conclusion}. We adopt a Planck 15 cosmology with $H_{\rm 0}$ = 67.7\,\kms\,Mpc$^{-1}$ and $\Omega_{\rm M}$ = 0.307 throughout this work \citep{planck2016}. The distances are given in proper units unless we use the prefix "c" before the unit, in which case it refers to co-moving distance.

%
%
\section{Galaxy and absorption line data}
\label{sec_data}
\input{overview_table}

An overview of the galaxy and absorption line data from the QSAGE and MAGG surveys that are used in this work is provided in Table~\ref{tab:overview_table}. A table listing the properties of the galaxies and the associated absorption is provided in the online Supporting Information. In the following sections we describe the data and the parameters derived from them for our analysis.

\subsection{QSAGE galaxy survey}
\label{sec_qsage_galaxy}

The QSAGE survey is based on 96 orbits of HST Wide-Field Camera 3 (WFC3) observations centred on 12 quasar fields, with 8 orbits per field (HST Cycle 24 Large Program 14594; PIs: R. Bielby, J. P. Stott). The WFC3 observations consist of NIR imaging using the F140W and the F160W filters and spectroscopy using the G141 grism. The description of the WFC3 data reduction is presented in \citet[][sections 2.1 and 2.2]{bielby2019} and \citet[][sections 2.1]{stott2020}. The F140W images of the 12 quasar fields are shown in figure 10 of \citet{stott2020}, and examples of the WFC3 grism spectra are shown in figure 3 of \citet{bielby2019}. The QSAGE NIR imaging data are 90\% complete down to F140W $\approx26$ mag, as inferred from the number counts of continuum-detected sources as a function of F140W magnitude. We estimate the stellar mass completeness limit following \citet{marchesini2009}. In brief, we use the 3D-HST photometric catalog in the Hubble Ultra Deep Field \citep[HUDF;][]{skelton2014}, which consists of 33 orbits in F140W, and we scale the stellar masses of the HUDF galaxies as if they were at the F140W completeness limit of our sample. The upper 90$^{\rm th}$ percentile of this distribution can be taken as the 90\% stellar mass limit of our sample, which is $\approx10^{8}$\,\msun\ and $\approx10^{9}$\,\msun\ for old and red galaxies at $z\approx0.5$ and $z\approx1.5$, respectively.

In addition to the WFC3 data, eight of the fields have optical IFU observation from the MUSE/VLT, from our own programs (94.B-0304 - PI: R. Bielby; 1100.A-0528 - PI: M. Fumagalli; 103.A-0389 - PI: R. Bielby), or from the archive (95.A-0200, 96.A-0222, 97.A-0089, 99.A-0159 - PI: J. Schaye; 96.A-0303 - PI: C. Peroux). The MUSE data were reduced following the description provided in section 2.3 of \citet{bielby2019}. Being assembled from different programs, the MUSE data are heterogeneous in terms of integration time and observing conditions. The on-source exposure times range from 1 to 5 h, and the seeing varies from 0.6 to 1.2 arcsec. 

All the fields have supporting optical ($g, r, i, z$) imaging data from VLT FOcal Reducer/low dispersion Spectrograph 2 \citep[FORS2;][]{appenzeller1998}, William Herschel Telescope (WHT) auxiliary-port camera \citep[ACAM;][]{benn2008}, Liverpool Telescope (LT) Infrared-Optical: Optical (IO:O) instrument \citep{steele2004}, and the Canada–France–Hawaii Telescope Legacy Survey (CFHTLS) Wide survey \citep{cuillandre2012}. These imaging data are obtained from a combination of our own and archival programs (W13BN5, PI: R. Crain; W14AN16, PI: R. Crain; W17AP6, PI: R. Bielby; 101.A-0815, PI: Bielby). The 80\% completeness estimates based on simulated point and disk-like sources placed in the images are $r\approx24.7-26.2$ mag and $r\approx24.3-25.9$ mag, respectively. For details of the optical imaging data reduction and properties we refer to section 2.4 of \citet{bielby2019} and Bielby et al., in prep.

The source catalog was generated based on the F140W stacked image using {\sc SExtractor} \citep{bertin1996}. The 1D spectra from both the WFC3 grism data and the MUSE cubes were extracted using the catalog and segmentation maps produced by {\sc SExtractor} \citep[see section 2.1.1 of][]{bielby2019}. The redshifts of the sources identified in MUSE data were derived based on the 1D spectra and spectral template fitting using {\sc marz} \citep{hinton2016}. A quality flag was assigned to each redshift based on the categorization given in section 2.3.2 of \citet{bielby2019}. In this work, we use only the sources with quality flag 3 and 4 (about 66\% of all the sources), i.e. sources with a single high signal-to-noise ratio (S/N) emission line with some low S/N features or multiple low S/N emission lines (flag 3), and sources with multiple high S/N emission and/or absorption lines (flag 4). 

The redshift identification of all the WFC3 sources was carried out using a line fitting algorithm and visual inspection as explained in section 2.2 of \citet{stott2020}. The sources were further given a quality flag from 1 to 4, going from poor to good quality spectra. About 74\% of the sources with redshifts have quality flag 3 and 4, i.e. with S/N of spectra $>$3 and $>$10, respectively. The full QSAGE source catalogue with redshifts will be released in Bielby et al., in prep. For the purpose of this work, we redid the redshift identification of the WFC3 sources, with quality flag 3 and 4 and with redshifts less than that of the quasar, using {\sc marz}, to have a uniform redshift identification across the full sample used here. We visually inspected each spectrum and checked the reliability of the redshift identifications. The new redshifts are consistent with the ones in the original catalogue within an average scatter of $\approx$500\,\kms.

The WFC3 grism spectra are low-resolution ($R\approx130$), with a nominal velocity uncertainty of $\approx$1000\,\kms\ at the median redshift of the galaxy sample, $z=1$ \citep{momcheva2016}, while the MUSE spectra have better velocity accuracy ($\approx50$\,\kms\ at $z=1$). Comparing the redshifts obtained from WFC3 grism and MUSE spectra for the subset of sources which have both, we find that in practice the velocity uncertainty of the grism redshifts is closer to $\approx500$\,\kms\ at $z=1$. Hence, we adopt this as the uncertainty in the redshifts of the WFC3 sources.

\begin{figure}
 \includegraphics[width=0.48\textwidth]{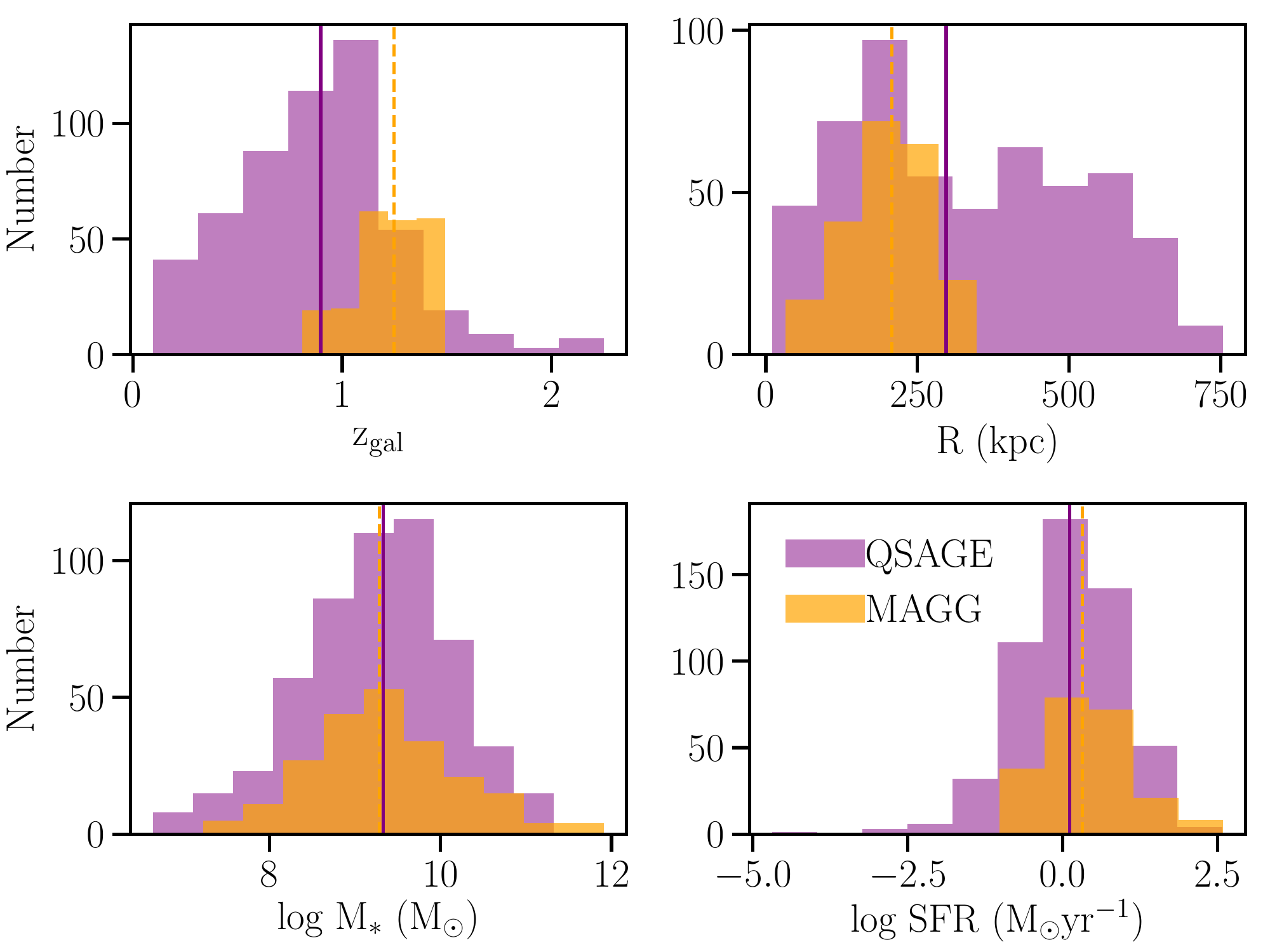}
 \caption{Comparison of the properties of the QSAGE and MAGG galaxies with coverage of associated \mgii\ absorption. The histograms show the distribution of redshift, impact parameter, stellar mass and SFR for the QSAGE (purple) and MAGG (orange) galaxies. The median values are marked by solid and dashed vertical lines for the QSAGE and MAGG samples, respectively. The QSAGE survey increases the sample size by more than a factor of three and extends the analysis to lower redshifts and higher impact parameters.
 }
 \label{fig:qsage_magg_galprop}
\end{figure}

Next, we derived the stellar mass and star formation rate (SFR) of the sources using the Monte Carlo Spectro-Photometric Fitter \citep[{\sc mc-spf};][]{fossati2018}. We follow the procedure described in \citet{fossati2019b} and \citet{dutta2020}. In brief, we use stellar population models from \citet{bruzual2003} at solar metallicity, nebular emission lines from the models of \citet{byler2018} and the initial mass function (IMF) from \citet{chabrier2003}. {\sc mc-spf} then fits the NIR and optical photometry, and also jointly fits the MUSE spectra whenever available. We correct for Galactic extinction following \citet{schlafly2011} and the \citet{fitzpatrick1999} extinction curve. In this work, we focus on the galaxies that are intervening, i.e. with redshifts less than ${\rm z}_{\rm QSO}  - (1 + {\rm z}_{\rm QSO})\times$5000\,\kms$/c$, and those with F140W magnitude below the 90\% completeness limit (26 mag). This sample spans the redshift range, $z\approx0.1-2.2$, stellar mass range, \mstar\ $\approx4\times10^{6}-2\times10^{11}$\,\msun, and SFR range $\approx0.001-400$\,\msunyr. The median redshift, stellar mass and SFR of this sample are 1.0, $2\times10^{9}$\,\msun\ and $1.5$\,\msunyr, respectively. The overall uncertainty in the logarithm of stellar mass is typically between 0.1 and 0.2\,dex \citep{mendel2014}. The halo mass is estimated from the stellar mass of the galaxies following the relation of \citet{moster2013}, and we adopt the radius within which the average mass density is 200 times the mean matter density of the Universe ($R_{\rm 200}$) as the virial radius ($R_{\rm v}$). Fig.~\ref{fig:qsage_magg_galprop} shows the distributions of the redshift, impact parameter, stellar mass and SFR of the QSAGE galaxies with coverage of \mgii\ absorption lines in the quasar spectra.

Note that over the redshift range $z\approx0.6-1.6$, we can estimate the SFR using the \ha\ fluxes measured from the grism spectra, while at lower redshifts we can use the \ha\ or \oii\ fluxes measured from the MUSE spectra whenever available. Overall, the SFR derived from the emission lines and that obtained from {\sc mc-spf} follow each other (scatter of $\approx$0.4\,dex). However, given that the \ha\ fluxes measured from the grism spectra are blended with contributions from \nii\ and that the \oii\ fluxes are affected by extinction, we adopt the SFR estimated by {\sc mc-spf} for the whole sample for the sake of uniformity in the analysis here. We explore the effect of using the SFR based on emission lines on our results in Appendix~\ref{appendix_sfr}. We find that the choice of indicator does not affect the results significantly.

\subsection{QSAGE quasar spectra}
\label{sec_qsage_qsospec}
\input{spectra_table}

The twelve quasars (\zqso\ = 1.2-2.4) in the QSAGE survey were all selected to have HST Space Telescope Imaging Spectrograph \citep[STIS;][]{kimble1998,woodgate1998,riley2018} archival spectra \citep[see for example figure 9 of][]{bielby2019}. We supplemented these with HST Cosmic Origins Spectrograph \citep[COS;][]{osterman2011,green2012} Far-Ultraviolet (FUV) and Near-Ultraviolet (NUV) spectra for four quasars, and high-resolution optical spectra for ten quasars. The details of the archival quasar spectra are listed in Table~\ref{tab:spectra_table}.

The UV spectra were retrieved from the Mikulski Archive for Space Telescopes (MAST). The STIS spectra were obtained using the E230M echelle spectrograph ($R$ = 30000). The COS spectra were obtained as part of the COS Absorption Survey of Baryon Harbors \citep[CASBaH;][]{tripp2011,burchett2019,prochaska2019,haislmaier2021}, using the FUV G130M and G160M gratings ($R$ = 12000-20000), and the NUV G185M and/or G225M gratings ($R$ = 16000-24000). The STIS and COS spectra were reduced using the {\sc calstis} and {\sc calcos} pipelines, respectively. The STIS data were reduced using the method described in \citet{tripp2001} and \citet{bielby2019}. The final STIS spectra were obtained from weighted co-addition of the overlapping regions of adjacent orders and of the individual exposures. The pipeline-reduced COS FUV 1D spectra were aligned and co-added using the IDL code\footnote{\url{https://casa.colorado.edu/~danforth/science/cos/costools}} developed by \citet{danforth2010}. The individual exposures were weighted by the integration time while co-adding. The COS FUV spectra are highly over-sampled, so we binned by three pixels to obtain the final spectra. The COS NUV spectra were reduced following the procedure given in \citet{haislmaier2021}. A smaller, 9-pixel, box was used to extract the spectra rather than the default pipeline extraction aperture, resulting in a small flux loss but improvement in the spectral S/N. All the spectra were normalized with the quasar continuum obtained by fitting a smooth spline to the absorption-free regions. The overlapping regions of the STIS and the COS NUV spectra were co-added with inverse variance weighting after interpolating to a common wavelength array.

The optical high-resolution spectra ($R\approx$ 40000) of the quasars were obtained using the High Resolution Echelle Spectrometer \citep[HIRES;][]{vogt1994} at Keck or the Ultraviolet and Visual Echelle Spectrograph \citep[UVES;][]{dekker2000} at the VLT. The reduced and continuum normalized HIRES 1D spectra were obtained from the Keck Observatory Database of Ionized Absorption toward Quasars \citep[KODIAQ;][]{omeara2015,omeara2017}. In all but one case, the reduced and continuum-normalized UVES 1D spectra were obtained from the Spectral Quasar Absorption Database \citep[SQUAD;][]{murphy2019}. For the quasar J012017$+$213346, we obtained the pipeline calibrated UVES spectrum from the European Southern Observatory (ESO) archive. The individual reduced spectra were converted into the heliocentric vacuum frame, and combined into a single spectrum by interpolating to a common wavelength array and weighting each pixel by the inverse variance using custom codes. The continuum normalization was carried out by fitting a smooth spline to the spectrum.

We conducted a blind search for \mgii\ and \civ\ absorption lines in the quasar spectra. We restricted our search up to 5000\,\kms\ blueward of the quasar emission redshift to avoid any absorption arising from the quasar proximity zone. We visually inspected the spectra to identify the \mgii\ $\lambda\lambda$2796, 2803 lines and the \civ\ $\lambda\lambda$1548, 1550 lines based on the doublet structure of the absorption profiles. We verified that the identified lines are not interloping absorption lines at other redshifts by checking that the equivalent width ratio for the unsaturated lines are approximately consistent with the expected value of $\approx$2:1, within the measurement uncertainty. Further, we checked for other common absorption lines such as \lya, \feii\ and \sifour\ at the redshift of the \mgii\ and \civ\ systems, whenever covered, to verify the line identification, though we do not require this to be a necessary criterion. 

We define absorption clumps arising within $\pm$500\,\kms\ of the line centroid as belonging to the same system. This value has been typically used in the literature to define absorption systems identified in high resolution quasar spectra, and is likely to encompass all the velocity components arising within the velocity dispersion of haloes of individual galaxies and small groups \citep[e.g.][]{churchill2000,hasan2020,dutta2020}. The redshift of the system is estimated as the optical depth weighted median redshift, where the flux is set to the error value for saturated pixels. We identified in total 28 \mgii\ systems across $z\approx0.1-1.3$, and 123 \civ\ systems across $z\approx0.1-2.4$. In this work, we characterize the absorption strength using equivalent width measurements, and we defer estimation of the physical properties of the gas using column density measurements to a future work. We measured the rest-frame equivalent width by integrating the spectra over the observed absorption profile of the stronger transition of the doublet. The equivalent widths of the \mgii\ systems (\wmgii) range between 0.004\,\AA\ to 2.3\,\AA, with a median value of 0.1\,\AA. For \civ, the equivalent widths (\wciv) range between 0.002\,\AA\ and 1\,\AA, with median \wciv\ = 0.14\,\AA. 
We determine the sensitivity to detect absorption lines by estimating the 3$\sigma$ upper limits on the rest-frame equivalent widths over 100\,\kms\ after masking out strong contaminating absorption lines. Based on the 90$^{\rm th}$ percentile of this distribution, we estimate the \wmgii\ sensitivity as $\approx$0.03\,\AA\ over the full redshift range probed. The overall \wciv\ sensitivity is $\approx$0.1\,\AA\, which is dominated by measurements based on the UV spectra at $z\le1$. The \wciv\ sensitivity is $\approx$0.03\,\AA\ at $z>1$, where the measurements are based on the more sensitive optical spectra.

\subsection{MAGG data}
\label{sec_magg_data}
\begin{figure*}
 \includegraphics[width=0.48\textwidth]{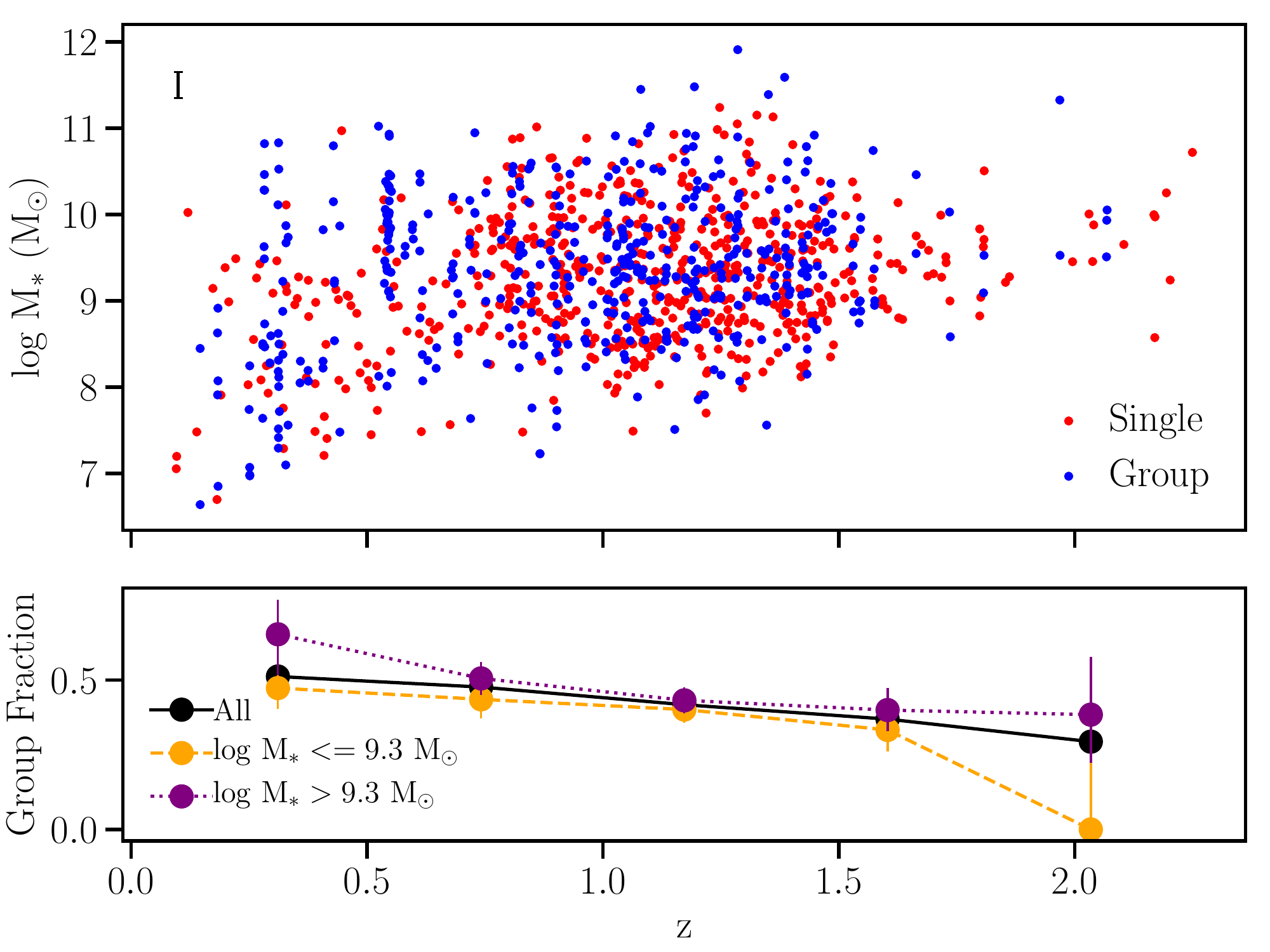}
 \includegraphics[width=0.48\textwidth]{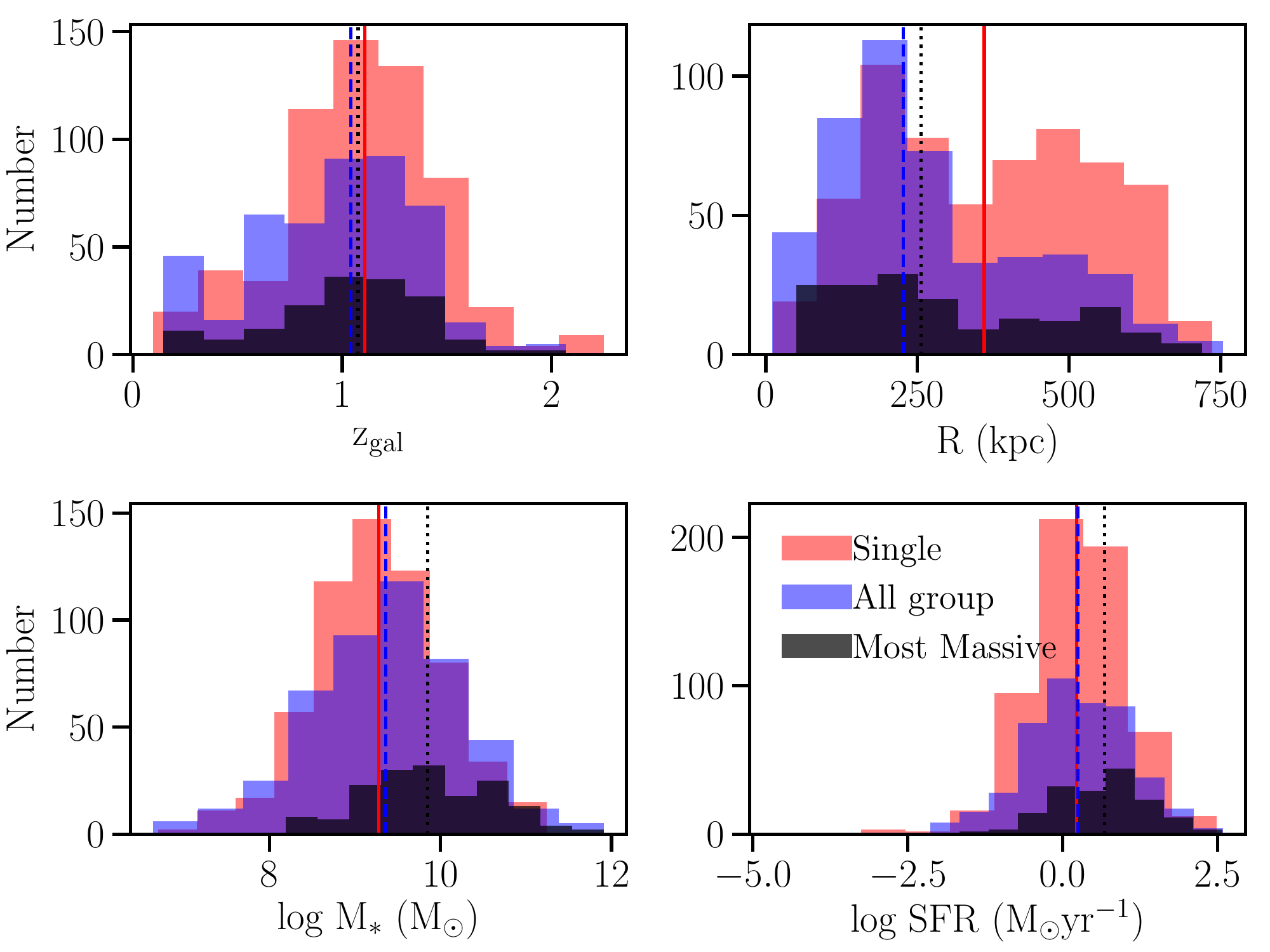}
 \caption{{\it Left:} The stellar mass of the galaxies in QSAGE and MAGG as a function of redshift is shown in the top panel. Galaxies identified in groups are shown as blue points and single galaxies are shown as red points. The typical uncertainty on the stellar mass estimates is shown in the top left corner of the plot. The errors on the redshifts are less than the size of the symbols. The bottom panel shows the fraction of galaxies that are in groups as a function of redshift. The black solid line is for all galaxies, while the orange dashed and purple dotted lines are for galaxies with stellar mass below and above $2\times10^9$\,\msun, respectively. 
 Overall 43\% of the galaxies are identified to be in groups, and the group fraction decreases slightly with increasing redshift and decreasing stellar mass.
 {\it Right:} Comparison of the properties (redshift, impact parameter, stellar mass, SFR) of single and group galaxies. The red, blue and black histograms show the distributions for single galaxies, all group galaxies and the most massive group galaxies, respectively. The solid, dashed and dotted lines mark the median properties of the single, all group and most massive group galaxies, respectively.
 The group and single galaxies have similar distributions of redshift, stellar mass and SFR, while group galaxies have smaller impact parameters on average.
 }
 \label{fig:groups_single_galprop}
\end{figure*}

The MAGG survey is primarily based upon MUSE/VLT observations of 28 fields, that are centred on quasars at \zqso\ = 3.2-4.5. The MUSE data are primarily from program ID: 197.A-0384 (PI: M. Fumagalli) with the addition of some fields from a guaranteed time observation programme \citep[PI: J. Schaye;][]{muzahid2021}. A detailed description of the reduction procedure of the MUSE data is presented in section 3.2 of \citet{lofthouse2020} and the details are tabulated in their table 1. Examples of the MUSE images and spectra can be seen in figures 7 and 10 of \citet{lofthouse2020} and figures 4 and 7 of \citet{dutta2020}. The MUSE data are 90\% complete for continuum detections down to an $r$ band magnitude of $\approx$26.3 \citep[section 5.1.1. of ][]{lofthouse2020}. These data are complemented by high-resolution ($R\approx$ 20000-50000) spectra of the background quasars obtained with UVES at the VLT, HIRES at Keck, or the Magellan Inamori Kyocera Echelle \citep[MIKE;][]{bernstein2003} at the Magellan Clay telescope, and medium resolution ($R\approx$ 4000-10000) spectra obtained with X-SHOOTER \citep{vernet2011} at the VLT or Echellette Spectrograph and Imager \citep[ESI;][]{sheinis2002} at Keck. Details of the reduction and properties of the archival quasar spectra are presented in section 3.1 and table 2 of \citet{lofthouse2020}.

The \mgii\ absorption line and galaxy samples from the MAGG survey used in this work are described in detail in sections 2.1 and 2.2 of \citet{dutta2020}, respectively. We use a sample of 218 galaxies at $0.8<z<1.5$ that are detected based on their continuum emission and \oii\ line emission in MUSE, and for which the \mgii\ absorption properties can be reliably measured from the quasar spectra. The survey does not exclude passive galaxies that can be identified via absorption lines, with a limiting stellar mass of $4\times10^9$\,\msun\ for an old and red galaxy at $z\approx1$, although we  note that emission line galaxies are more easily identified in MUSE data.
The lower redshift limit in the MAGG \mgii\ sample is set by the quasar \lya\ forest, which hinders reliable line identification. We have updated here the sample of continuum-detected galaxies used in \citet{dutta2020} with four new galaxies for which we could estimate upper limits on the \mgii\ equivalent width following a re-assessment of the spectra. We do not include here the 14 galaxies identified purely based on \oii\ line emission in the MUSE data, around the redshifts of known \mgii\ absorption systems, to have a blind galaxy sample which is consistent with the QSAGE sample.

The physical properties of the galaxies like stellar mass and SFR are estimated from the MUSE photometry and spectra, corrected for Galactic extinction, using the code {\sc mc-spf} as described in Section~\ref{sec_qsage_galaxy}. The MAGG \mgii\ sample consists of 27 systems at $0.8<z<1.5$ with \wmgii\ $\approx0.016-3.23$\,\AA\ \citep[see for an example figure 2 of][]{dutta2020}. The 90$^{\rm th}$ percentile of the distribution of 3$\sigma$ upper limits of \wmgii\ is $\approx$0.03\,\AA\ \citep[see figure 1 of][]{dutta2020}. The line identification and measurement of absorption properties are carried out in the same way as described in Section~\ref{sec_qsage_qsospec} for the QSAGE sample. In Fig.~\ref{fig:qsage_magg_galprop}, we compare the properties of the MAGG and QSAGE galaxy samples for which we can study the corresponding \mgii\ absorption. Recall that the MAGG and QSAGE samples are independent, i.e. there is no overlap in fields between the two surveys. The QSAGE survey increases the size of the sample available to study \mgii-galaxy correlation by more than a factor of three. Further, the QSAGE sample extends the analysis to lower redshifts and higher impact parameters, and also enables study of \civ-galaxy correlation at $z<2$. 

\subsection{Group identification}
\label{sec_groups}

To identify groups in the QSAGE and MAGG galaxy samples, we run a Friends-of-Friends (FoF) algorithm \citep{knobel2009,knobel2012,diener2013}. Note that we use the term "group" here to refer to an association of two or more galaxies, without any specification on the group halo mass, and these structures may not necessarily be virialized. The FoF method links galaxies into structures by finding all the galaxies that are connected within linking lengths in transverse physical distance ($\Delta r$) and redshift space ($\Delta v$). We use $\Delta r$ = 500\,kpc and $\Delta v$ = 500\,\kms\ to link the galaxies together into groups \citep[e.g.][]{knobel2009,diener2013}. Recall that we find the average uncertainty in redshifts measured from WFC3 grism spectra, on comparison with MUSE spectra, to be $\approx$500\,\kms\ (Section~\ref{sec_qsage_galaxy}), and our choice of $\Delta v$ is based on this. We also experimented with group identification using $\Delta v$ = 700\,\kms\ and 1000\,\kms, and $\Delta r$ = 400\,kpc, and found that the results presented in this work do not change significantly. The groups identified by FoF range from galaxy pairs to groups with 33 members, with a median halo mass of \mhalo\ $\approx5\times10^{11}$\,\msun. To derive the halo mass of the groups, we use the stellar mass of the most massive galaxy in each group \citep[adopted as the central galaxy; see][]{yang2008} and the redshift-dependent stellar-to-halo mass relation of \citet{moster2013}. 

The top left panel of Fig.~\ref{fig:groups_single_galprop} shows the distribution of stellar mass as a function of redshift of the galaxies, color-coded by whether they belong to a group (blue) or not (red) at the depth of our observations. About 43\% of the galaxies are determined to belong to groups. The bottom left panel shows the fraction of galaxies in groups as a function of redshift for all the galaxies and for the galaxies below and above median stellar mass $2\times10^9$\,\msun. The group fraction declines slightly with increasing redshift ($\approx48$\% group fraction at $z\le1$ compared to $\approx40$\% at $z>1$), which could be due to the increasing limiting stellar mass of the observations with redshift. Further, the group fraction increases slightly with increasing stellar mass, with overall $\approx46$\% of the more massive galaxies being in groups compared to $\approx41$\% of the less massive galaxies. The right panel of Fig.~\ref{fig:groups_single_galprop} compares the properties of the single galaxies with those of all the groups galaxies and the most massive galaxies in each group. The group galaxies have similar distributions of redshift, stellar mass and SFR as the single galaxies, but they tend to be at smaller impact parameters on average. When it comes to the most massive galaxies in groups, they tend to have higher median values of stellar mass and SFR compared to the single galaxies. 

We point out the caveat that the group identification is subject to the field-of-view (FoV) and limiting stellar mass of the observations. It is likely that some of the galaxy members of the groups at high impact parameters are outside the FoV, and that galaxies near the edge of the FoV, which are classified as single, may actually be part of groups. We might also be missing passive galaxies that belong to groups due to limitations of the redshift identification, which is more complete for emission line galaxies. Since the definition of a group depends on the depth and completeness of the galaxy sample, it is not trivial to quantitatively compare the group fraction estimated in our relatively deep galaxy data to that found in studies in the literature with different observational parameters. However, we mention here the groups fractions estimated in the literature for different galaxy samples for reference. At low redshift ($z\le0.2$), \citet{eke2004} found that $\approx55$\% of galaxies in the 2dFGRS survey are in groups of at least two members using FoF algorithm. Similarly, \citet{tempel2012} estimated that $\approx46$\% galaxies are in groups of two or more members at $z\le0.2$, using a FoF group catalog based on Sloan Digital Sky Survey (SDSS). In the zCOSMOS 20k group catalog, the fraction of galaxies in groups of two or more members, based on FoF algorithm, is found to be $\approx30-40$\% at $z\le0.6$, and this drops to $\le10$\% at $z\approx1$ \citep{knobel2012}. On the other hand, in the DEEP2 survey, using the Voronoi-Delaunay method, \citet{gerke2012} estimated a group fraction of $\approx14$\% over $z\approx0.7-1.5$. The higher fraction of galaxies in groups that we find at $z\ge1$ could be due to the relatively greater depth of our observations, which enable us to identify groups down to \mhalo\ $\approx10^{11}$\,\msun, as well as due to different methods used to link galaxies into groups in the literature.

%
%
\section{Correlation of metal absorbers and galaxies}
\label{sec_galaxies_groups}

Leveraging the larger sample afforded by QSAGE, we can extend the investigation of the gas-galaxy connection presented in \citet{dutta2020} using MAGG data alone to lower redshifts and the more highly ionized gas phase traced by \civ. We follow a galaxy-centric approach in this work. We do not place any constraint on the impact parameter, i.e. we consider the full FoV of MUSE (up to $\approx$0.5\,arcmin or $\approx$250\,kpc at $z=1$) and HST/WFC3 (up to $\approx$1.5\,arcmin or $\approx$750\,kpc at $z=1$). We associate an absorption system to a galaxy if its redshift falls within $\pm$500\,\kms\ (the average redshift uncertainty of the WFC3 sources) of the galaxy redshift. In case there is no associated absorption in the above line-of-sight (LoS) velocity window, we place a $3\sigma$ upper limit on the rest-frame equivalent width per 100\,\kms\ after masking any contaminating strong absorption lines in the quasar spectra.
We checked that our results do not change significantly if we use a LoS velocity window of $\pm$1000\,\kms. We expect that the effect of the redshift uncertainty of WFC3 sources on the results will be within the statistical uncertainties. The LoS velocity window used here is similar to values adopted in the literature to associate absorbers with galaxies \citep[e.g.][]{tumlinson2013,burchett2016,chen2018,dutta2020}.
We note that based on the {\sc eagle} cosmological simulations, \citet{ho2020} have found that a LoS velocity cut of $\pm$500\,\kms, as used here, at an impact parameter of 100\,kpc around galaxies with stellar masses of $10^{9-9.5}$\,\msun\ (similar to the median \mstar\ of our sample) at $z\approx0.3$ will select detectable \mgii\ gas (log~$N$[\mgii/\cms] $\ge$ 11.5) that lies beyond the virial radius $\approx$80\% of the time. 

We focus on the cool, metal-enriched, low-ionization gas phase as traced by \mgii\ absorption, and the more highly ionized gas phase as traced by \civ\ absorption. For the analysis of the \mgii-galaxy connection, we use the combined sample of MAGG and QSAGE, while in the case of \civ, we use only the QSAGE sample. Note that in the case of MAGG, the \lya\ forest makes it difficult to reliably identify \civ\ absorption at $z<2$ in the quasar spectra. First, in Section~\ref{sec_metals_galaxies}, we examine the dependence of the \mgii\ and \civ\ equivalent width and covering fraction on galaxy properties. The covering fraction here is defined as the ratio of the number of galaxies that show absorption above a certain equivalent width limit to the total number of galaxies with sensitivity to detect absorption above this limit \citep[see e.g.][]{dutta2020}. Then we move to a group-centric approach and compare the metal distribution in groups and single galaxies in Section~\ref{sec_metals_groups}. We further investigate the absorber-galaxy connection using the angular cross-correlation function in Section~\ref{sec_cross_corr}.

\subsection{Dependence of metal absorption on galaxy properties}
\label{sec_metals_galaxies}

\begin{figure}
 \includegraphics[width=0.48\textwidth]{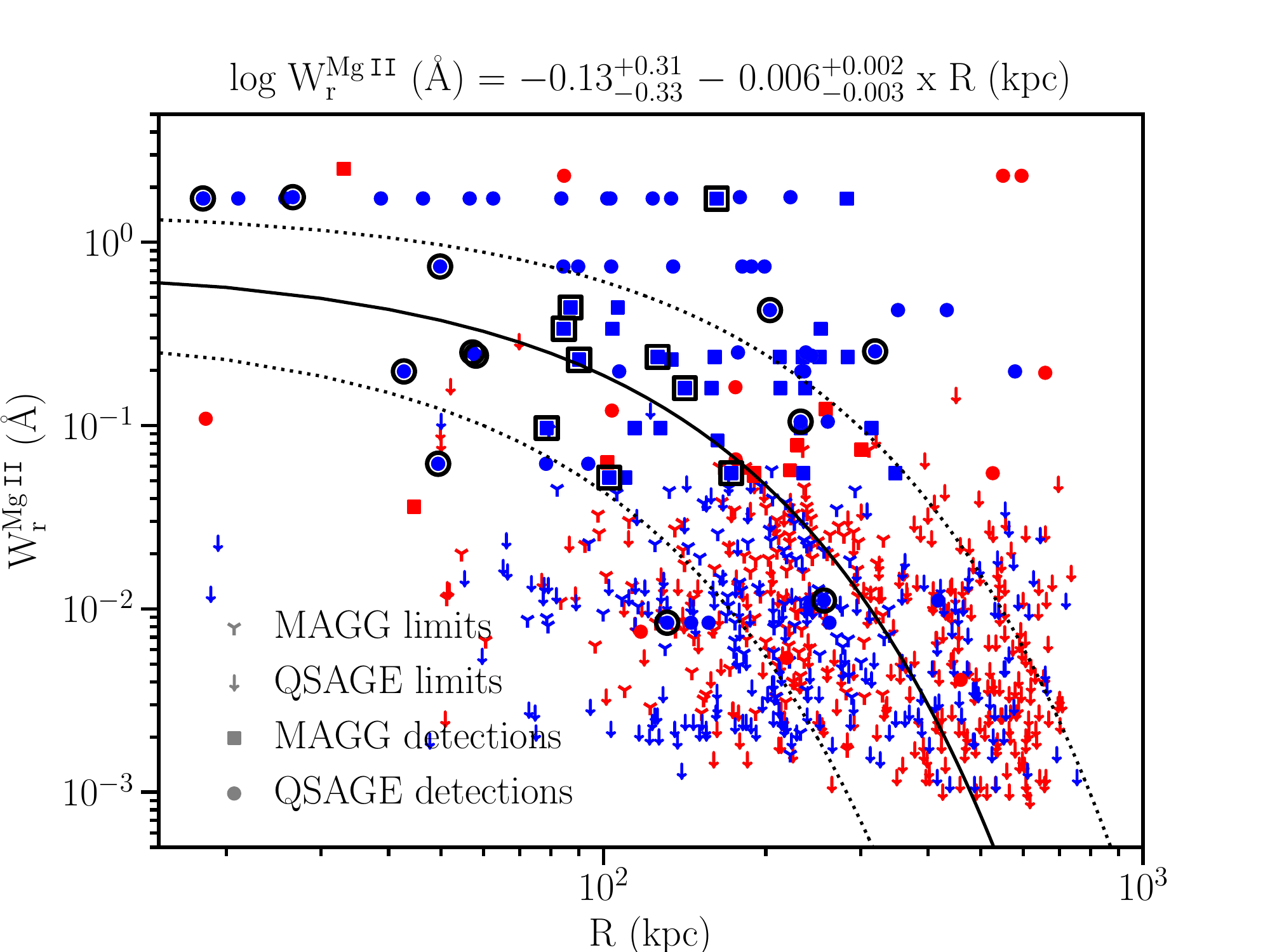}
 \caption{Rest-frame equivalent width of \mgii\ absorption as a function of galaxy impact parameter. Galaxies with detection of \mgii\ absorption from the MAGG survey are shown as squares, while those from the QSAGE survey are shown as circles. Non-detections, represented by 3$\sigma$ upper limits, are shown as arrows of different types as indicated in the figure legend. Galaxies in groups are marked in blue and single galaxies are marked in red. Note that all the galaxies in each group are associated with the same \wmgii. The galaxy with smallest impact parameter in a group in case of \mgii\ detection is marked by an outer black circle/square. The best-fit log-linear relation between \wmgii\ and impact parameter of closest galaxies is given at the top of the plot and plotted using a solid line (1$\sigma$ uncertainty with dotted lines).
 The average \wmgii\ declines with increasing impact parameter when considering the closest galaxies, although there is a considerable scatter in the relation.
 }
 \label{fig:mgii_ew_imp}
\end{figure}

The anti-correlation between the \mgii\ equivalent width, \wmgii, and the galaxy impact parameter, $R$, is a well-established relation in the literature \citep[e.g.][]{lanzetta1990,chen2010,nielsen2013}. However, as pointed out in \citet{dutta2020}, recent IFU galaxy surveys that are more flux complete and spatially uniform, are beginning to identify multiple galaxies that can be associated with a single \mgii\ system, and not just the brightest or closest ones. This is leading to ambiguity in defining a galaxy-absorber pair and larger scatter in the equivalent width--impact parameter plane of \mgii. This is reflected in Fig.~\ref{fig:mgii_ew_imp}, which shows the \wmgii\ versus impact parameter relation for the systems in MAGG and QSAGE. The points are color-coded as blue for galaxies identified to be in groups based on FoF and red for single galaxies. The \wmgii\ shows a declining trend with increasing $R$, which is more pronounced when restricting to the closest galaxies in a group. We fit a linear relation between log~\wmgii\ and $R$ of the closest galaxies in the combined MAGG and QSAGE sample using a Bayesian method that accounts for the upper limits \citep[see for details][]{chen2010,rubin2018,dutta2020}. The best fit is plotted in Fig.~\ref{fig:mgii_ew_imp}. This fit is consistent within the 1$\sigma$ uncertainties with the fit obtained by \citet{dutta2020} using only the MAGG sample at $z\approx1$, although we obtain a slightly shallower fit now due to the detection of galaxies at larger impact parameters enabled by the larger FoV of the QSAGE HST observations.

\begin{figure*}
 \subfloat[]{\includegraphics[width=0.48\textwidth]{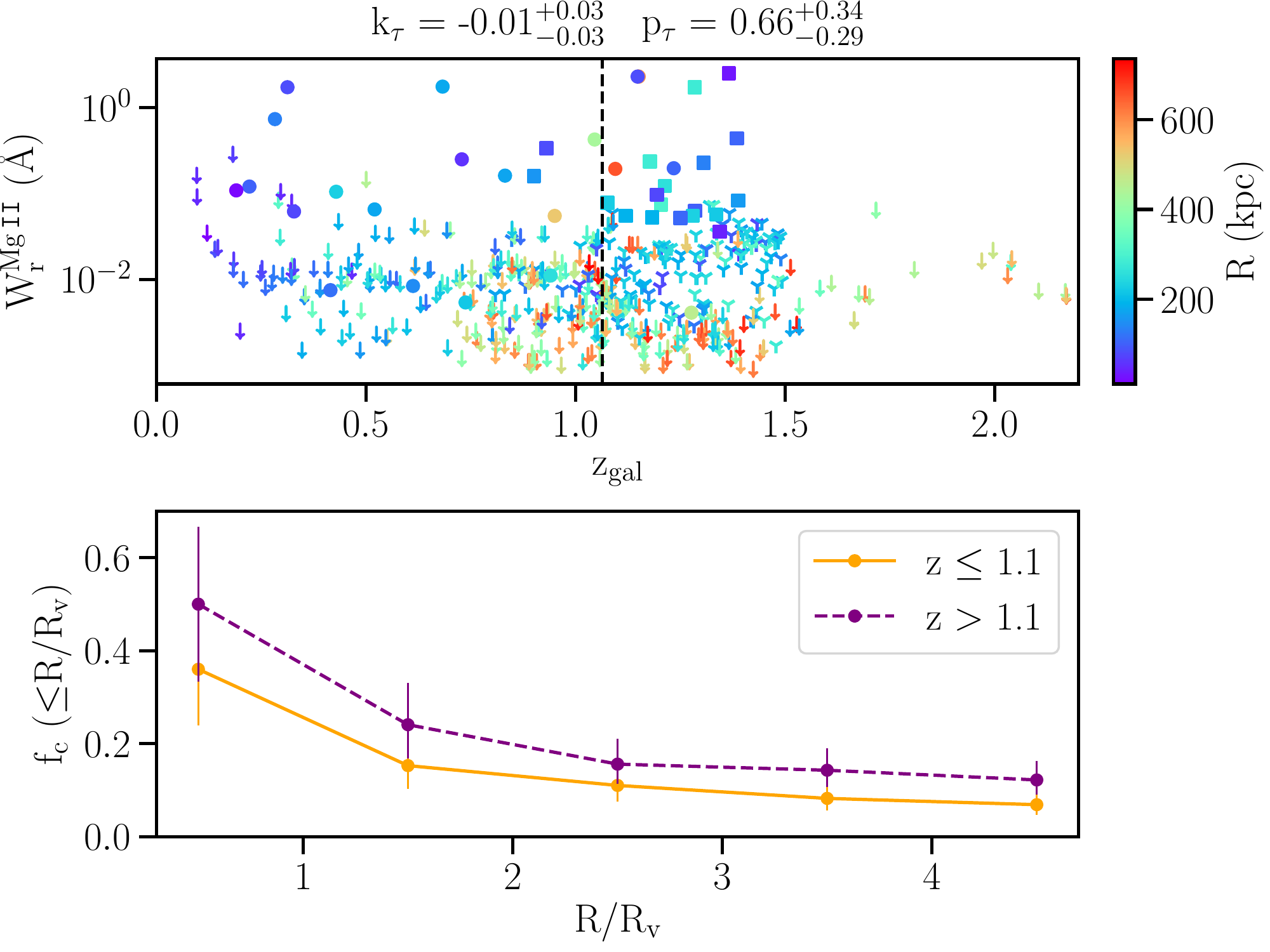}}
 \subfloat[]{\includegraphics[width=0.48\textwidth]{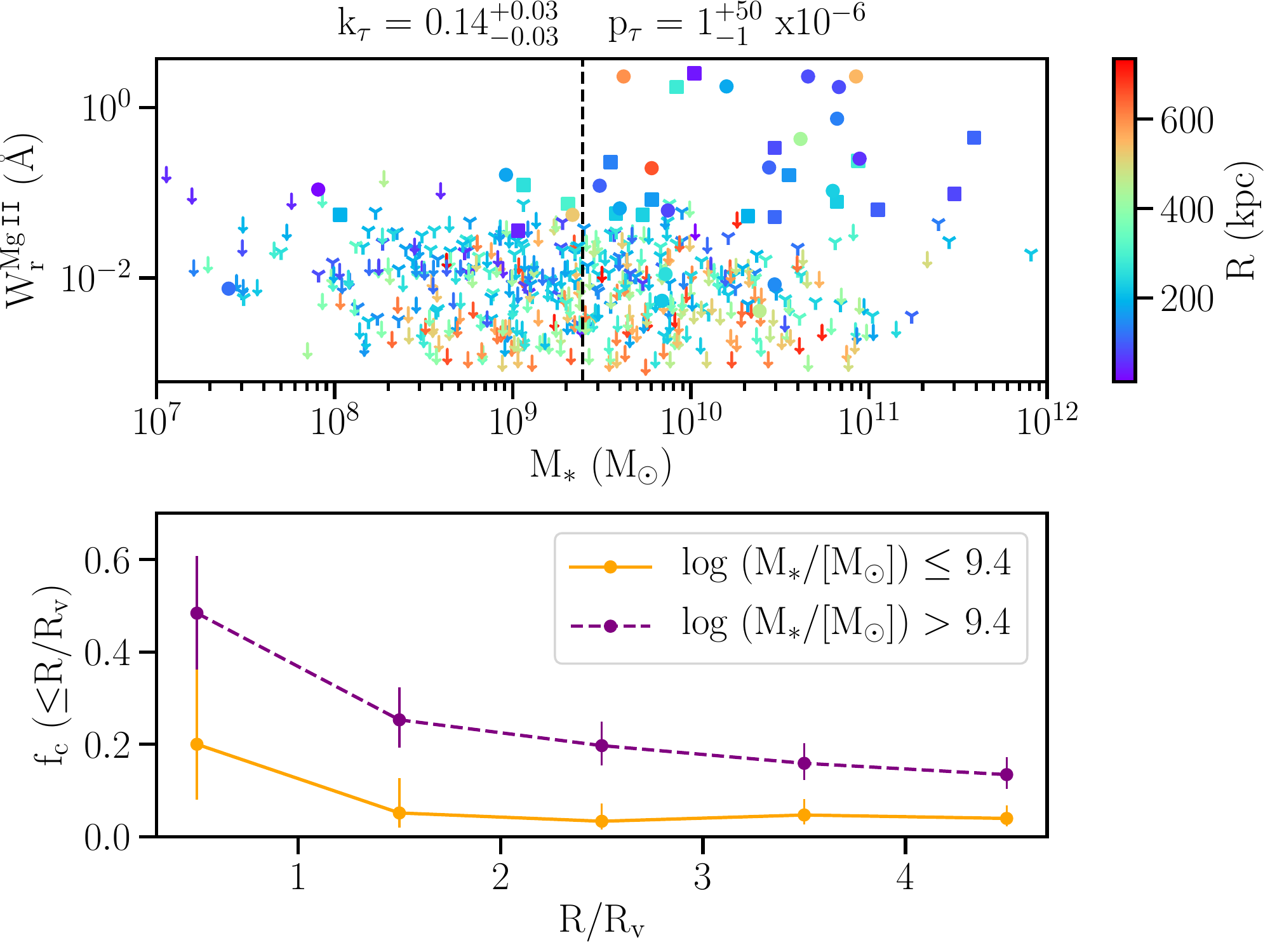}} 
 \hspace{0.01cm}
 \subfloat[]{\includegraphics[width=0.48\textwidth]{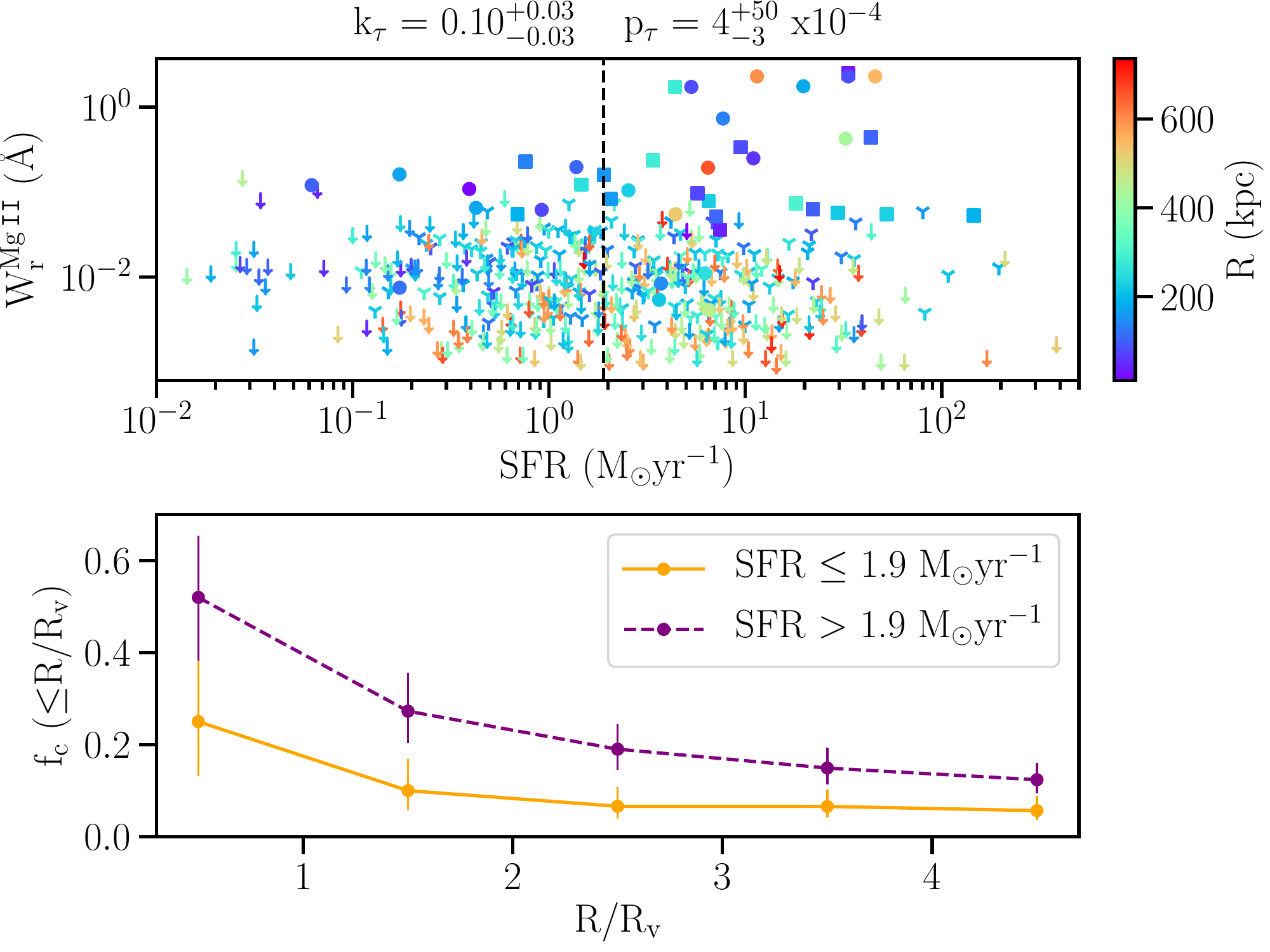}}
 \subfloat[]{\includegraphics[width=0.48\textwidth]{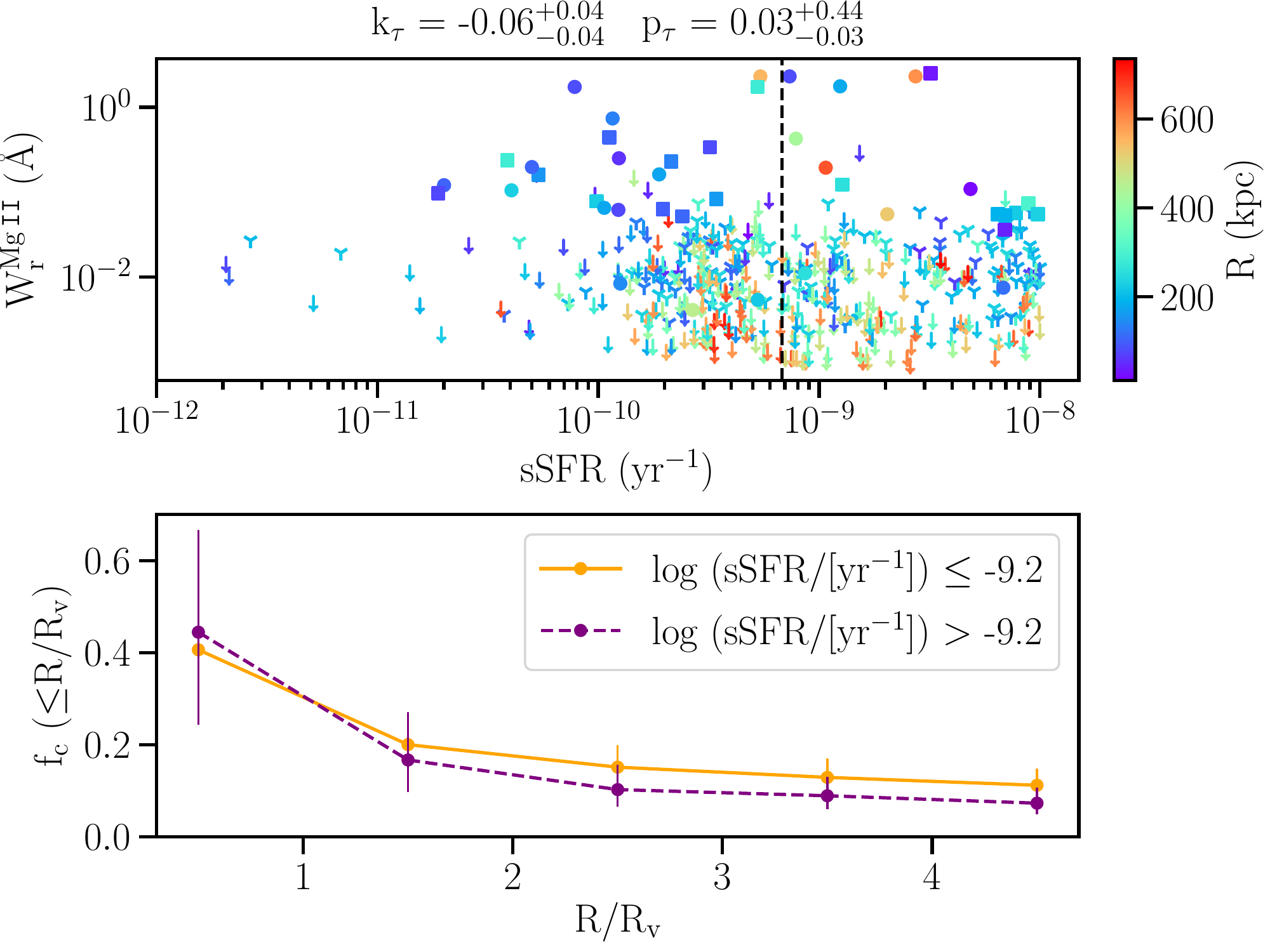}}
 \caption{Dependence of \mgii\ absorption on galaxy properties [(a) redshift, (b) stellar mass, (c) SFR, (d) specific SFR] in the combined MAGG and QSAGE sample. The top plot in each subset shows \wmgii\ as a function of the galaxy property. The symbols are the same as in Fig.~\ref{fig:mgii_ew_imp}, i.e. detections from MAGG and QSAGE are shown as squares and circles, respectively, while the upper limits are shown as arrows. The symbols are color-coded by the galaxy impact parameter. The median value of the galaxy property ($z=1.1$, \mstar\ = $3\times10^9$\,\msun, SFR = 1.9\,\msunyr, sSFR = $6\times10^{-10}$\,yr$^{-1}$) is marked by vertical dashed lines. The Generalized Kendall's k$_\tau$ and p$_\tau$ values for the correlation between \wmgii\ and the galaxy property are given on top of each plot (a smaller value of p$_\tau$ indicates a more significant correlation). The bottom panel shows the cumulative covering fraction estimated for a sensitivity limit of 0.03\,\AA\ as a function of the impact parameter normalized by the virial radius for the sub-samples below (orange solid line) and above (purple dashed line) the median galaxy properties. The error bars represent 1$\sigma$ Wilson score confidence intervals.
 The most significant correlations of \wmgii\ and covering fraction of \mgii\ are with stellar mass and SFR.
 }
 \label{fig:mgii_gal_prop}
\end{figure*}

We first look at the dependence of the cool gas distribution, as traced by \mgii\ absorption, on various galaxy properties. In Fig.~\ref{fig:mgii_gal_prop}, we show the dependence of \wmgii\ and \mgii\ covering fraction (\fc) on the galaxy redshift, stellar mass, SFR and specific SFR (sSFR = SFR/\mstar). To study the correlation with the galaxy properties, we perform the Generalized Kendall's non-parametric rank correlation test, which accounts for upper limits \citep[see][for details]{brown1974,isobe1986}. We give the Kendall's $\tau$ correlation, k$_\tau$, and the probability of no correlation, p$_\tau$, on top of each plot. Note that a small value of p$_\tau$ indicates significant correlation. We estimate the 1$\sigma$ uncertainties on the correlation coefficients and p-values using bootstrapping. To estimate the covering fraction, we use a \wmgii\ limit of 0.03\,\AA, to which $\approx$90\% of the measurements are sensitive. The covering fraction is shown as a function of the impact parameter normalized by the virial radius. The relative trends are the same if instead we plot it as a function of the impact parameter. In the case of galaxies in groups, we have considered the most massive galaxy, which we assume to be the central galaxy. We find that the trends discussed below hold for the individual MAGG and QSAGE samples as well, though at a lesser significance than for the combined sample.

The equivalent width of \mgii\ shows no strong trend with redshift, but its covering fraction is $\approx$1.5-2 higher at higher redshifts, i.e. $z>1$. This is consistent with the evolution in covering fraction of strong absorbers (\wmgii\ $>1$\,\AA) over $0.4<z<1.3$ found by \citet{lan2020} based on SDSS data. Using MUSE observations, \citet{schroetter2021} have found a similar evolution over $1.0<z<1.5$ for \wmgii\ $\ge0.8$\,\AA. The dependence on redshift could also be related to the dependence on stellar mass (see below), since the sample lacks low-mass galaxies (\mstar\ $<10^8$\,\msun) at higher redshifts ($z>1$; see Fig.~\ref{fig:groups_single_galprop}). To check this, we estimated the covering fractions considering only galaxies with stellar mass above $10^9$\,\msun, where the distribution across redshifts is more consistent. We find that the average covering fraction appears higher at $z>1$, but this difference is not statistically significant as the two are consistent within the 1$\sigma$ uncertainty.

Both the \wmgii\ and \fc\ exhibit a significant dependence on the stellar mass and SFR, with the more massive and higher star forming galaxies showing $\approx$2 times stronger and more prevalent absorption, even beyond the virial radius. Further, we find that these trends are present in sub-samples at $z<1$ and at $z>1$, albeit at lower significance (p$_\tau$ $\approx10^{-2}-10^{-4}$). Several studies in the literature, based on different methodology including stacking and background galaxy spectroscopy in addition to quasar spectroscopy, have reported dependence of \mgii\ absorption properties on galaxy mass, color, and SFR \citep[e.g.][]{chen2010,bordoloi2011,nielsen2013,lan2014,rubin2018}. Compared with some of the above studies, we find weaker dependence on the specific SFR. Since the stellar mass and SFR of the galaxies in our sample are correlated, to distinguish between the effects of the two, we divide the sample into two sub-samples based on the median stellar mass of $3\times10^9$\,\msun. We find that the \wmgii\ and \fc\ do not show a significant dependence on either SFR or sSFR in these sub-samples. This seems to imply that the stellar mass or halo mass plays a more dominant role in determining the cool gas distribution around galaxies, as also noted by \citet{dutta2020}. However, we do note that the present sample is dominated by star forming galaxies, lacking a significant fraction of passive galaxies, with less than 10\% of the galaxies having sSFR $<10^{-10}$\,yr$^{-1}$. 

\begin{figure}
 \includegraphics[width=0.48\textwidth]{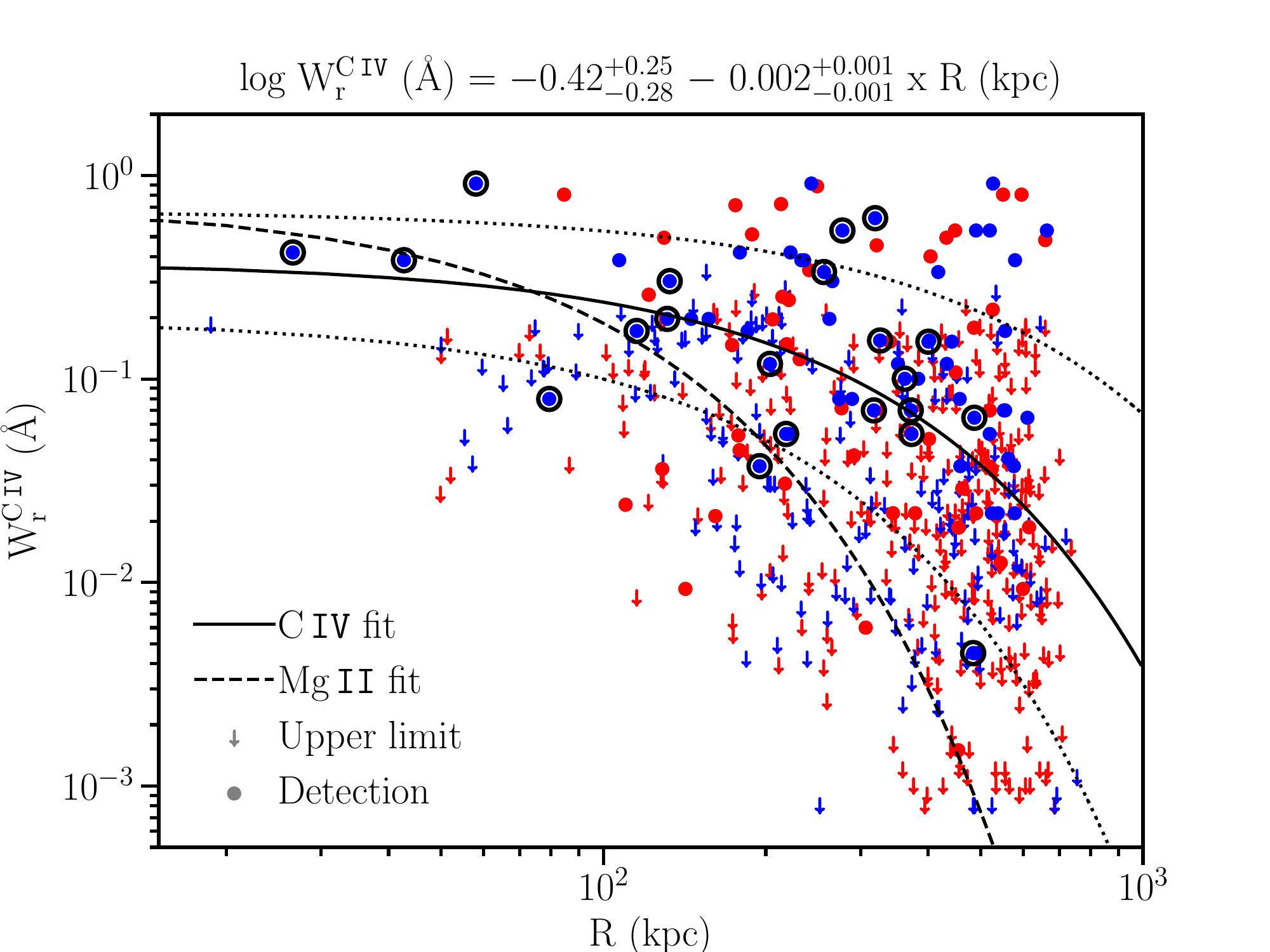}
 \caption{Rest-frame equivalent width of \civ\ absorption versus the galaxy impact parameter. Detections are shown as circles and 3$\sigma$ upper limits corresponding to non-detections are shown as arrows. Galaxies in groups are shown in blue, while single galaxies are shown in red. The galaxy with the smallest impact parameter in each group is marked by an outer black circle in case of a \civ\ detection. The best log-linear fit between \wciv\ and impact parameter of closest galaxies is plotted with a solid line, with the 1$\sigma$ uncertainty indicated by dotted lines. The fit is also given at the top of the plot. The best fit between \wmgii\ and impact parameter from Fig.~\ref{fig:mgii_ew_imp} is shown with a dashed line for comparison.
 The average \wciv\ declines with increasing impact parameter, but the relation is shallower than in the case of \mgii, implying a more extended distribution of \civ\ gas.
 }
 \label{fig:civ_ew_imp}
\end{figure}

Now we move to the more highly ionized phase of the gas as traced by \civ\ absorption in the QSAGE survey. Fig.~\ref{fig:civ_ew_imp} shows the equivalent width of \civ, \wciv, as a function of the galaxy impact parameter. We follow the same color scheme as for \mgii. We find that in the case of \civ\ as well, a single absorption system can be associated with multiple galaxies in a group out to large distances. On average, the \civ\ equivalent width declines with increasing distance from galaxies. As in the case of \mgii, we fit a log-linear relation between \wciv\ and impact parameter (see Fig.~\ref{fig:civ_ew_imp}). The fit to the \civ\ absorption is shallower than that to the \mgii\ absorption around galaxies. This could imply that the \civ\ absorbing gas has a more extended radial profile around galaxies than the \mgii\ absorbing gas. This would support the picture of the low-ionization gas being more centrally concentrated than the high-ionization gas phase. Note that the shallower fit to the \civ\ absorption could also arise due to a lack of strong upper limits on the equivalent width. The \civ\ upper limits at $z<1$ are estimated from the UV spectra, in contrast to the upper limits for \mgii\ which are estimated from the more sensitive optical spectra.

Most studies of galaxies associated with \civ\ absorption at $z<1$ in the literature have been based on imaging and long/multi-slit spectroscopy \citep[e.g.][]{chen2001,bordoloi2014,liang2014,burchett2016}. All these studies report an anti-correlation between \civ\ equivalent width and impact parameter. However, \citet{burchett2016}, using HST-COS observations of \civ\ absorbers at $z<0.05$ and galaxies in SDSS down to \mstar\ $\approx10^8$\,\msun, have also found \civ\ absorption at large impact parameters ($\approx$400\,kpc). Based on MUSE observations, \citet{muzahid2021} have found no strong dependence of \civ\ absorption on transverse distance over $\approx$50-250\,kpc around $z\approx3$ \lya\ emitters. In our sample, we find that \civ\ absorption can be present at large impact parameters around galaxies, beyond their virial radii. At these large distances, it becomes more difficult to discern whether and how the \civ\ absorption is physically connected to the galaxies.

\begin{figure*}
 \subfloat[]{\includegraphics[width=0.48\textwidth]{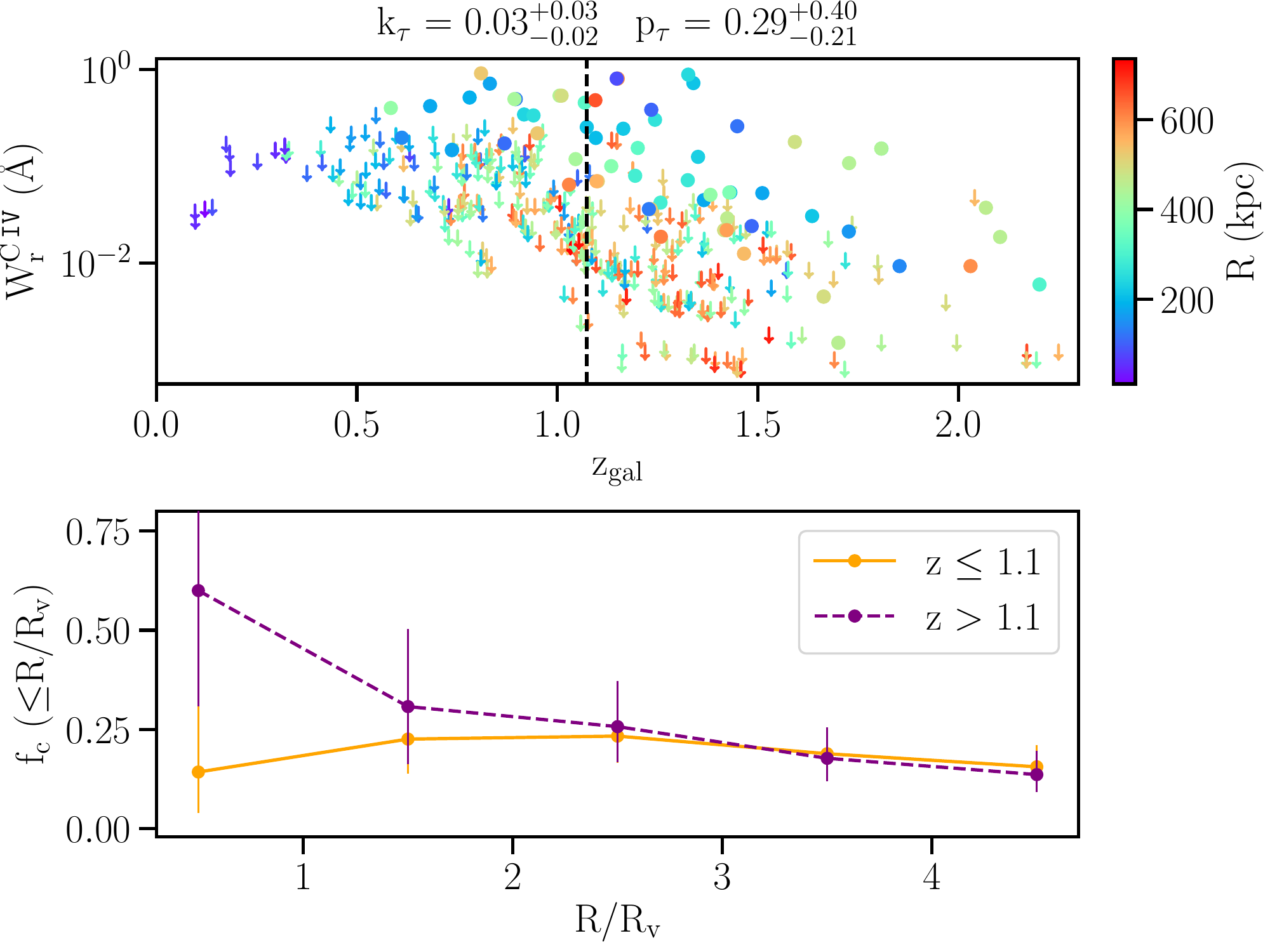}}
 \subfloat[]{\includegraphics[width=0.48\textwidth]{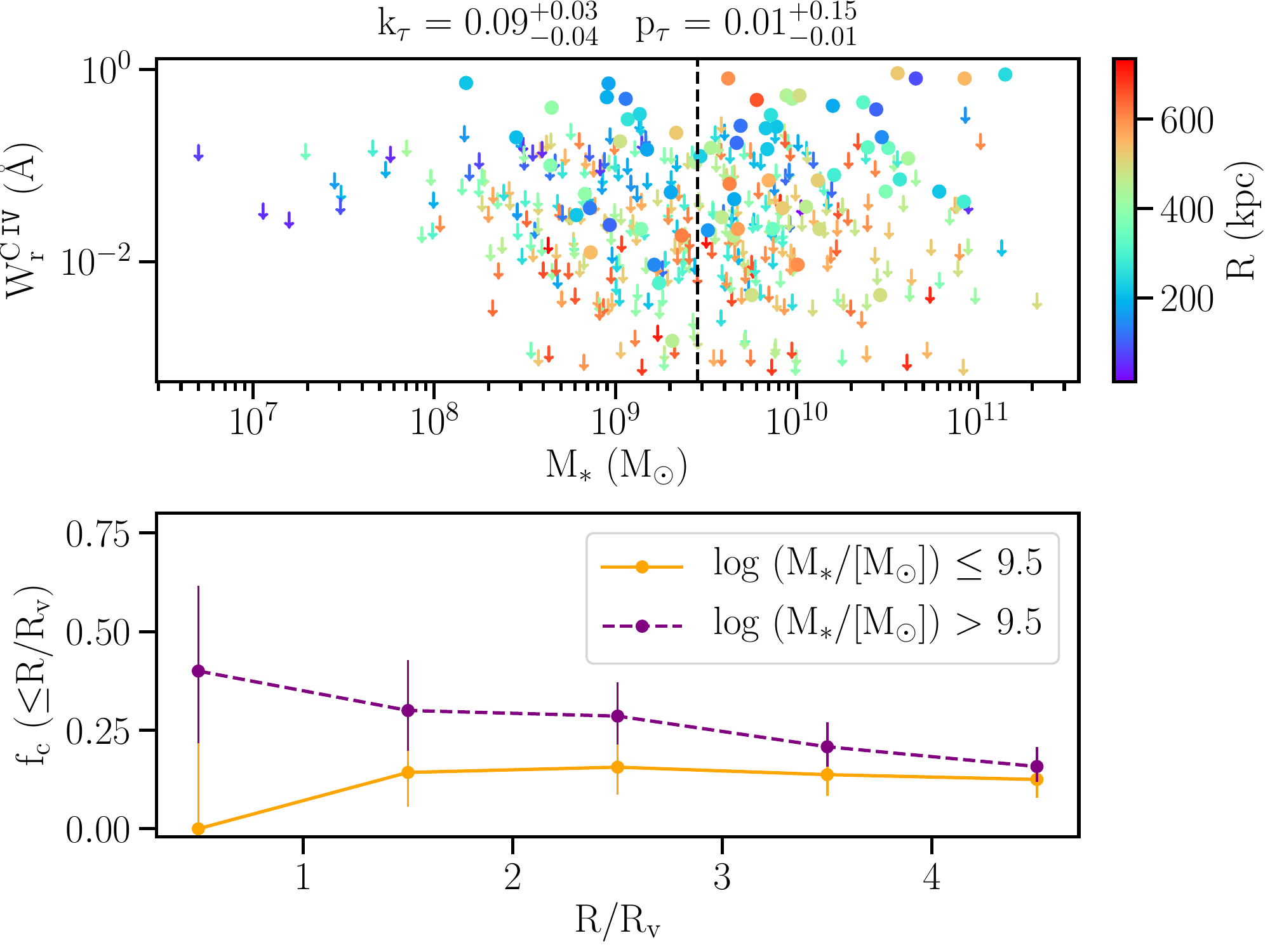}} 
 \hspace{0.01cm}
 \subfloat[]{\includegraphics[width=0.48\textwidth]{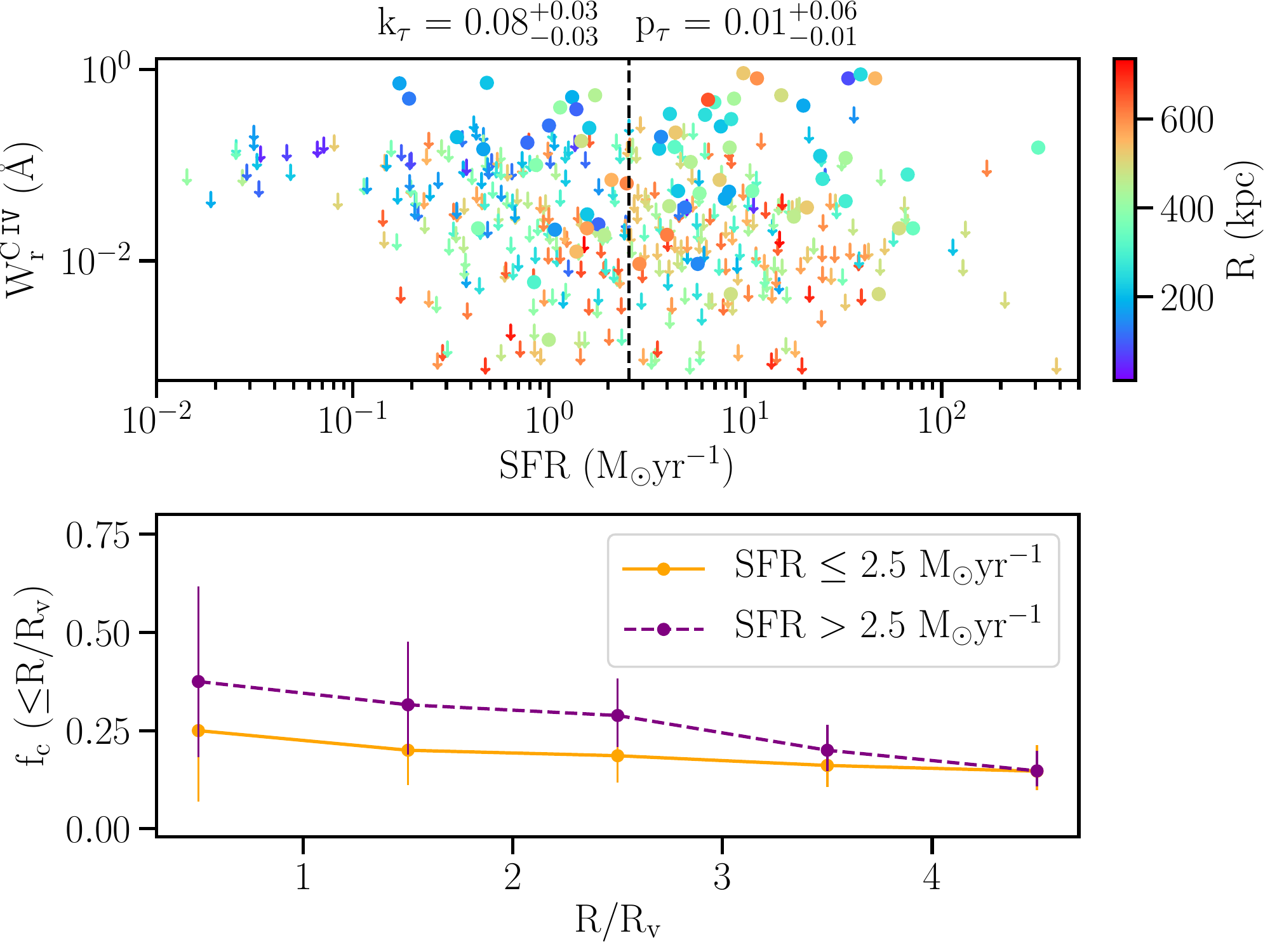}}
 \subfloat[]{\includegraphics[width=0.48\textwidth]{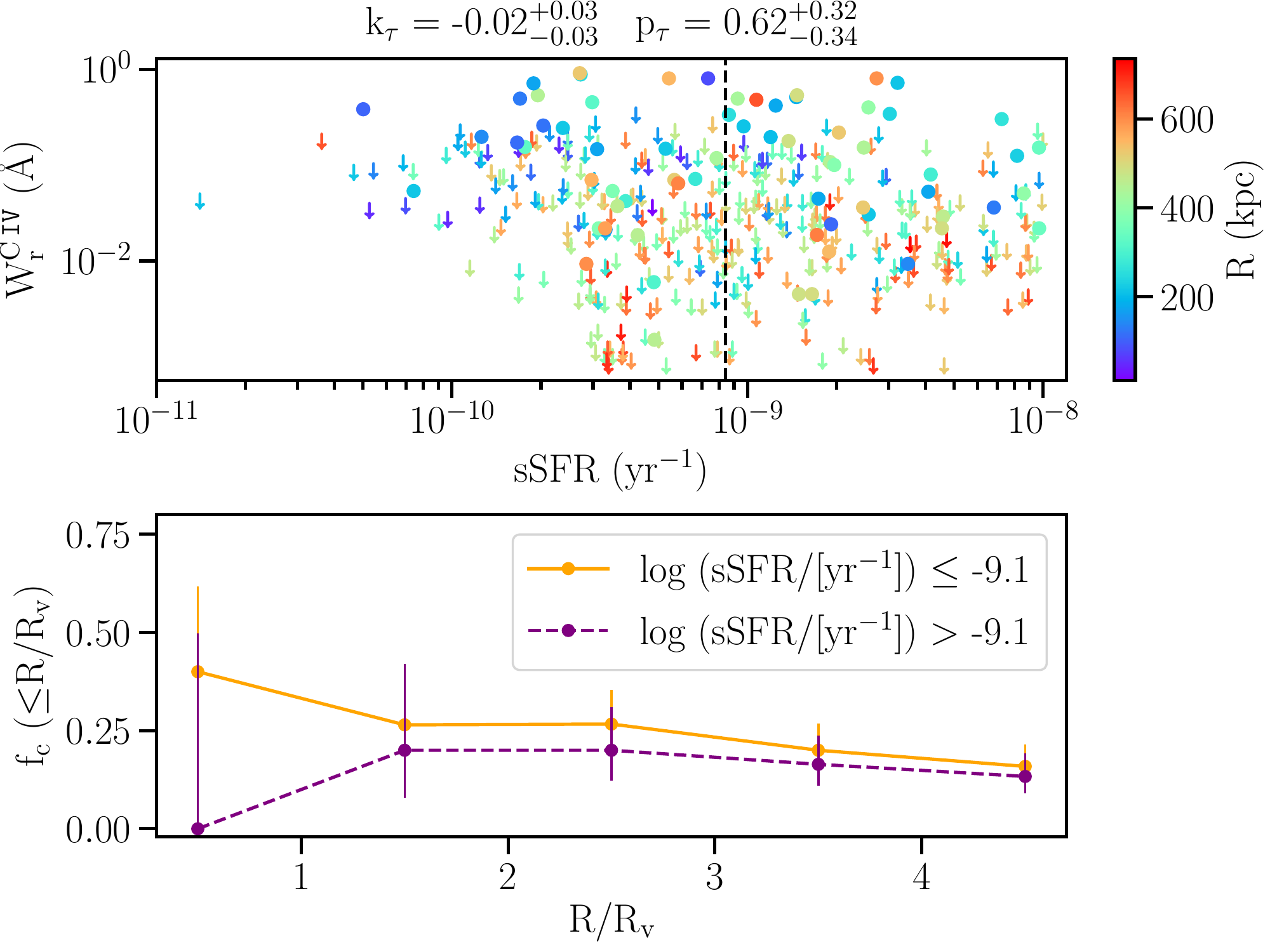}}
 \caption{Same as Fig.~\ref{fig:mgii_gal_prop} but showing the dependence of \civ\ absorption on galaxy properties in the QSAGE sample. The symbols are the same as in Fig.~\ref{fig:civ_ew_imp}. Here the covering fraction has been estimated for a sensitivity limit of 0.1\,\AA.
 The covering fraction of \civ\ tends to increase on average with redshift, stellar mass and SFR.
 }
 \label{fig:civ_gal_prop}
\end{figure*}

To try to address this connection, as done for \mgii, we look at the dependence of \wciv\ and \civ\ covering fraction on galaxy properties in Fig.~\ref{fig:civ_gal_prop}. For the covering fraction analysis we use a \wciv\ limit of 0.1\,\AA, based on the 90\% sensitivity limit of the sample. Note that at lower redshifts, i.e. $z\lesssim1$, the \civ\ lines are detected in the UV spectra as opposed to the optical spectra at higher redshifts. Due to the higher S/N and resolution of the optical spectra, we are able to probe to deeper \wciv\ at higher redshifts. The equivalent width of \civ\ shows no strong redshift evolution. However, the covering fraction of \civ\ is $\approx$4 higher at $z>1$ within the virial radius, beyond which there is no significant difference. The dependence of covering fraction on redshift is similar when considering only the galaxies with \mstar\ $\ge10^9$\,\msun. We note that \citet{schroetter2021} have found no significant evolution in \civ\ covering fraction over $z\approx1.0-1.5$ for \wciv\ $\ge$0.7\,\AA\ using MUSE observations.

Similar to the \mgii\ gas, the covering fraction of \civ\ gas shows increasing trends with stellar mass and SFR ($\approx$1.5-2 times higher \fc\ around the more massive and star-forming galaxies). The \civ\ equivalent width is only weakly dependent on the stellar mass and SFR, and the dependence is much less significant than in the case of \mgii. The results are similar for the sub-samples at $z<1$ and at $z>1$. There are comparatively few studies in the literature on the association of \civ\ with galaxy properties, especially at $z\approx0.5-2$, as most of the studies are at $z\lesssim0.1$ \citep{borthakur2013,bordoloi2014,burchett2016} or at $z\ge2$ \citep{adelberger2005,prochaska2014,muzahid2021}. \citet{burchett2016} have found that the \civ\ equivalent width and covering fraction within the virial radius are significantly higher for the more massive galaxies with \mstar\ $>3\times10^9$\,\msun, which agrees with our results. On the other hand, \citet{borthakur2013} and \citet{bordoloi2014} have found a tentative correlation between \civ\ equivalent width and specific SFR. At $z\approx3$, \citet{muzahid2021} have not found a significant dependence of the stacked \civ\ absorption around \lya\ emitters on their SFR. We do not find a dependence of \civ\ absorption on the specific SFR in the QSAGE sample, with the caveat that there is a dearth of passive galaxies with sSFR $<10^{-10}$\,yr$^{-1}$. Further, we find no significant dependence of \wciv\ and \fc\ on either SFR or sSFR when considering sub-samples above and below the median stellar mass. Overall, the distribution of the more highly ionized gas phase around galaxies seems also to be determined primarily by the stellar mass, although the strength of absorption from the gas phase traced by \civ\ appears to be less tightly correlated to the properties of the individual galaxies than the less ionized phase traced by \mgii.  

Recall that we have used different equivalent width limits of 0.03\,\AA\ and 0.1\AA\ for estimating the covering fractions of \mgii\ and \civ\ gas, respectively, based on sensitivity of the quasar spectra, to make optimal use of the available data. If instead we use a limit of 0.1\AA\ for estimating the \mgii\ covering fraction, to match that used for \civ, we find that the trends discussed above and shown in Fig.~\ref{fig:mgii_gal_prop} are valid at similar significance levels. For the sake of clarity, we do not show the covering fraction for different equivalent width limits in Figs.~\ref{fig:mgii_gal_prop} and \ref{fig:civ_gal_prop}. However, as expected, we find that the covering fraction at a fixed impact parameter decreases as we go for a higher cut in equivalent width, i.e. the incidence of stronger absorbers at a given distance from galaxies decreases. We note that the same equivalent width would correspond to different gas densities for \mgii\ and \civ\ because of differences in oscillator strength, rest wavelength and relative solar abundance. However, to empirically compare the distribution of the different absorbing gas at the same sensitivity of the data, we look at the covering fractions of \mgii\ and \civ\ gas at the same equivalent width limit. We find that the covering fraction profiles of \civ\ gas are generally flatter than those of the \mgii\ gas. While the average covering fraction of \mgii\ gas within two to four virial radii drops below 10\%, that of \civ\ gas stays around 15-20\%. This is in line with the more extended distribution of \civ\ absorption around galaxies compared to \mgii\ absorption hinted at by the equivalent width versus impact parameter plots (see discussion above).

\subsection{Dependence of metal absorption on environment}
\label{sec_metals_groups}
\begin{figure*}
 \includegraphics[width=0.48\textwidth]{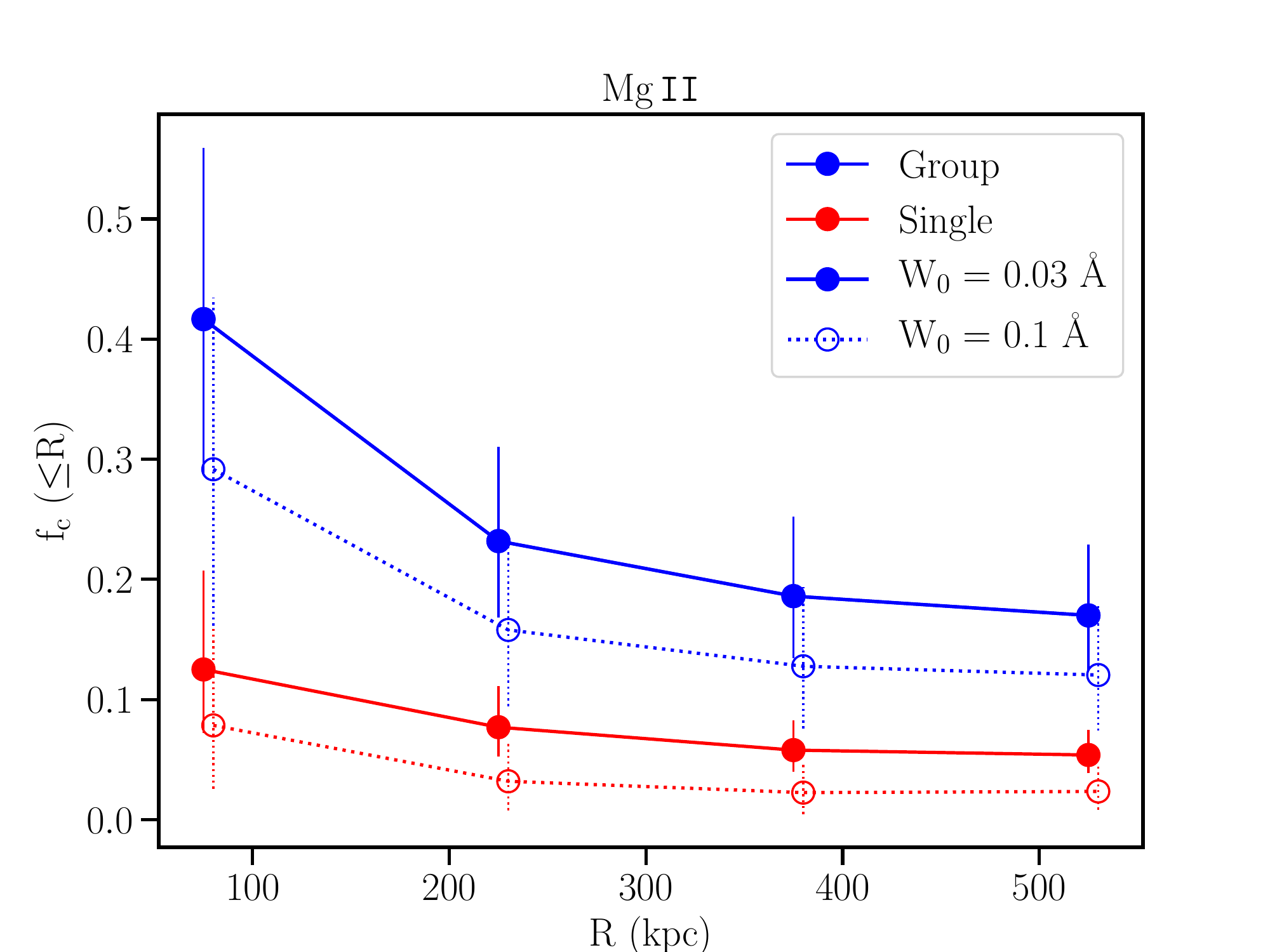}
 \includegraphics[width=0.48\textwidth]{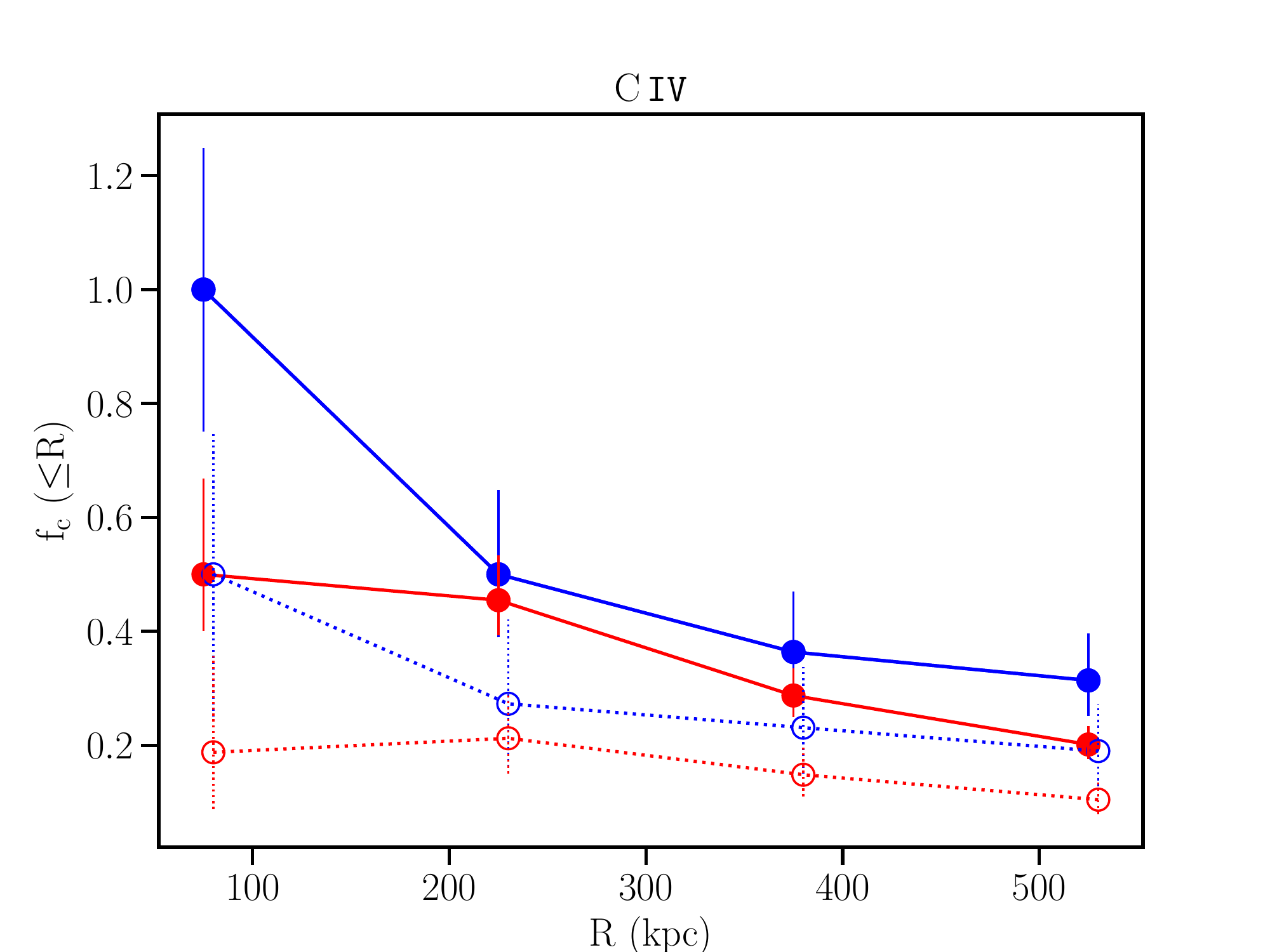}
 \includegraphics[width=0.48\textwidth]{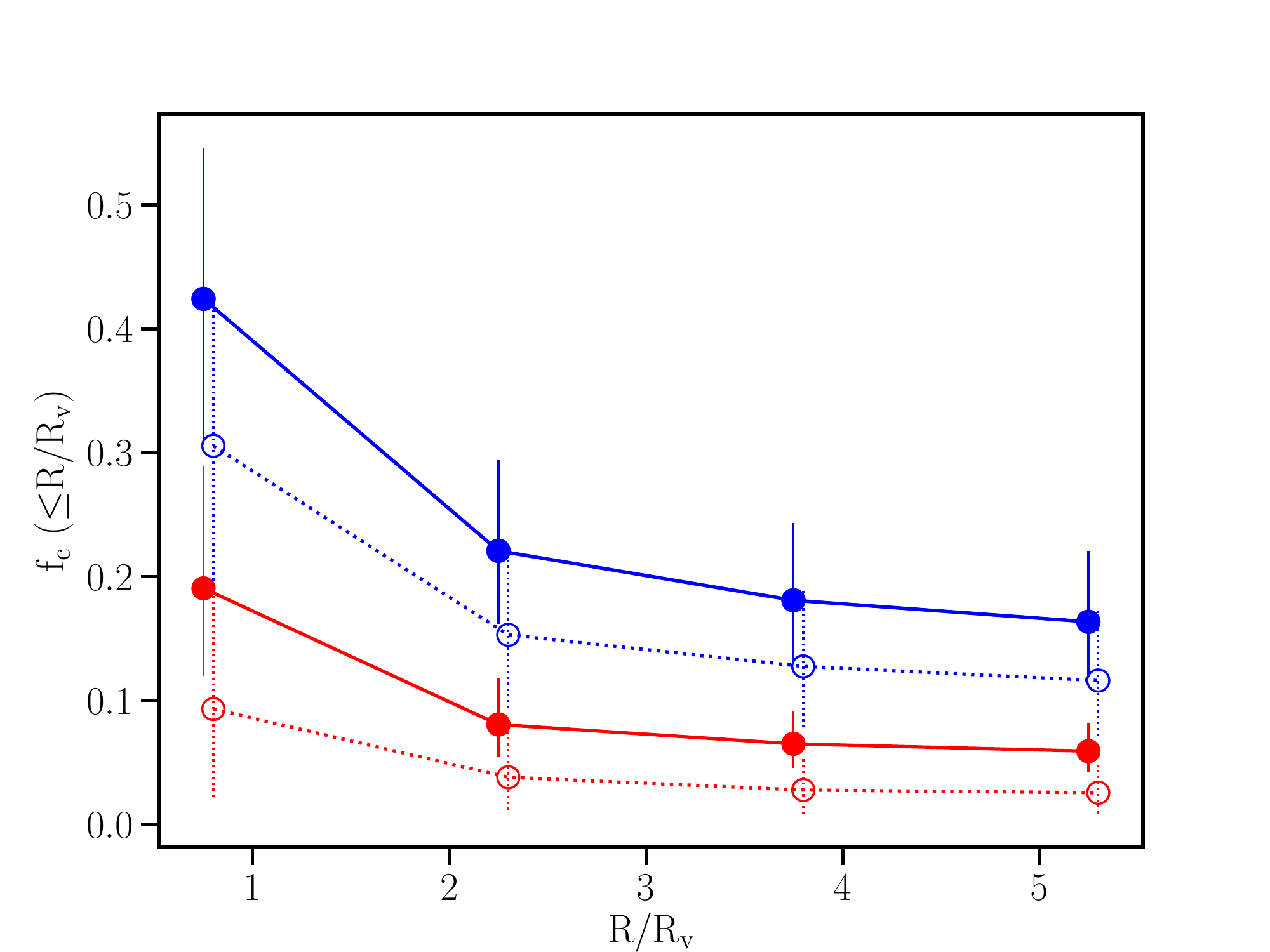}
 \includegraphics[width=0.48\textwidth]{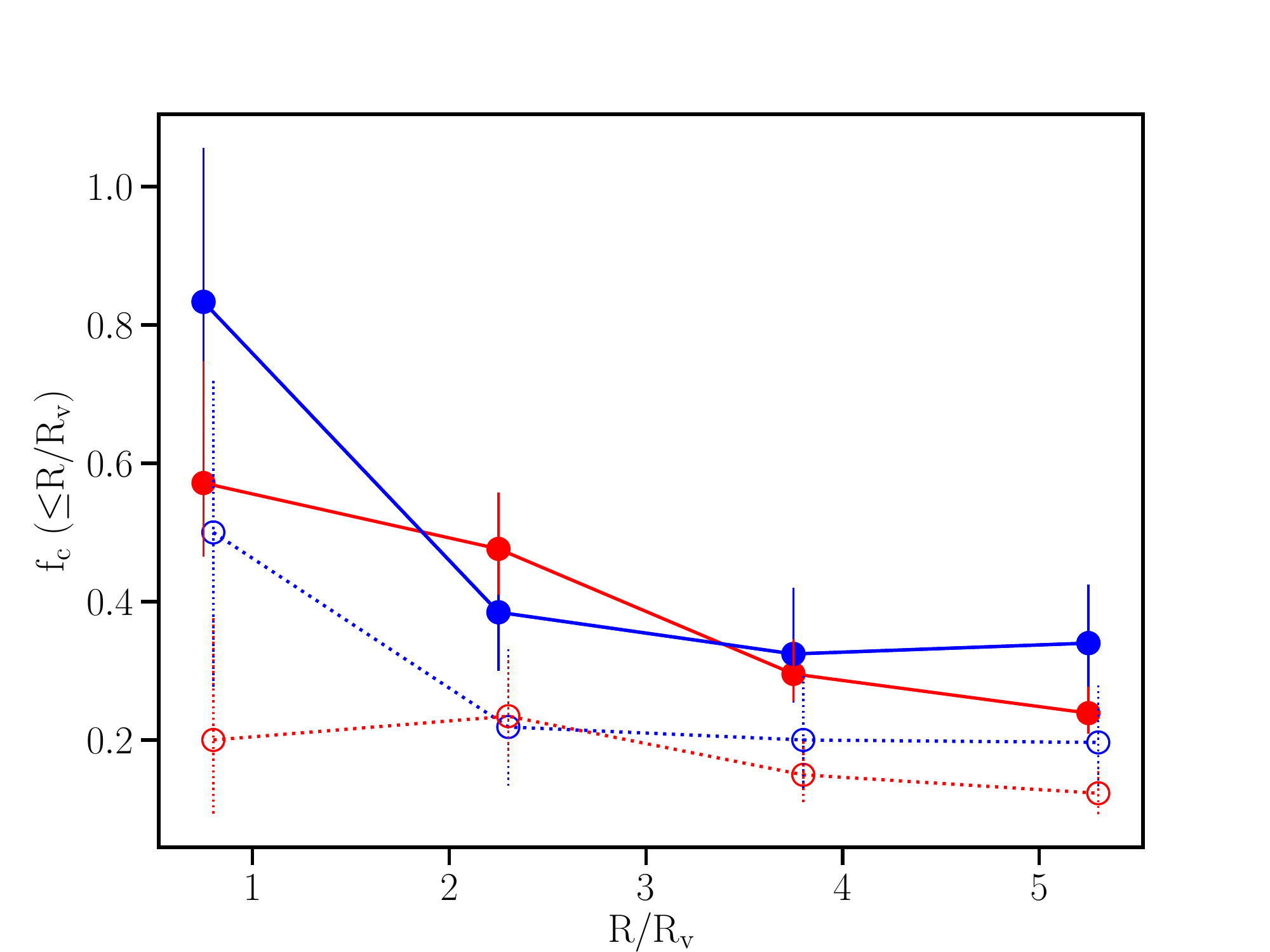}
 \caption{Covering fraction of \mgii\ (left) and \civ\ (right) absorption as a function of impact parameter (top) and impact parameter normalised by the virial radius (bottom), for most massive galaxies in groups in blue and single galaxies in red. The solid and dotted lines show the covering fraction estimated for a sensitivity limit of 0.03\,\AA\ and 0.1\,\AA, respectively. Some of the points have been shifted slightly along the x-axis for the purpose of clarity. The error bars represent 1$\sigma$ Wilson score confidence intervals. 
 Groups show $\approx2-3$ times higher covering fraction of \mgii\ than single galaxies at a given distance and sensitivity, while their \civ\ covering fractions are not significantly different.
 }
 \label{fig:cf_grp}
\end{figure*}

We have come to appreciate that the star formation activity and morphology of galaxies depend on the local environment, with the fraction of star-forming spiral galaxies decreasing from the field to the cores of clusters, compensated by an opposite increase of passive elliptical galaxies \citep[e.g.][]{dressler1980,balogh2004,kauffmann2004,baldry2006,fossati2017}. Similarly, it is now becoming evident from recent studies that, in order to understand what shapes the distribution of gas and metals around galaxies, the galaxy environment needs to be taken into account in addition to internal galaxy properties. In particular, there have been recent IFU observations of group environment giving rise to \mgii\ absorption \citep[e.g.][]{bielby2017a,peroux2019,chen2019}, and studies have found evidence that the properties of the cool, low-ionization gas in more overdense environments are different from those seen in more isolated environments \citep[e.g.][]{nielsen2018,fossati2019b,dutta2020,lundgren2021}. 
Using the MAGG sample, \citet{dutta2020} found that the \mgii\ equivalent width and covering fraction are higher around group galaxies than around single galaxies. They argue that tidal interactions between group members or ram pressure stripping by the intra-group medium could be dislodging the metals from haloes of individual galaxies and enhancing the overall strength and cross-section of cool gas in denser environments. Such environmental interactions are analogous to what have been seen in MUSE observations of extended \ha\ and \oiii\ emission around galaxies in groups and clusters at $z\lesssim0.5$ \citep[e.g.][]{fumagalli2014,fossati2016,fossati2019a,johnson2018,chen2019}. 

\begin{figure*}
 \includegraphics[width=0.48\textwidth]{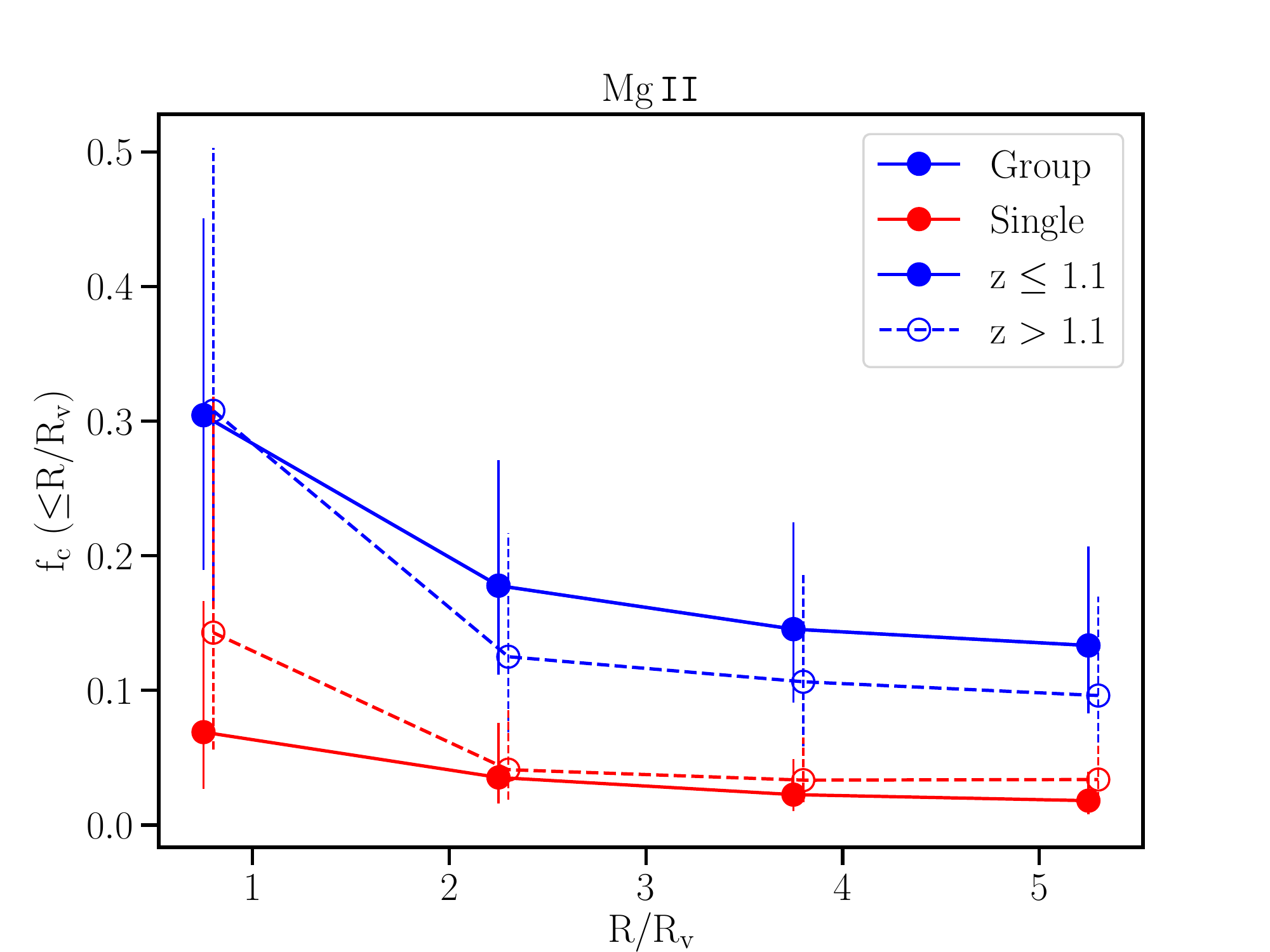}
 \includegraphics[width=0.48\textwidth]{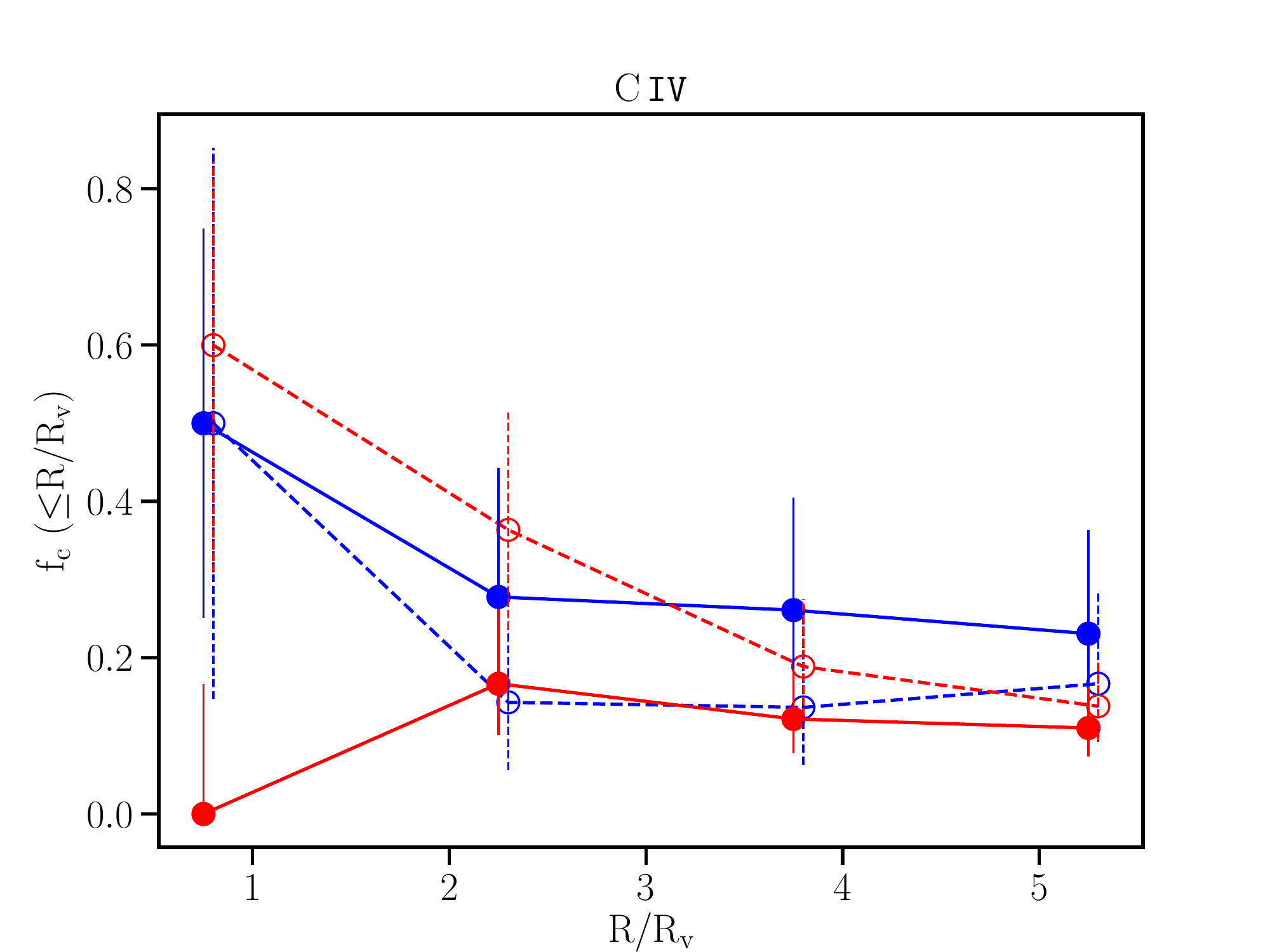}
 \caption{Covering fraction of \mgii\ (left) and \civ\ (right) absorption as a function of normalized impact parameter, for most massive galaxies in groups (blue) and single galaxies (red), estimated for a sensitivity limit of 0.1\,\AA. The covering fraction profiles at $z\le1.1$ are shown in solid lines and those at $z>1.1$ are shown in dashed lines. Some of the points have been shifted slightly along the x-axis for the purpose of clarity. The error bars represent 1$\sigma$ Wilson score confidence intervals.
 Groups show higher covering fraction of \mgii\ at all redshifts probed here, while there is no clear trend in the case of \civ.
 }
 \label{fig:cf_grp_zgal}
\end{figure*}

The combined sample of MAGG and QSAGE is three times larger and probes a wider galaxy parameter space (see Fig.~\ref{fig:qsage_magg_galprop}), so in this work we are able to not just test the results of \citet{dutta2020} but also to extend the scope of their analysis. Fig.~\ref{fig:cf_grp} (left panels) compares the covering fraction of \mgii\ gas in groups and single galaxies. The top panel shows the covering fraction as a function of impact parameter and the bottom panel shows it as a function of impact parameter normalized by the virial radius. Here we have considered the most massive galaxy in each group. The results are similar when we consider instead the closest galaxy in a group, the group geometric centre or the mass-weighted group centre. Group galaxies show higher covering fraction of \mgii\ gas for both equivalent width thresholds of 0.03\,\AA\ and 0.1\,\AA. Galaxies in groups show twice the covering fraction of single galaxies within the virial radius, and $\approx3$ times the covering fraction of single galaxies within twice the virial radius. This is consistent with the results of \citet{nielsen2018} and \citet{dutta2020}. Further, we divide the sample into sub-samples based on the median redshift, stellar mass and SFR. The difference in covering fraction of group and single galaxies tend to be more prominent at low redshift ($z\le1$), and for more massive (\mstar\ $>3\times10^9$\,\msun) and more star-forming (SFR $>$1.9\,\msunyr) galaxies, although the differences are consistent within the large uncertainties. In Fig.~\ref{fig:cf_grp_zgal}, we show the covering fraction profiles for the low and high redshift sub-samples, estimated for equivalent width threshold of 0.1\,\AA.

\begin{figure*}
 \includegraphics[width=0.48\textwidth]{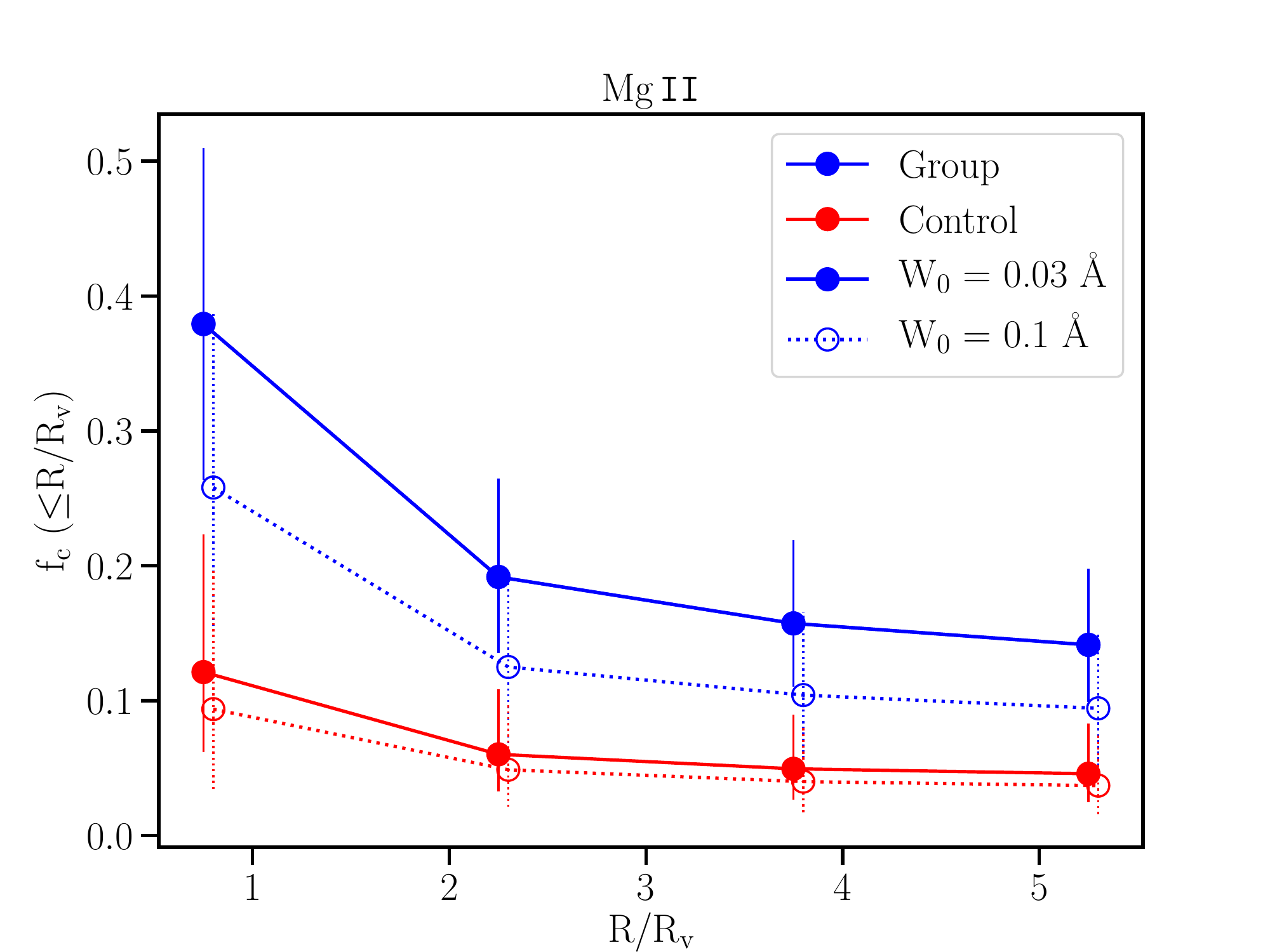}
 \includegraphics[width=0.48\textwidth]{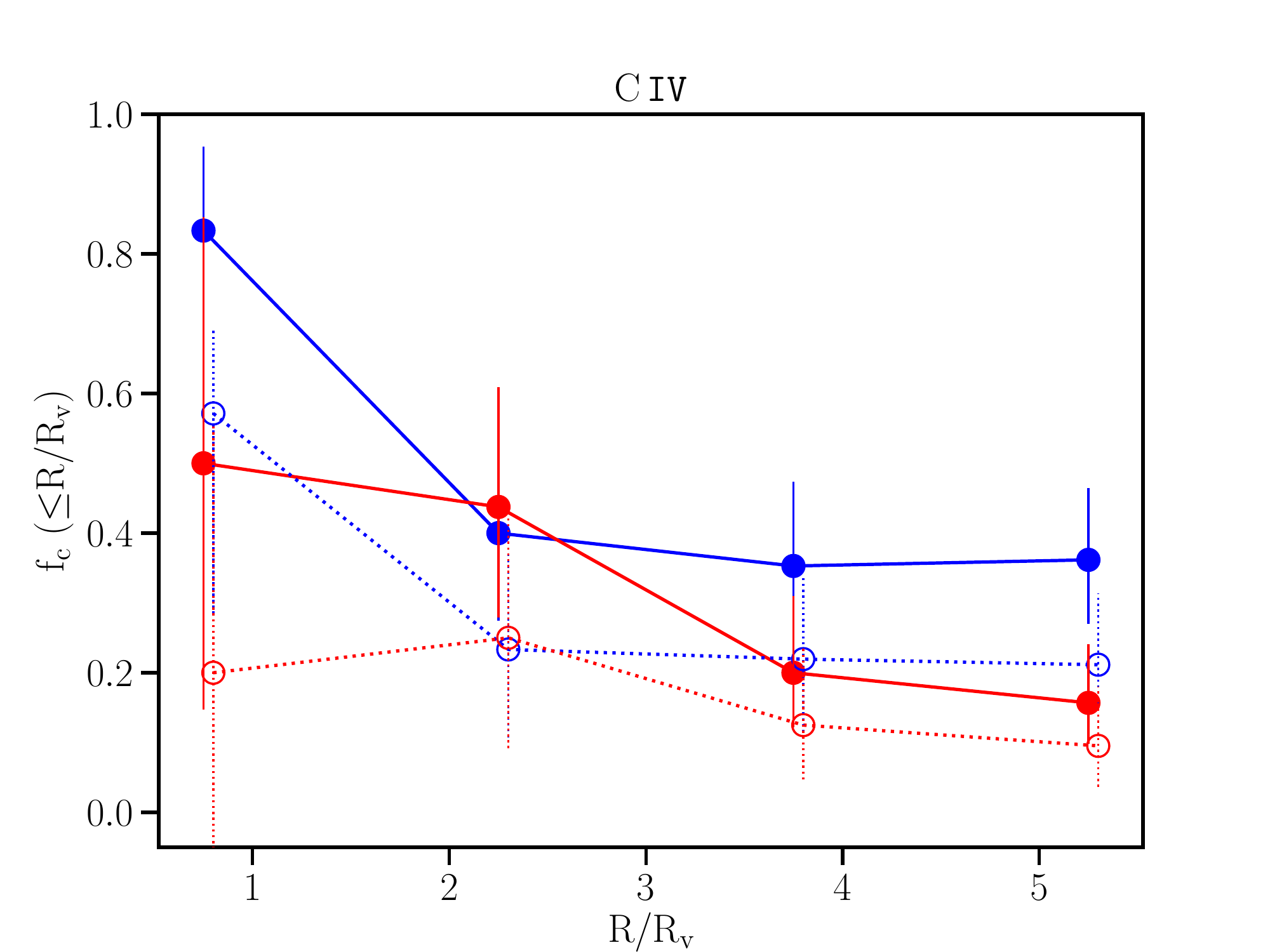}
 \caption{Covering fraction of \mgii\ (left) and \civ\ (right) absorption as a function of normalized impact parameter, for most massive galaxies in groups (blue) and control sample (red) of single galaxies matched in impact parameter ($\pm$50\,kpc), stellar mass ($\pm$0.3\,dex) and redshift ($\pm$0.3). The solid and dotted lines show the covering fraction estimated for a sensitivity limit of 0.03\,\AA\ and 0.1\,\AA, respectively. Some of the points have been shifted slightly along the x-axis for the purpose of clarity. The error bars represent 1$\sigma$ Wilson score confidence intervals.
 The covering fraction of \mgii\ remains elevated in groups even after controlling for other galaxy parameters. There is no significant difference in covering fraction of \civ\ between group and control samples.
 }
 \label{fig:cf_grp_control}
\end{figure*}

In Fig.~\ref{fig:groups_single_galprop}, we have seen that the group galaxies in our sample have lower impact parameters on average than the single galaxies, and that the most massive galaxies in each group have higher average stellar mass and SFR compared to the isolated ones. In Fig.~\ref{fig:mgii_gal_prop}, we have shown that the \mgii\ equivalent width and covering fraction depend on the stellar mass and SFR, and the \mgii\ covering fraction to a lesser extent depends on the redshift. To check whether the difference we find between groups and single galaxies is due to factors other than the environment, we form a control sample of single galaxies. For each of the most massive group galaxies, we select at random a single galaxy from the sample that is within $\pm$50\,kpc in impact parameter, $\pm$0.3\,dex in stellar mass and $\pm$0.3 in redshift. We repeat this process 1000 times and take the galaxy whose \wmgii\ is closest to the median value of this matched sample. We are able to form a control sample for $\approx$95\% of the group galaxies in this way. The properties of the group and control samples are compared in Table~\ref{tab:grp_control_comp}. The distribution of \mgii\ equivalent widths in groups is different from that in the control sample of single galaxies, with groups showing two times higher \wmgii\ on average. The covering fractions of the group and control samples are compared at fixed normalized impact parameters in Fig.~\ref{fig:cf_grp_control}. We find that group galaxies still show similarly enhanced covering fractions compared to the control sample. The results are similar when we compare the profiles at fixed impact parameters as well. We further tried a more restrictive control sample, where we select a single galaxy within $\pm$25\,kpc in impact parameter, $\pm$0.2\,dex in stellar mass and $\pm$0.2 in redshift of a group galaxy. We are able to construct a control sample for $\approx$70\% of the group galaxies in this way. Comparing to this more restrictive control sample, we find that the distribution of \wmgii\ of group galaxies is no longer significantly different ($P$ = 0.1). However, the group galaxies show larger covering fraction by a factor of two at a given radius compared to the more restrictive control sample. The covering fraction of group galaxies remains enhanced even if we go more for a stringent matching in impact parameter, down to $\pm$10\,kpc, although the difference is no longer statistically significant. Therefore, the difference in \mgii\ gas covering fraction between group and single galaxies is most likely due to environmental processes.

\begin{table}
\caption{Comparison of different properties of the group and control samples (matched within $\pm$50\,kpc in $R$, $\pm$0.3\,dex in \mstar\ and $\pm$0.3 in $z$) for \mgii\ and \civ\ absorption. The table lists the maximum difference between the cumulative distributions of the two samples ($D$), and the probability of the two distributions arising from the same parent distribution ($P$) estimated from a two-sided Kolmogorov–Smirnov test. The distributions of equivalent widths of \mgii\ and \civ\ are different among the group and control samples.}
\centering
\begin{tabular}{ccccc}
\hline
Property & \multicolumn{2}{c}{\mgii} & \multicolumn{2}{c}{\civ} \\
         & $D$ & $P$ & $D$ & $P$ \\
\hline         
$z$         & 0.07 & 0.87  & 0.11 & 0.70 \\
$R$         & 0.07 & 0.94  & 0.06 & 1.00 \\
\mstar\     & 0.10 & 0.65  & 0.11 & 0.72 \\
SFR         & 0.08 & 0.86  & 0.17 & 0.19 \\
W$_{\rm r}$ & 0.24 & 0.003 & 0.19 & 0.09 \\
\hline
\end{tabular}
\label{tab:grp_control_comp}
\end{table}


When it comes to \civ, there are only few studies that have looked in detail at the galaxy environment around \civ\ absorbers. \citet{burchett2016}, based on number counts and group halo masses, find a dearth of \civ\ absorption with a column density detection threshold of $N$(\civ) = $3\times10^{13}$\cms\ within $R<160$\,kpc in denser environments at $z<0.05$, with no \civ\ absorption detected at \mhalo\ $>5\times10^{12}$\,\msun. \citet{manuwal2021} similarly find no \civ\ absorption at $N$(\civ) $\le 3\times10^{13}$\cms\ within 160\,kpc of $L>0.13L^*$ galaxies at $z<0.05$, which have 7 or more neighbouring galaxies in SDSS within 1.5\,Mpc and 1000\,\kms. The highest group halo mass at which \civ\ absorption is detected in our sample is $3\times10^{12}$\,\msun. There are only two groups in the sample with \mhalo\ $>10^{13}$\,\msun, and neither show associated \civ\ absorption (\wciv\ $\le0.004-0.2$\,\AA). This is consistent with the above literature results. We note however that \citet{muzahid2021} have found significantly stronger \civ\ absorption around groups of multiple \lya\ emitters detected in MUSE at $z\approx3$, based on a stacking analysis.

Fig.~\ref{fig:cf_grp} (right panels) shows the covering fraction of \civ\ gas around groups and single galaxies in the QSAGE sample for two different equivalent width thresholds of 0.03\,\AA\ and 0.1\,\AA. The \civ\ covering fraction at $R\le$100\,kpc of group galaxies is two times higher than that of single galaxies for \wciv\ = 0.03\,\AA, and consistent within the uncertainties for \wciv\ = 0.1\,\AA. At larger impact parameters, the \civ\ covering fraction around group galaxies is consistent within the uncertainties with that around single galaxies. The covering fractions around group and single galaxies are also consistent at a given normalized impact parameter. We further look at the \civ\ covering fraction difference between group and single galaxies in sub-samples based on low/high redshift, stellar mass and SFR. We do not find any clear trends such as found in the case of \mgii\ gas. Groups tend to show higher covering fraction at low redshifts and at high stellar mass, while the opposite trend is found at high redshifts and low mass. The right panel of Fig.~\ref{fig:cf_grp_zgal} compares the \civ\ covering fractions in the low and high redshift sub-samples.

Next, we construct a control sample of single galaxies, that are matched in impact parameter ($\pm$50\,kpc), stellar mass ($\pm$0.3\,dex) and redshift ($\pm$0.3) with those of the most massive galaxies in groups, as done in the case of \mgii\ gas. The distributions of \civ\ equivalent width in the group and control samples are slightly different (see Table~\ref{tab:grp_control_comp}), with groups showing 1.5 times higher \wciv\ on average. The \civ\ covering fractions of the group and control single galaxies are consistent within the large uncertainties (right panel of Fig.~\ref{fig:cf_grp_control}). The covering fraction and \wciv\ distribution of group galaxies are consistent with those of a more restrictive control sample ($\pm$25\,kpc in $R$, $\pm$0.2\,dex in \mstar\ and $\pm$0.2 in $z$). In general, the ionized gas traced by \civ\ absorption appears to be less influenced by the galaxy environment than the low-ionization gas traced by \mgii. Additionally, at a given equivalent limit and distance from galaxy, the covering fraction of \civ\ gas is $\approx$2 times higher compared to that of \mgii\ gas for both group and single galaxies, consistent with what we found in Section~\ref{sec_metals_galaxies}.

\subsection{Cross-correlation of metals and galaxies}
\label{sec_cross_corr}
\begin{figure}
 \includegraphics[width=0.48\textwidth]{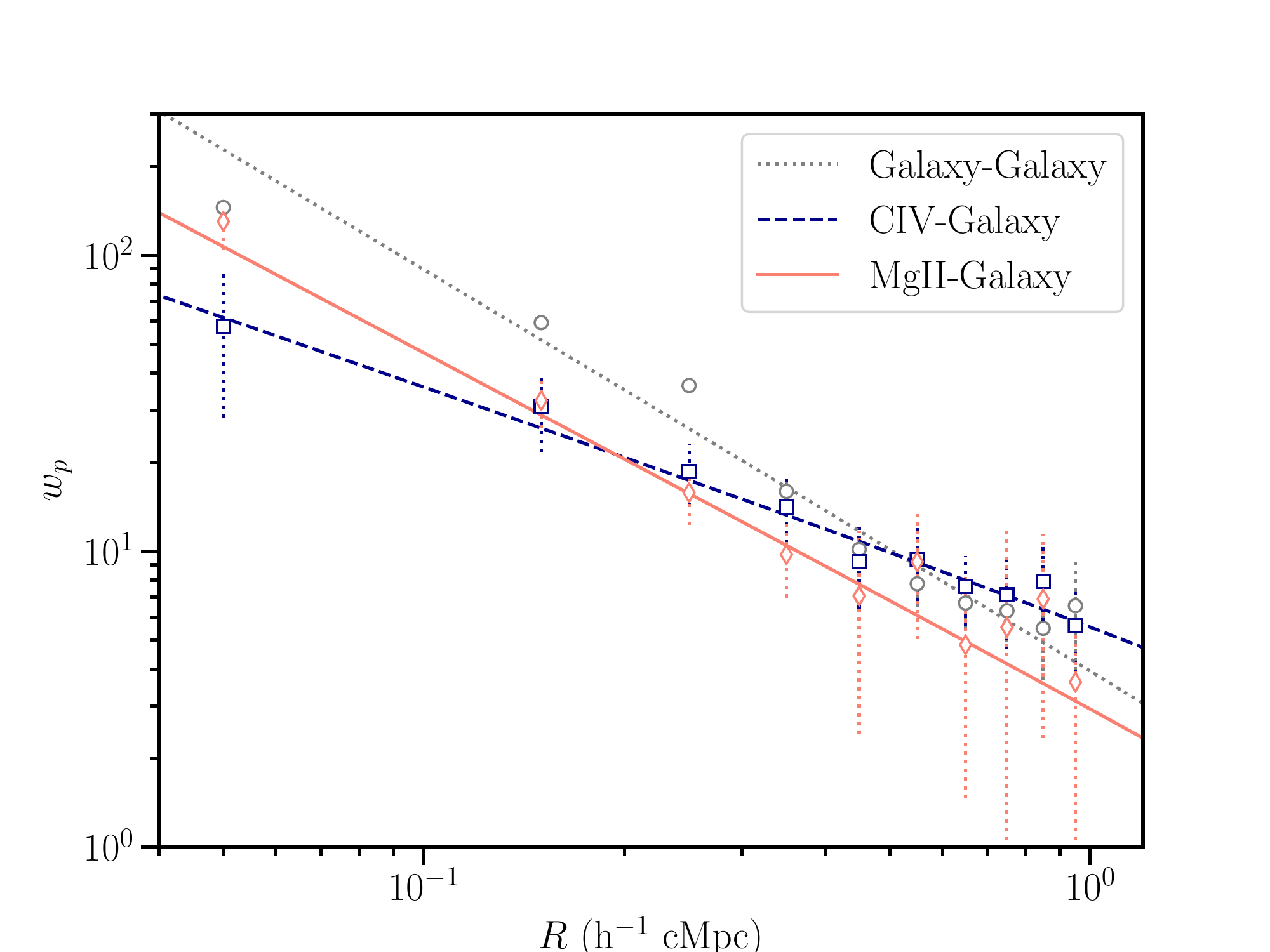}
 \caption{The absorber-galaxy angular cross-correlation estimates for \mgii\ (diamonds) and \civ\ systems (squares) along with the galaxy-galaxy auto-correlation (circles). The solid, dashed and dotted lines are fits to the \mgii-galaxy, \civ-galaxy and galaxy-galaxy correlation functions, respectively (see Section~\ref{sec_cross_corr} for details). The uncertainties on data points are from Poisson statistics.
 The \mgii-galaxy cross-correlation function is similar to the galaxy-galaxy auto-correlation, while the \civ-galaxy cross-correlation function has a larger correlation length and shallower slope.
 }
 \label{fig:cross_corr}
\end{figure}

Motivated by the dependence of the metal distribution, particularly \mgii\ gas, on the galaxy environment, and to further compare the distributions of \mgii\ and \civ\ gas around galaxies, we next examine the absorber-galaxy cross-correlation function. Several studies have investigated the large-scale (up to $\gtrsim$10\,h$^{-1}$\,cMpc) cross-correlation between metal absorbers and galaxies, e.g. \mgii\ absorbers and luminous red galaxies at $z\approx0.5$ \citep{bouche2006,gauthier2009,lundgren2009}, \civ\ absorbers and Lyman break galaxies at $z\approx2.5$ \citep{adelberger2005,crighton2011,turner2014}, \civ\ absorbers and quasars at $z\approx2$ \citep{prochaska2014}, \ovi\ absorbers and galaxies at $z<1$ \citep{prochaska2019}. Here we look at the small-scale clustering (up to 1\,h$^{-1}$\,cMpc) of galaxies with \mgii\ and \civ\ absorbers in our sample. Due to the relatively small angular scales probed by our observations, we do not attempt to interpret the absorber correlation length in the framework of bias models, rather we qualitatively compare the cross-correlation of \mgii\ and \civ\ absorbers with galaxies.

The absorber-galaxy cross-correlation function, which is an estimate of the clustering properties of galaxies with absorbers, is defined as the excess conditional probability of finding a galaxy within a volume $dV$ at a position $r_2$ given that there is an absorber at position $r_1$, $P(r_2|r_1)dV = P_0[1 + \xi(r)]dV$, where $r = |r_1 - r_2|$ and $P_0 dV$ is the probability of finding a galaxy at an average place in the Universe. The three-dimensional cross-correlation function is usually assumed to take a power-law form, $\xi(r) = (r/r_0)^{-\gamma}$, where $r_0$ is the comoving distance at which the local number density of galaxies is twice that in an average place in the Universe, and $\gamma$ is the slope parameter.

However, the three-dimensional correlation function, $\xi(r)$, is difficult to directly measure in practice, and we compute here instead the projected angular cross-correlation function, $w_p(R)$, within a velocity window around the absorbers \citep[see][]{davis1983,trainor2012}. To estimate the correlation function, we count the number of galaxies within a velocity window of $\pm$1000\,\kms\ (based on  the WFC3 redshift uncertainty) around the absorber redshifts and within projected circular annuli centred on the quasars. We then compare this with the average number of galaxies expected in each of these volumes based on the volume density of galaxies at that redshift in the full survey. To obtain this, we fit a smooth third order polynomial to the volume density of galaxies as a function of redshift in the MAGG and QSAGE surveys. 

Following \citet{trainor2012}, we parameterize the correlation function in the following form:
$$
    w_p(R) = (r_0/R)^\gamma \times {}_{2}F_1(1/2,\gamma/2,3/2,-z_0^2/R^2)
$$
where $R$ is the 2D projected comoving separation between the absorbers and galaxies, ${}_{2}F_1$ is the Gaussian hypergeometric function and $z_0$ is the half-width of the redshift window. We show the absorber-galaxy cross-correlation functions along with the best-fit relations in Fig.~\ref{fig:cross_corr}. The uncertainties on the measurements are estimated from Poisson statistics and the uncertainties on the fit parameters are obtained from bootstrap resampling with repetitions. We find a positive correlation for both \mgii\ and \civ\ absorbers, indicating significant clustering of galaxies around the absorbers at small-scales. Based on the fits to the absorber-galaxy correlation functions, we obtain for \mgii: $r_0$ = 4.3$^{+2.9}_{-1.4}$\,h$^{-1}$\,cMpc, $\gamma$ = 2.2$\pm$0.4, and for \civ: $r_0$ = 7.7$^{+5.8}_{-2.0}$\,h$^{-1}$\,cMpc, $\gamma$ = 1.8$^{+0.3}_{-0.4}$. We further estimate the galaxy-galaxy auto-correlation function by centering the circular annuli on the galaxies instead of the quasars, and obtain $r_0$ = 4.6$^{+0.5}_{-0.4}$\,h$^{-1}$\,cMpc and $\gamma$ = 2.3$\pm$0.1. The \mgii\ correlation length and slope are similar to that of the galaxy auto-correlation function, while the \civ\ correlation length is larger. We tested that the above parameters do not change beyond the uncertainties if we exclude the measurement in the innermost radial bin from the fit. We note that here we have considered absorption clumps arising within $\pm$500\,\kms\ of the peak absorption as a single absorption system. The way the absorbers are classified into systems could have an effect on the cross-correlation function, and may explain the difference we see between the absorber-galaxy and galaxy-galaxy correlation measurements at $R\lesssim0.3$\,h$^{-1}$\,cMpc.

We can compare with studies of the angular correlation function for different types of galaxies in the literature. The form of the \mgii-luminous red galaxy cross-correlation function is found to be similar to that of the luminous red galaxy auto-correlation function over scales $\sim$1-10\,h$^{-1}$\,cMpc at $z<$1 \citep{bouche2006,gauthier2009,lundgren2009}. \citet{adelberger2005} measured the correlation function between \civ\ and Lyman break galaxies at $z\approx2.5$ and found a correlation length of $r_0$ = 4$\pm$0.6\,h$^{-1}$\,cMpc for a fixed $\gamma$ of 1.6 and $N$(\civ) $\ge3\times10^{12}$\cms\ at scales $\lesssim$5\,h$^{-1}$\,cMpc, similar to the galaxy auto-correlation length in their study, implying that \civ\ absorbers and galaxies reside in similar structures. \citet{prochaska2014} measured the \civ-quasar cross-correlation function at $z\approx2$ and obtained $r_0$ = 7.5$^{+2.8}_{-1.4}$\,h$^{-1}$\,cMpc and $\gamma$ = 1.7$^{+0.1}_{-0.2}$, inferring that the \civ\ gas traces massive ($10^{12}$\,\msun) haloes. \citet{bielby2014}, from a wide area (2.4\,deg$^2$) galaxy survey, estimated the galaxy auto-correlation length to be $r_0$ = 3.9$\pm$0.3\,h$^{-1}$\,cMpc and $\gamma$ = 1.90$\pm$0.04 at $z$=1.1 for galaxies in the similar stellar mass range as in our sample. These values are similar to the correlation lengths and slopes we find for \mgii-galaxy and galaxy-galaxy correlation in our sample. 

We note that the fit to the \mgii\ correlation function has a slightly steeper slope than that to the \civ\ correlation function. This could tie in with the results obtained in the previous sections which support a more centrally-peaked distribution of the low-ionization gas traced by \mgii\ compared to the high-ionization gas traced by \civ. On the other hand, the clustering scale length is larger in case of \civ\ absorbers compared to \mgii. Although we note that the $r_0$ and $\gamma$ measurements for \mgii\ and \civ\ are consistent within the uncertainties. If we fix the $\gamma$ parameter in the fit to the \mgii\ correlation function to be 1.8, as found for the \civ\ correlation function, the $r_0$ value for \mgii\ absorbers increases to 6.6$\pm$0.6\,h$^{-1}$\,cMpc. However in this case the fit is poor and underestimates the observations at the inner radii. Similarly, if we fix $\gamma$ = 2.2 in the fit to the \civ\ correlation function, $r_0$ decreases to 5.0$\pm$0.3\,h$^{-1}$\,cMpc, but the fit overestimates the observations at the inner radii. Hence, if we fix the slope to be the same for both \mgii\ and \civ, the correlation length for both will be similar. However, the observations are more consistent with a shallower slope and a larger correlation length for \civ\ absorbers. We point out that the estimates at present are dominated by large uncertainties and the small number of absorber-galaxy pairs prevents us from reliably studying the dependence of the cross-correlation function on redshift, galaxy properties and absorber strength. Larger samples of absorber-galaxy pairs are required to confirm these clustering trends on small-scales and in different sub-samples of galaxies. 

%
%
\section{Linking metals and galaxy overdensity}
\label{sec_overdensity}

While the degree to which the galaxy environment affects the different gas phases in the halo may differ, it is becoming clear from wide-field, complete galaxy surveys that one-to-one association of galaxies and absorption systems is not always straightforward. Rather, this is subject to the sensitivity and impact parameter probed by the galaxy survey, the physical search window used for cross-matching galaxies and absorption lines, and the definition of group/isolated galaxies. In the above analysis, we have used physical linking lengths to identify groups using the Friends-of-Friends algorithm independent of galaxy stellar mass, and we have included galaxy pairs under groups. Groups can be alternatively identified via various other algorithms as well, including adaptive matched filters, halo-based group finders, overdensities in position and color spaces and the Voronoi-Delaunay Method \citep[e.g.][]{gerke2005,miller2005,yang2005,gal2006}. Furthermore, the choice of which galaxy in a group to consider for absorber correlation also needs to be made in these kinds of studies. 

Using large samples of hydrogen and metal absorption lines in quasar spectra at $z\ge2$, several studies have modeled the gaseous halo or the absorption component in a "continuous" and statistical manner with the pixel optical depth method \citep{rakic2012,turner2014,chen2020b}. Now that we have sufficiently large galaxy samples available in quasar fields, we can describe the galaxy component in "continuous" terms as well using local densities. This way, we can address the galaxy environment and gaseous halo connection in a complementary and statistical manner, which also overcomes some of the limitations of galaxy-absorber association mentioned above. 

There are several methods of quantifying the galaxy environment \citep{muldrew2012,haas2012}, with the two most common ones being the number of galaxies within fixed or adaptive apertures \citep{hogg2003,croton2005} and the distance to the $N$th nearest neighbour \citep{dressler1980,baldry2006}. The number density within a fixed aperture method has been found to be more sensitive to high mass overdensities, better represent the actual 3D overdensity and be less affected by viewing angle and redshift \citep{shattow2013}. Here we apply the method of estimating galaxy overdensity within a fixed aperture to the MAGG and QSAGE samples. 

We first estimate the number of galaxies within fixed cylindrical volumes, with the projected circular area centred on the quasar in each field. We use Kernel Density Estimation (KDE) to better represent the underlying probability distribution of the galaxy redshifts. Along each quasar line-of-sight (LoS) up to ${\rm z}_{\rm QSO}  - (1 + {\rm z}_{\rm QSO})\times$5000\,\kms$/c$, we perform one-dimensional KDE to obtain the probability density function of the galaxy number counts within a certain projected radius (depending on the survey and redshift range as detailed below). We use a Gaussian kernel and the Improved Sheather-Jones algorithm \citep{sheather1991,wand1994,botev2010} to select the bandwidth. Then by integrating this 1D KDE, we estimate the number density in equal physical volumes along the quasar LoS, defined by the radius and a LoS rest-frame velocity of $\delta v =\pm$1000\,\kms. Note that we divide the survey volume into equal physical volumes in this way in order to characterize the local galaxy environment around the sightline in a continuous manner and cross-match the galaxy densities with the absorption properties. In case there are fewer than ten galaxies along a quasar LoS within the redshift range and radius considered, we directly count the number of galaxies within the volumes. We similarly estimate the stellar mass, SFR and sSFR densities in these volumes by integrating the weighted KDE. 

To check the robustness of the galaxy densities estimated in the above way, we performed various tests including using a different kernel and bandwidth for KDE, a different LoS velocity window for the volumes, a different way of partitioning the LoS into volumes, and directly counting galaxies in volumes instead of using KDE. The galaxy densities estimated through different ways correlate with each other, and we find that overall the results discussed in this section do not change significantly. A more detailed discussion on these tests and their effect on the results is given in Appendix~\ref{appendix_density}.

\begin{figure*}
 \subfloat[]{\includegraphics[width=0.48\textwidth]{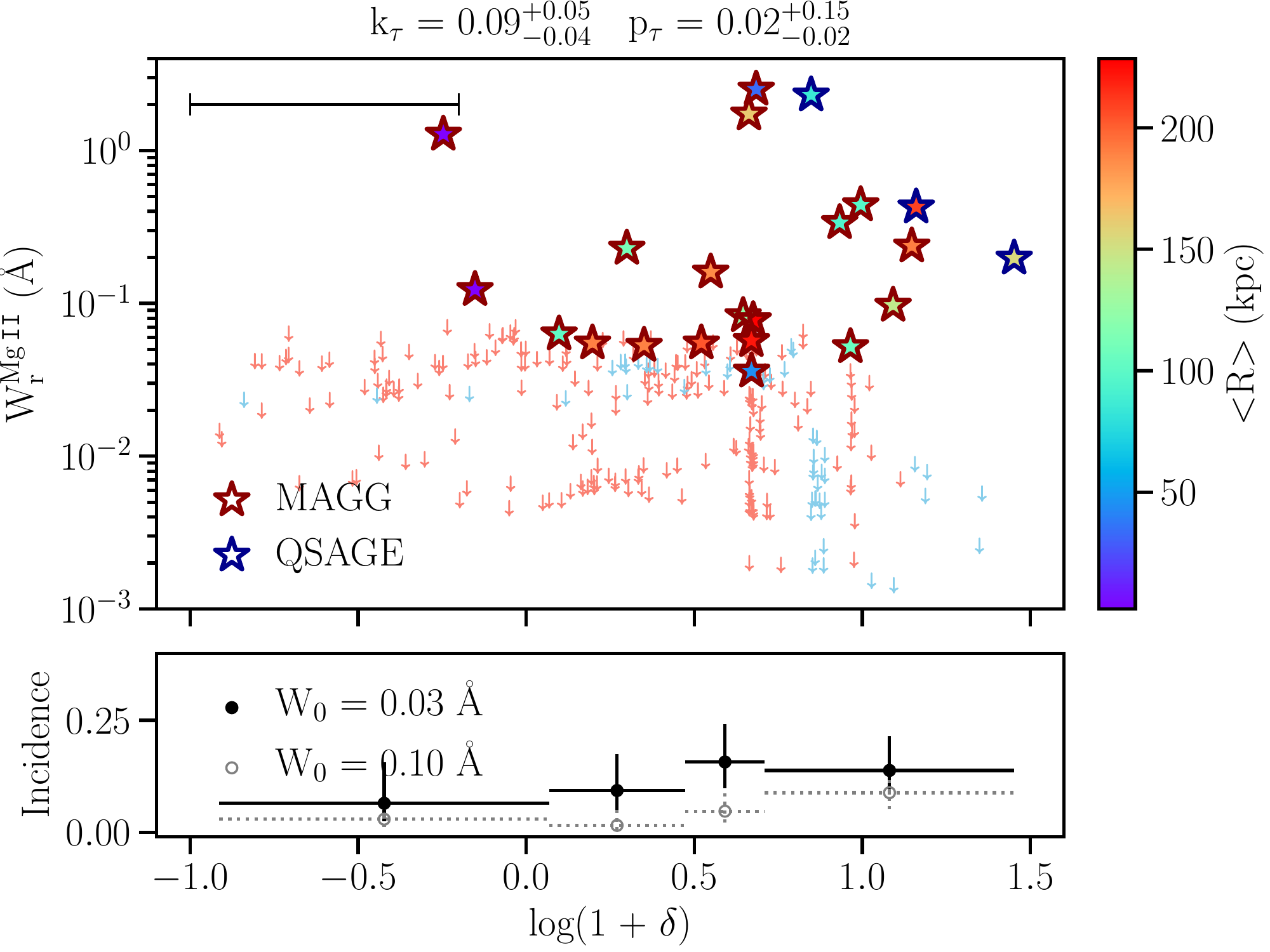}}
 \subfloat[]{\includegraphics[width=0.48\textwidth]{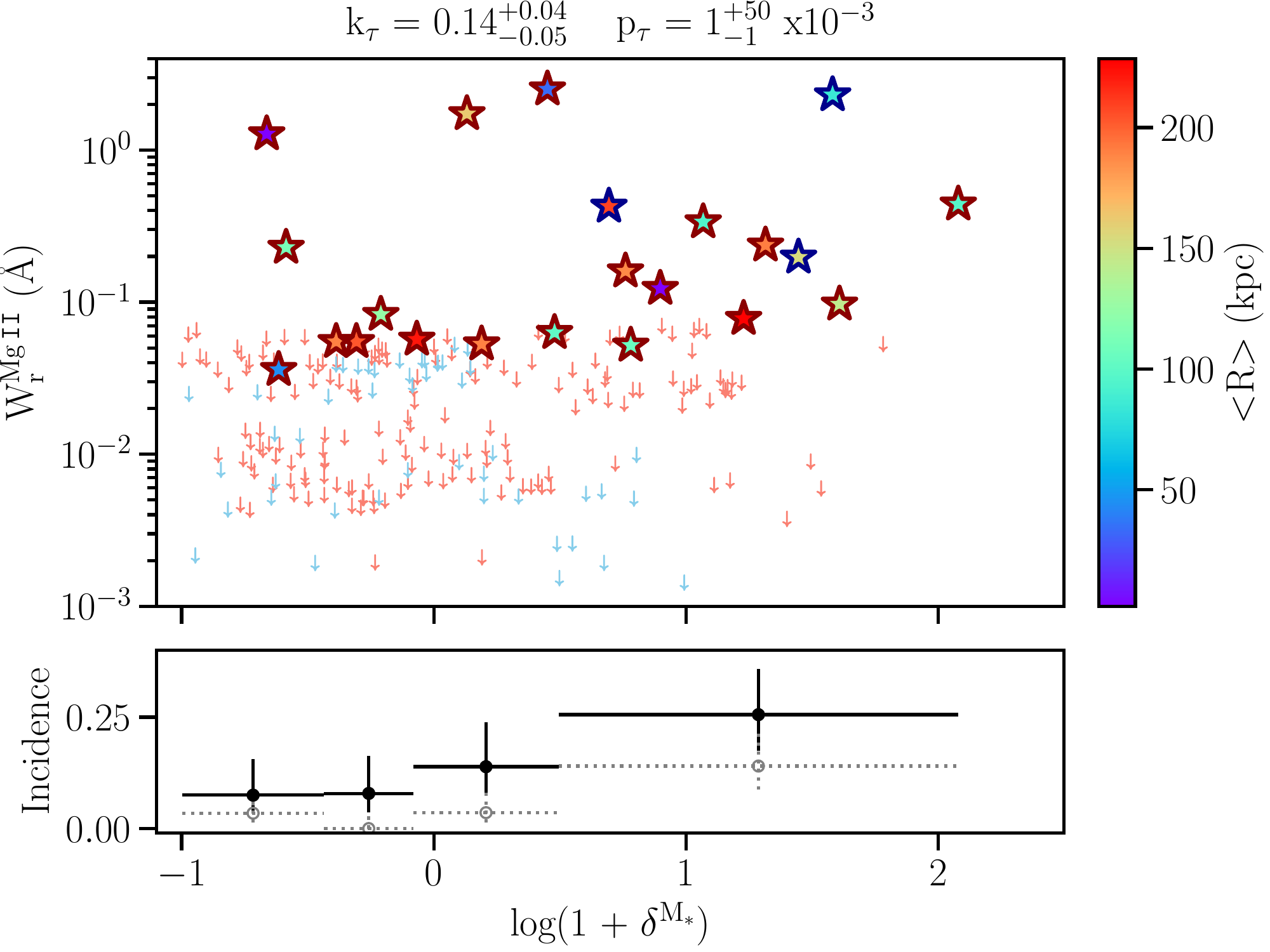}}
 \hspace{0.01cm}
 \subfloat[]{\includegraphics[width=0.48\textwidth]{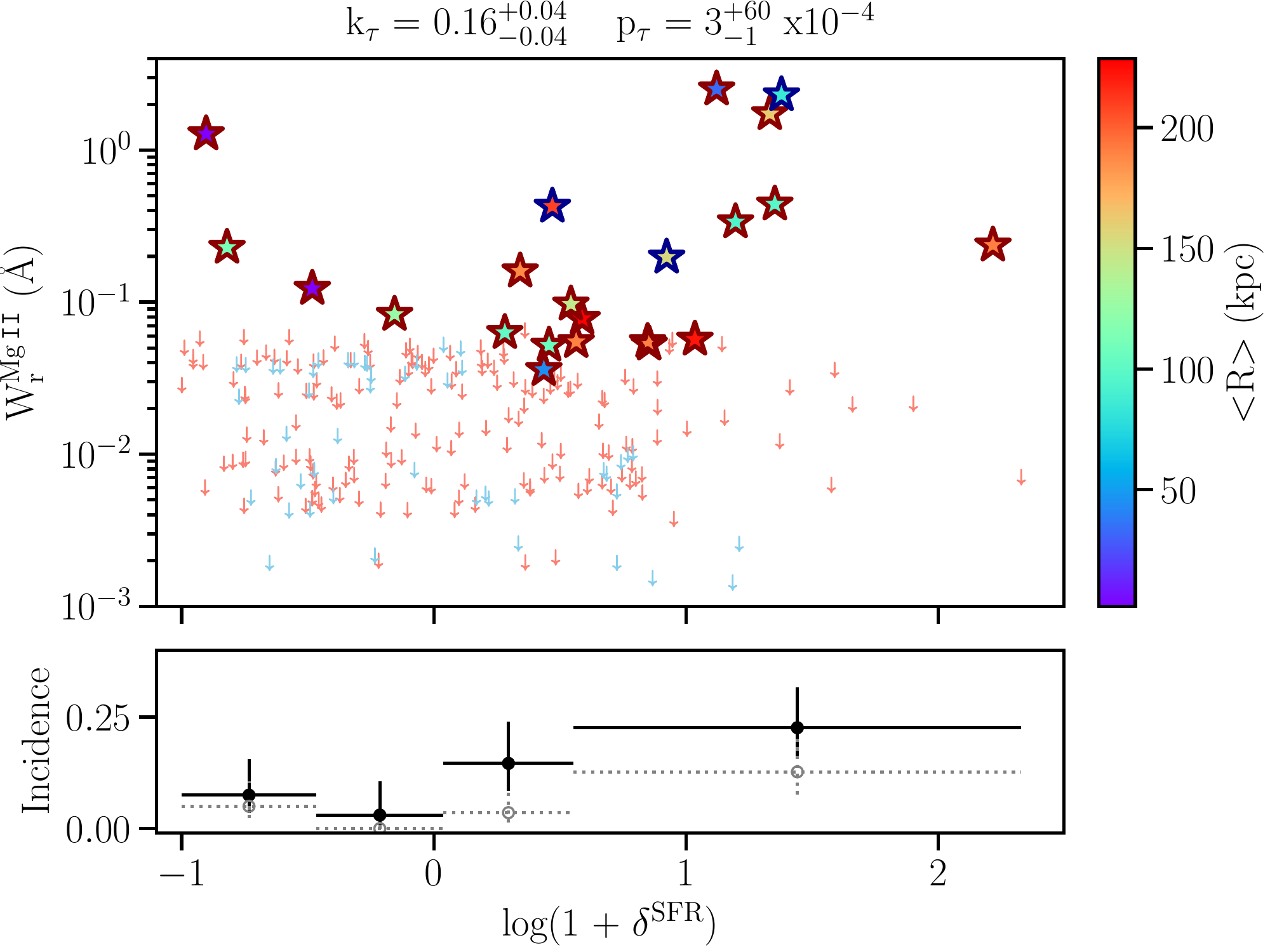}}
 \subfloat[]{\includegraphics[width=0.48\textwidth]{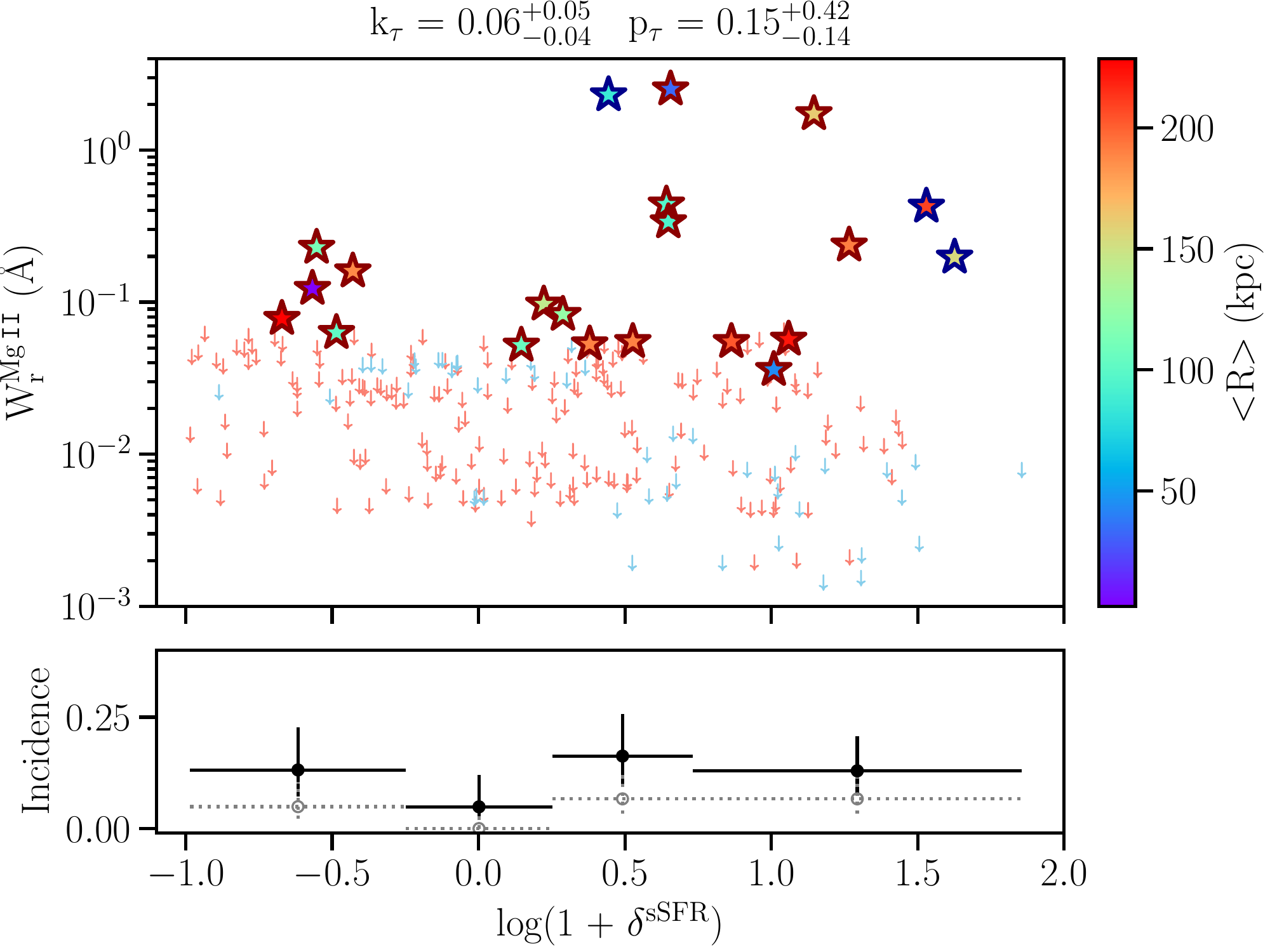}}
 \caption{Rest-frame equivalent width of \mgii\ as a function of the galaxy number (a), stellar mass (b), SFR (c), and sSFR (d) overdensity over $0.9<z<1.6$ within a radius of 240\,kpc in the combined MAGG and QSAGE sample. In case of \mgii\ detection, the \wmgii\ measurements are shown as stars, while the 3$\sigma$ upper limits are shown as downward arrows in the case of non-detections. Data from the MAGG sample are shown as red stars/arrows, and those from the QSAGE sample are shown in blue. The detections are color-coded by the average galaxy impact parameter. The average uncertainty in the overdensity measurements from Poisson statistics is shown at the top left of plot (a). Results from correlation analysis between \wmgii\ and the overdensity are given at the top of each plot, where a small value of p$_\tau$ indicates significant correlation. To the bottom of the plots, the incidence or detection rate of \mgii\ absorption is shown for each quartile of the respective overdensity, for \wmgii\ sensitivity limits of 0.03\,\AA\ (solid black line) and 0.1\,\AA\ (dashed grey lines). The uncertainties on the detection rates are 1$\sigma$ Wilson score confidence intervals.
 \wmgii\ and incidence of \mgii\ show increasing trends with overdensities in number, stellar mass and SFR of galaxies over $0.9<z<1.6$.
 }
 \label{fig:mgii_ovden_combined}
\end{figure*}

Once we have the galaxy densities, we estimate the overdensity relative to the field density as $\delta = (\rho - \rho_\mathrm{field}(z)) / \rho_\mathrm{field}(z)$. Since we have large and complete galaxy surveys at hand, we estimate the field density from the respective surveys, MAGG and QSAGE, themselves. To obtain the average number density of galaxies at a given redshift in a survey, we compute the volume density of galaxies in that whole survey and parameterize its redshift dependence with a smooth third order polynomial. In this work, we use the term overdensity to refer to the logarithmic density contrast defined as log$(1 + \delta)$, where the logarithm is to the base 10. 

Finally, to associate the galaxy overdensities with the absorption, we search for \mgii/\civ\ absorption lines within $\delta v =\pm$1000\,\kms\ of each cylindrical volume in which we estimate the overdensity along the quasar LoS. We take the total equivalent width of all the absorption components found within this $\delta v$. If no absorption is present in a volume, we estimate a 3$\sigma$ upper limit on the rest-frame equivalent width for a linewidth of 100\,\kms, based on the noise in the quasar spectra in that region, after masking out any strong contaminating lines. Note that the choice of $\delta v$ is motivated by the redshift uncertainties of the WFC3 grism sources. Different $\delta v$ values of $\pm$500\,\kms\ and $\pm$700\,\kms\ were tried and found not to have significant effect on the results (see Appendix~\ref{appendix_density}).

We determine the radius of the fixed volumes in which to estimate the overdensities based on the optimal compromise between the maximum physical FoV accessible over the entire redshift range considered and the maximum number of absorption systems that can be associated with the overdensities. For the MAGG and QSAGE combined \mgii\ sample over the redshift range $0.9<z<1.6$, we consider a radius of 240\,kpc to make full use of the MUSE FoV. For the analysis based only on the QSAGE sample, we consider a radius of 200\,kpc up to $z=1$, and 600\,kpc beyond that to get the full FoV of WFC3. We restrict ourselves to redshifts $<1.6$, beyond which the \ha\ emission line falls out of the wavelength coverage of the WFC3 grism, and the \oiii\ and/or the \oii\ emission line is used instead for redshift identification. This way we ensure that the average number density varies as a smooth function of redshift.

\begin{figure*}
 \subfloat[]{\includegraphics[width=0.48\textwidth]{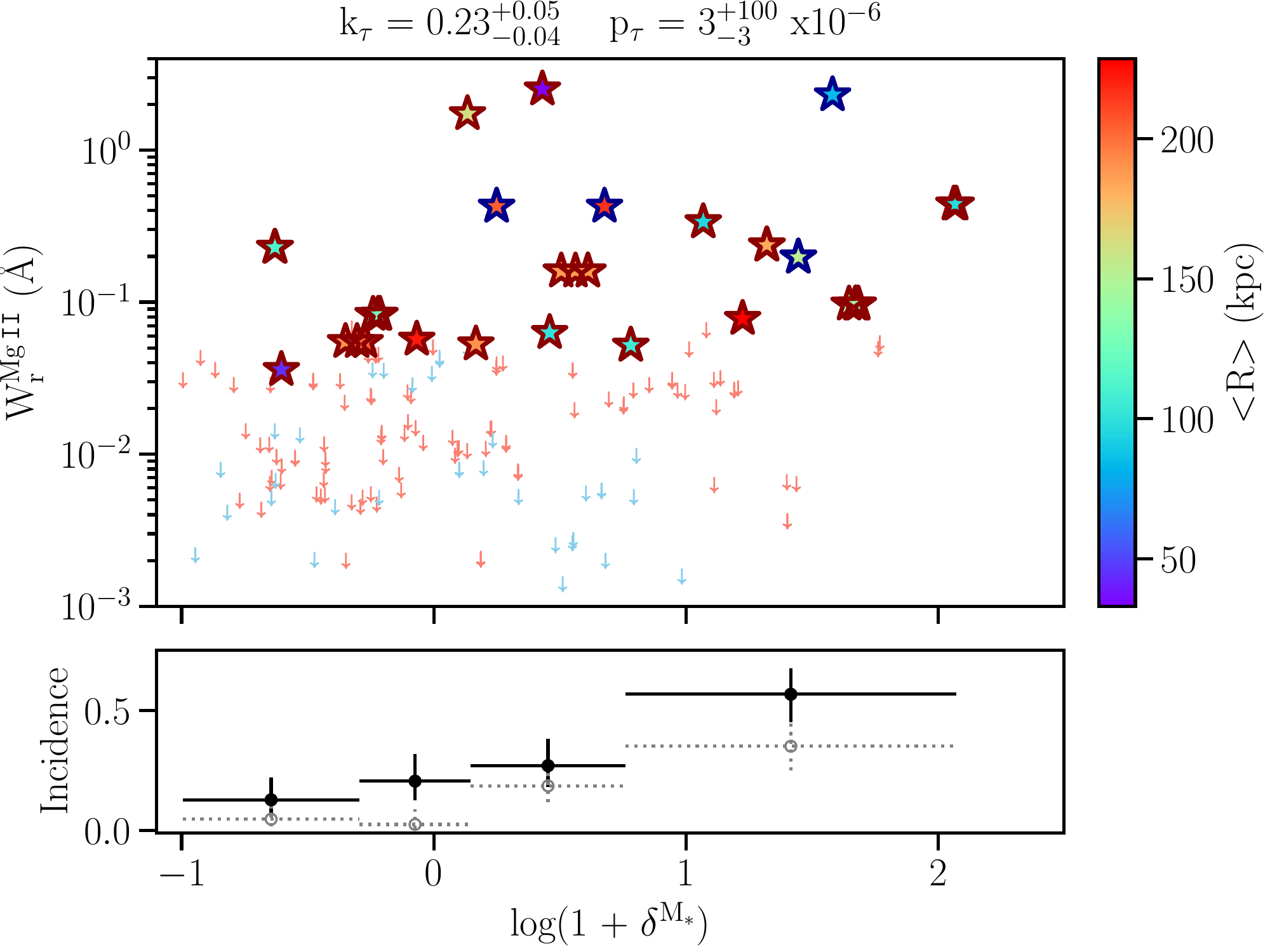}}
 \subfloat[]{\includegraphics[width=0.48\textwidth]{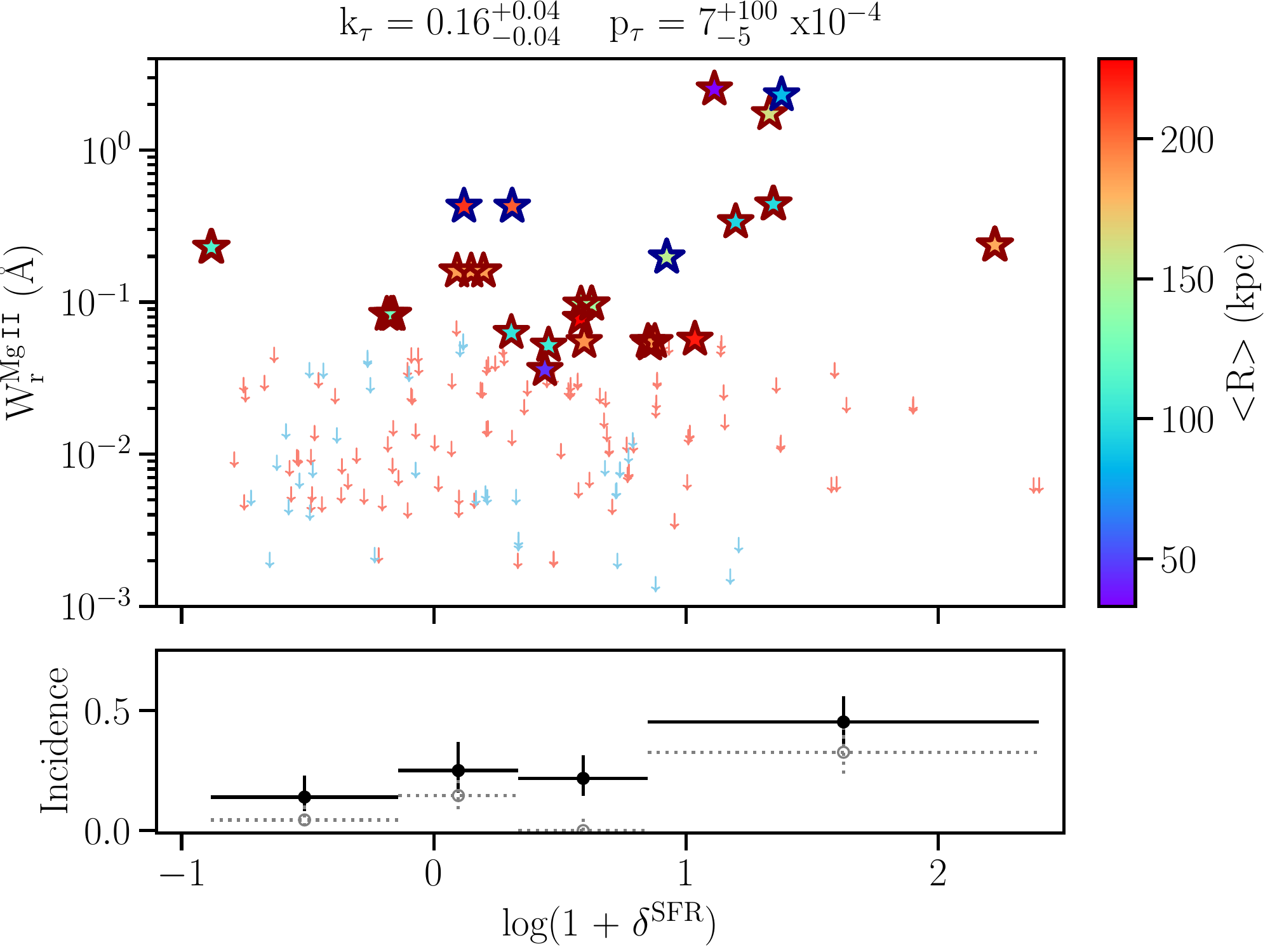}}
 \caption{Rest-frame equivalent width (top) and incidence (bottom) of \mgii\ as a function of the galaxy stellar mass (a) and SFR (b) overdensity over $0.9<z<1.6$ within a radius of 240\,kpc in the combined MAGG and QSAGE sample. The plots are same as in panels (b) and (c) of Fig.~\ref{fig:mgii_ovden_combined}, except that here the overdensities have been estimated in volumes centred on galaxy redshifts.
 The positive correlations of \wmgii\ and \mgii\ incidence with stellar mass and SFR overdensities around galaxies underlines the connection between the strongest and most prevalent \mgii\ absorption and regions of greater overall stellar mass and star formation activity.
 }
 \label{fig:mgii_ovden_galaxy}
\end{figure*}
\subsection{Dependence of \mgii\ absorption on overdensity}
\label{sec_mgii_overdensity}

Over the redshift range $0.9<z<1.6$, we use the combined sample of MAGG and QSAGE to study the connection of \mgii\ absorbing gas and galaxy overdensities. While the two galaxy surveys have different observing conditions and strategies, we find that the trends discussed here are also valid for the individual samples, and they become more significant when the two samples are combined due to the larger size. Here we first present the results from the combined sample. Then we move to the results based on the QSAGE sample alone, which makes it possible to extend the analysis to lower redshifts.

In Fig.~\ref{fig:mgii_ovden_combined}, we show the dependence of \mgii\ equivalent width as a function of the galaxy number, mass, SFR and sSFR overdensities in the combined MAGG and QSAGE sample for $0.9<z<1.6$, considering a radius of 240\,kpc. The k$_\tau$ and p$_\tau$ values from the Generalized Kendall's correlation test between \wmgii\ and the respective overdensities are provided in the plots, along with their $1\sigma$ uncertainties from bootstrap analysis. We also plot the detection rate or incidence of \mgii\ absorption as a function of the overdensity. The trends in incidence are similar for the two different sensitivity limits of \wmgii\ = 0.03\,\AA\ and 0.1\,\AA. We estimate the incidence in each 25$^{\rm th}$ percentile of the overdensities. The general trends of incidence also hold if we instead estimate the incidence in fixed overdensity bins. 

The average equivalent width of \mgii\ increases with the number overdensity, i.e. relatively more of the strongest \mgii\ absorbers tend to arise in the most overdense regions in our sample. The prevalence of \mgii\ absorbers increases with the overdensity, with $\approx$15\% incidence above overdensities of 0.5 for \wmgii\ = 0.03\,\AA. This ties in with our results from Section~\ref{sec_metals_groups}, that galaxies in group environments give rise to stronger and more prevalent \mgii\ absorption. Note that in two cases, \mgii\ absorption is detected but no galaxy is detected in the given volume [i.e., they are plotted as arising in underdense regions with log$(1 + \delta) < 0$]. For one of them, we detect associated galaxies beyond the aperture of 240\,kpc considered here. We note that there could be additional galaxies associated with these absorbers within the considered volume that lie below the sensitivity limit of the data, or just below the quasar point spread function (PSF; $\lesssim$10\,kpc), where residual quasar emission after PSF subtraction limits our ability to securely identify galaxies. We estimate an average 3$\sigma$ limit on flux $\le3\times10^{-17}$\,\ergscm\ around $\pm$100\,\kms\ of the expected location of \oii\ emission in the MUSE spectra of the quasars, which corresponds to unobscured SFR $\le$1\,\msunyr\ at $z=1.2$.

\begin{figure*}
 \subfloat[]{\includegraphics[width=0.48\textwidth]{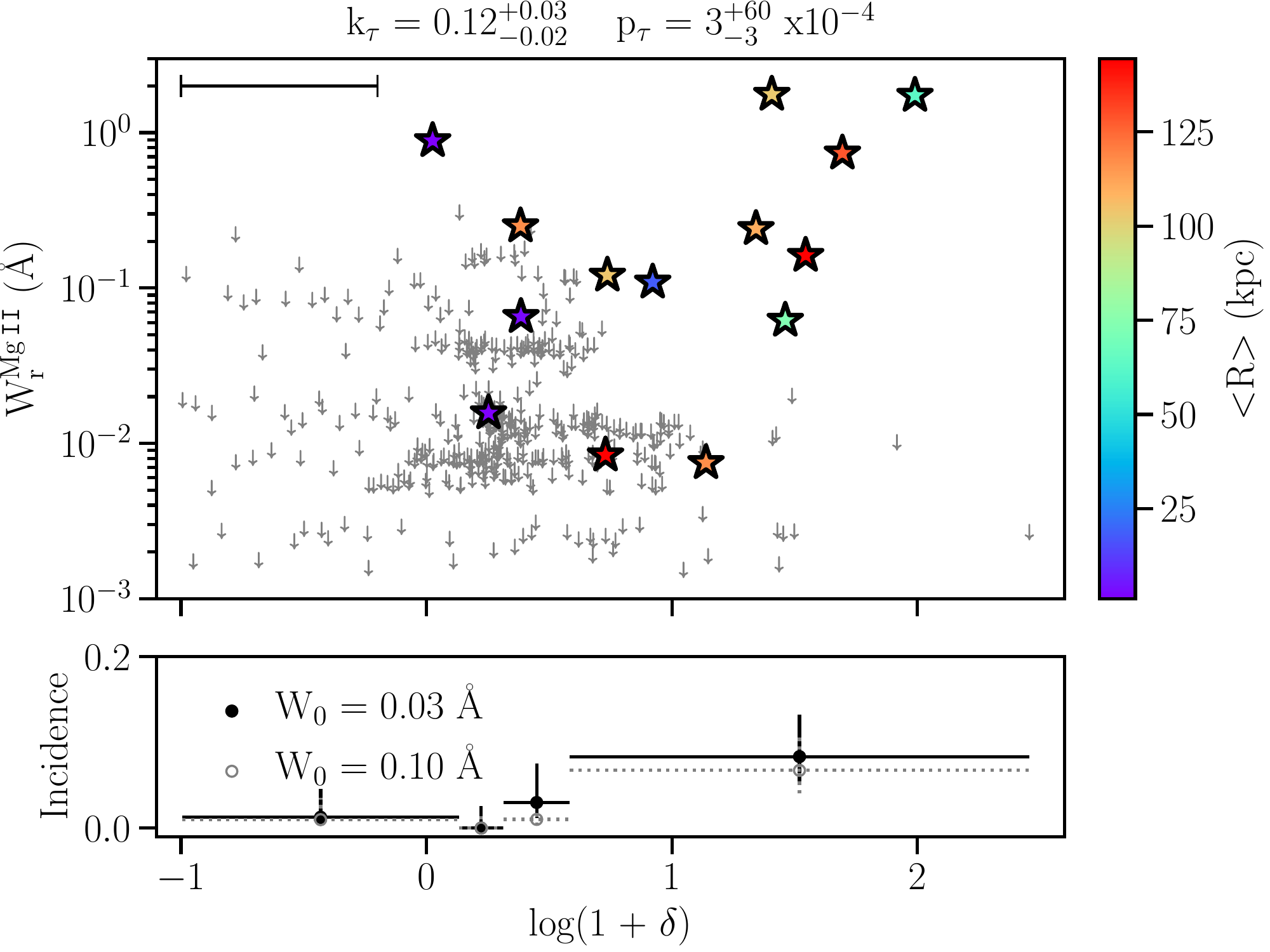}}
 \subfloat[]{\includegraphics[width=0.48\textwidth]{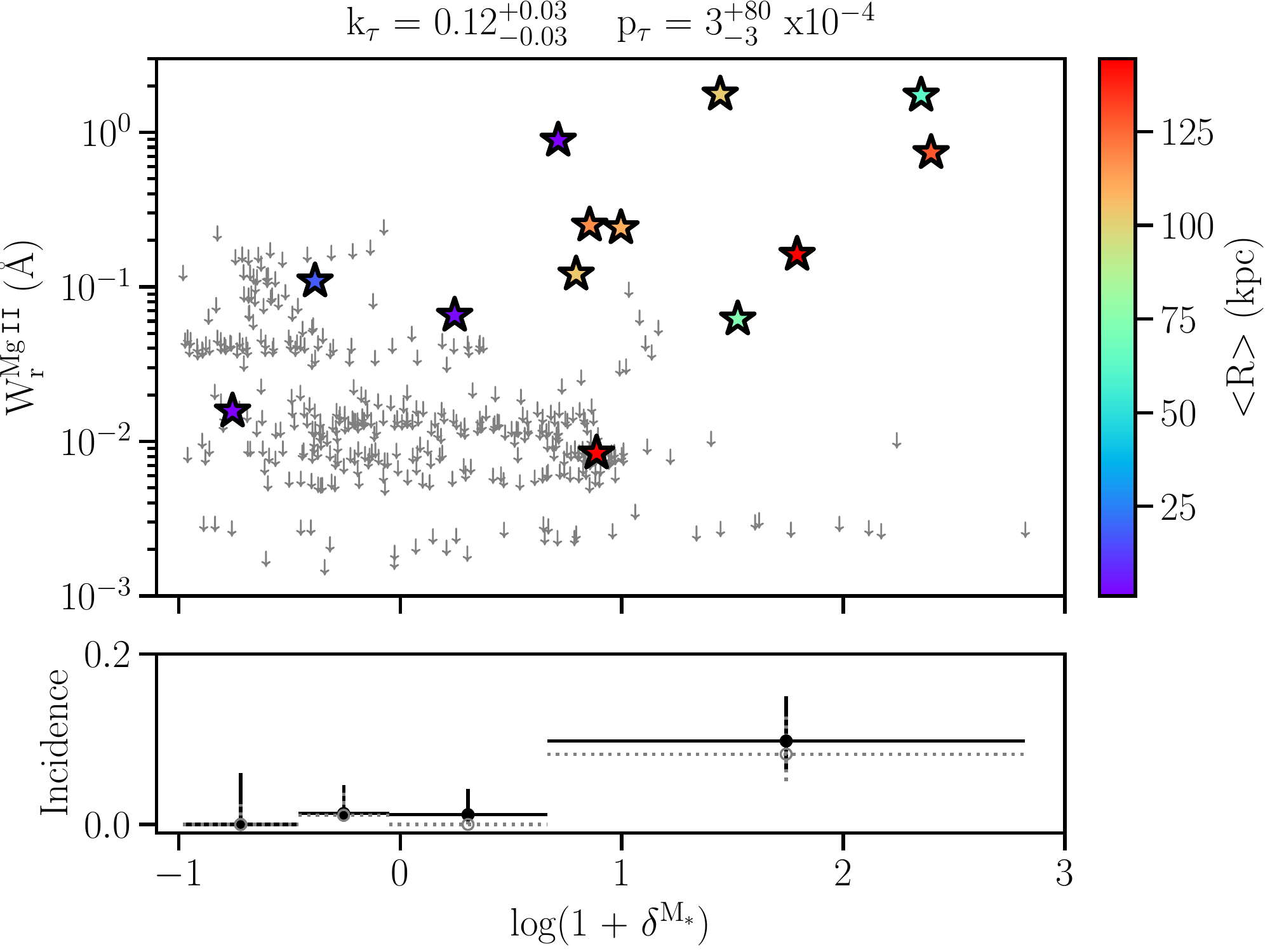}}
 \hspace{0.01cm}
 \subfloat[]{\includegraphics[width=0.48\textwidth]{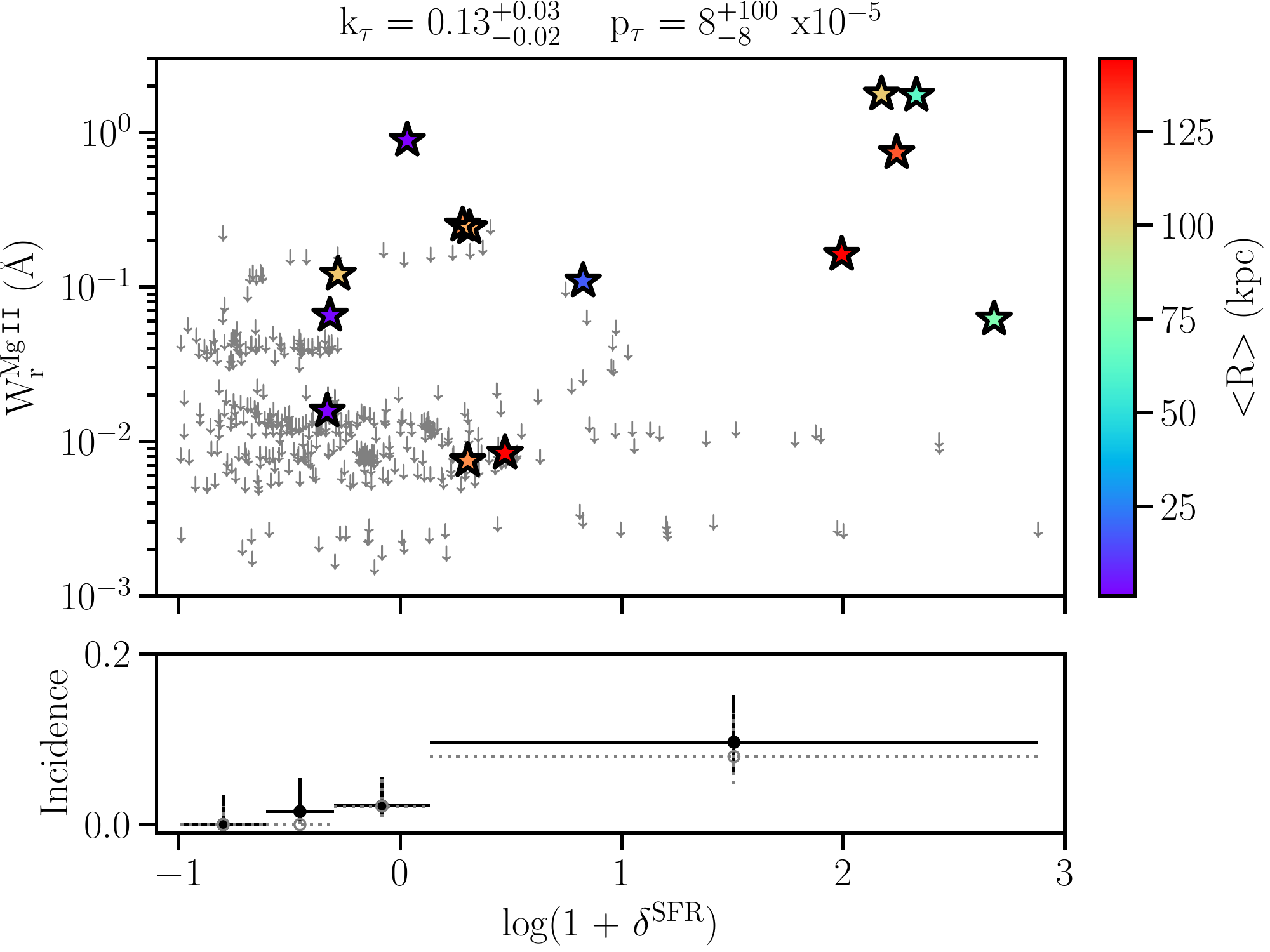}}
 \subfloat[]{\includegraphics[width=0.48\textwidth]{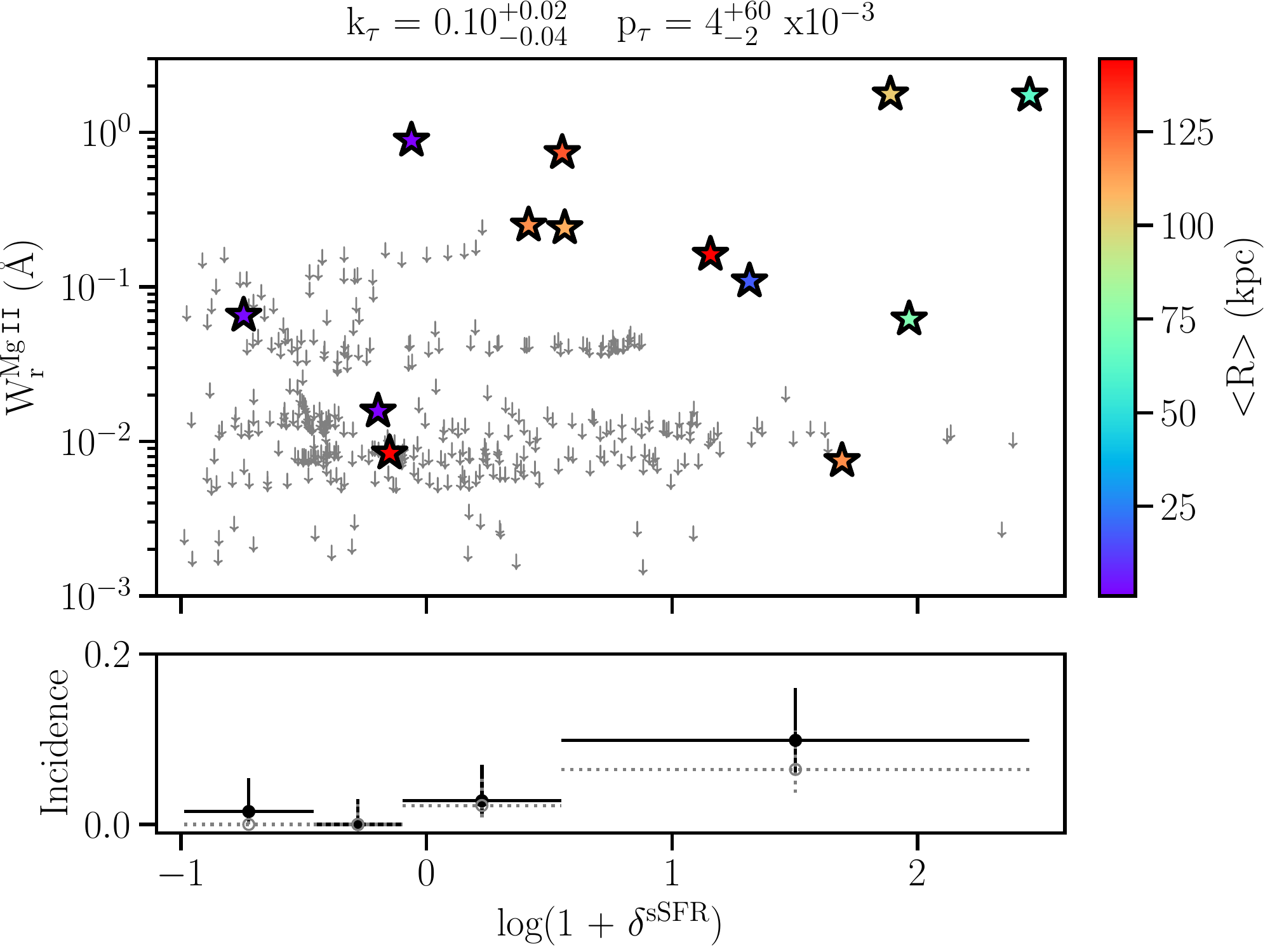}}
 \caption{Same as in Fig.~\ref{fig:mgii_ovden_combined}, showing the dependence of \wmgii\ and \mgii\ incidence on galaxy overdensity over $0.1<z<1$ within a radius of 200\,kpc in the QSAGE sample.
 \wmgii\ and incidence of \mgii\ show increasing trends with overdensities in number, stellar mass, SFR and sSFR of galaxies at $z<1$.
 }
 \label{fig:mgii_ovden_qsage}
\end{figure*}

Besides the number density, \mgii\ equivalent width and incidence exhibit increasing trends with the mass and SFR overdensities as well. \mgii\ absorption is most prevalent at the highest mass and SFR overdensities, with incidences of $\approx$25\% above overdensities of 0.5 for \wmgii\ = 0.03\,\AA. Since the stellar mass and SFR are themselves correlated, it is hard to disentangle the dependence of the absorption properties on these two quantities. However, there is no significant dependence of \mgii\ absorption and incidence on sSFR overdensities. Based on this and the results in Section~\ref{sec_metals_galaxies}, we infer that the stellar mass could be the dominant factor behind the SFR dependence. 

We note that the uncertainties in the overdensity measurements are large due to small number statistics, and that there is a wide scatter in \wmgii\ across two orders of magnitude of overdensity, with the correlations mainly driven by the relative lack of upper limits at overdensities greater than 1. However, the above analysis does point towards a scenario where strong \mgii\ absorption preferentially resides not just in the most overdense regions, but those comprising the most massive and star-forming galaxies. To investigate this further, we repeat the above analysis but instead of estimating the overdensities in all volumes along the quasar LoS, we restrict the analysis to volumes centred on the galaxy redshifts. Fig.~\ref{fig:mgii_ovden_galaxy} shows the dependence of \wmgii\ and \mgii\ incidence on the stellar mass and SFR overdensities estimated this way. Both the \mgii\ equivalent width and incidence show strong increasing trends with the mass and SFR overdensities around galaxies. The incidence of \mgii\ absorption becomes $\approx$60\% and $\approx$45\% (for \wmgii\ = 0.03\,\AA) for mass and SFR overdensities greater than 1, respectively. We do not find any dependence on the sSFR overdensities in this case. We note that the overdensities estimated in this way are not independent when there are multiple galaxies in an overdense region and will be biased high. Nevertheless, this analysis reinforces the connection between the strongest and most prevalent \mgii\ absorption and the highest mass and SFR overdensities. 

Next, we extend our study to lower redshifts using the QSAGE sample. Specifically, we probe the galaxy overdensities over the redshift range $0.1<z<1$, considering a radius of 200\,kpc. Fig.~\ref{fig:mgii_ovden_qsage} shows the dependence of \mgii\ equivalent width and incidence on the various overdensities at these redshifts. We again find that the strongest \mgii\ absorbers are found in the most overdense regions, with the average strength of the absorption increasing with increasing overdensity. As noted above for $z>1$, there are fewer non-detections at overdensities greater than 1 at $z<1$ as well. However, we note that the highest overdensity (log$(1 + \delta) \approx 2.5$) in this case does not show detectable \mgii\ absorption. This overdensity comprises of 22 galaxies at $z\approx0.5$, and is further identified by the FoF algorithm (see Section~\ref{sec_qsage_galaxy}) as belonging to a large group of 33 galaxies with a velocity dispersion of $\approx$700\,\kms. The lack of strong \mgii\ absorption could be due to several reasons including a lack of cool gas in this group \citep[e.g.][]{lopez2008} or the LoS probing only a tiny portion of a patchy region. In general though, the more overdense regions have a higher detection rate of \mgii\ absorption ($\approx$10\% at overdensities greater than 0.5 for \wmgii\ = 0.03\,\AA). The increasing trend in incidence with overdensities also holds true when we use a limit of 0.1\,\AA.

\begin{figure*}
 \subfloat[]{\includegraphics[width=0.48\textwidth]{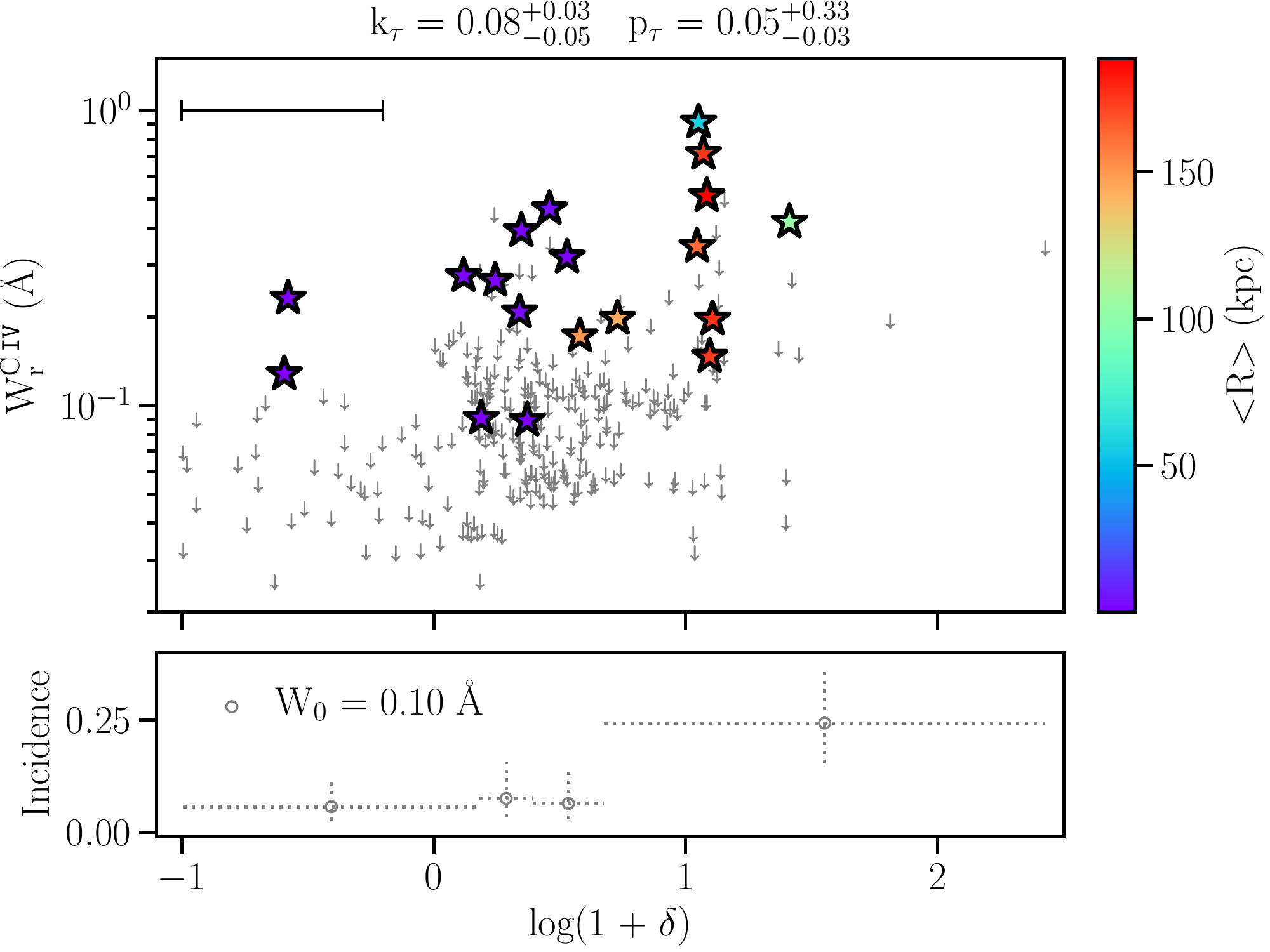}}
 \subfloat[]{\includegraphics[width=0.48\textwidth]{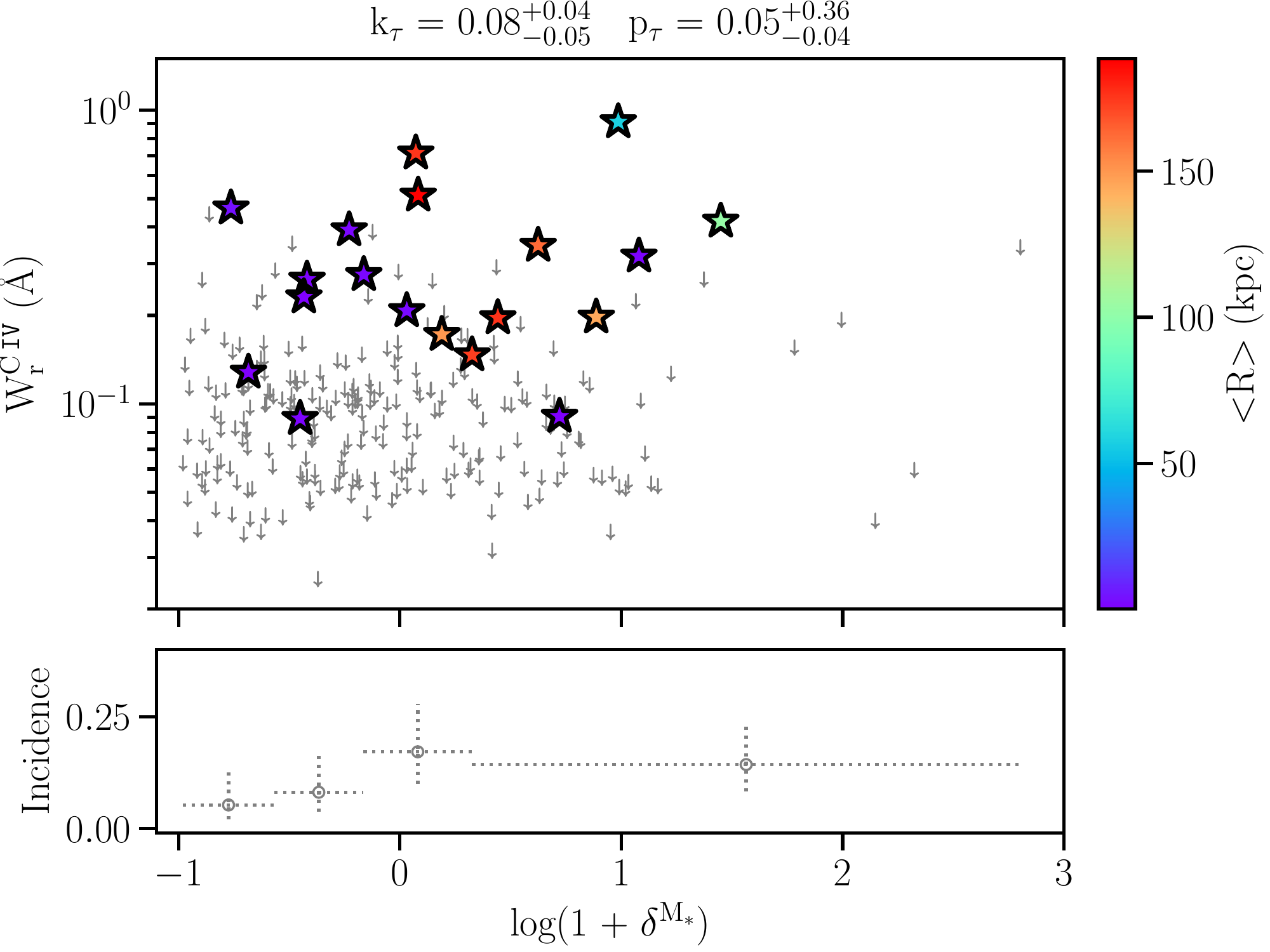}}
 \hspace{0.01cm}
 \subfloat[]{\includegraphics[width=0.48\textwidth]{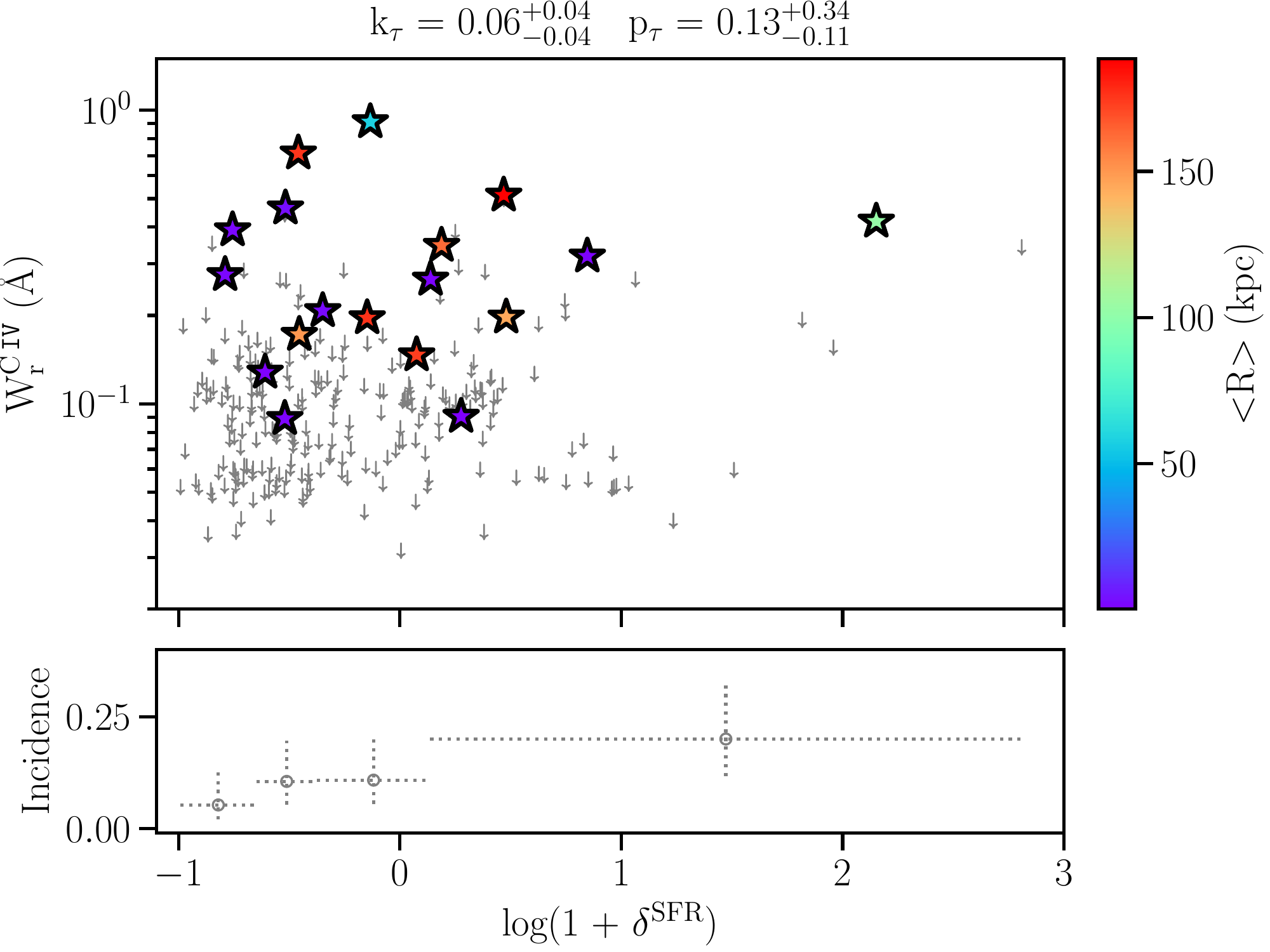}}
 \subfloat[]{\includegraphics[width=0.48\textwidth]{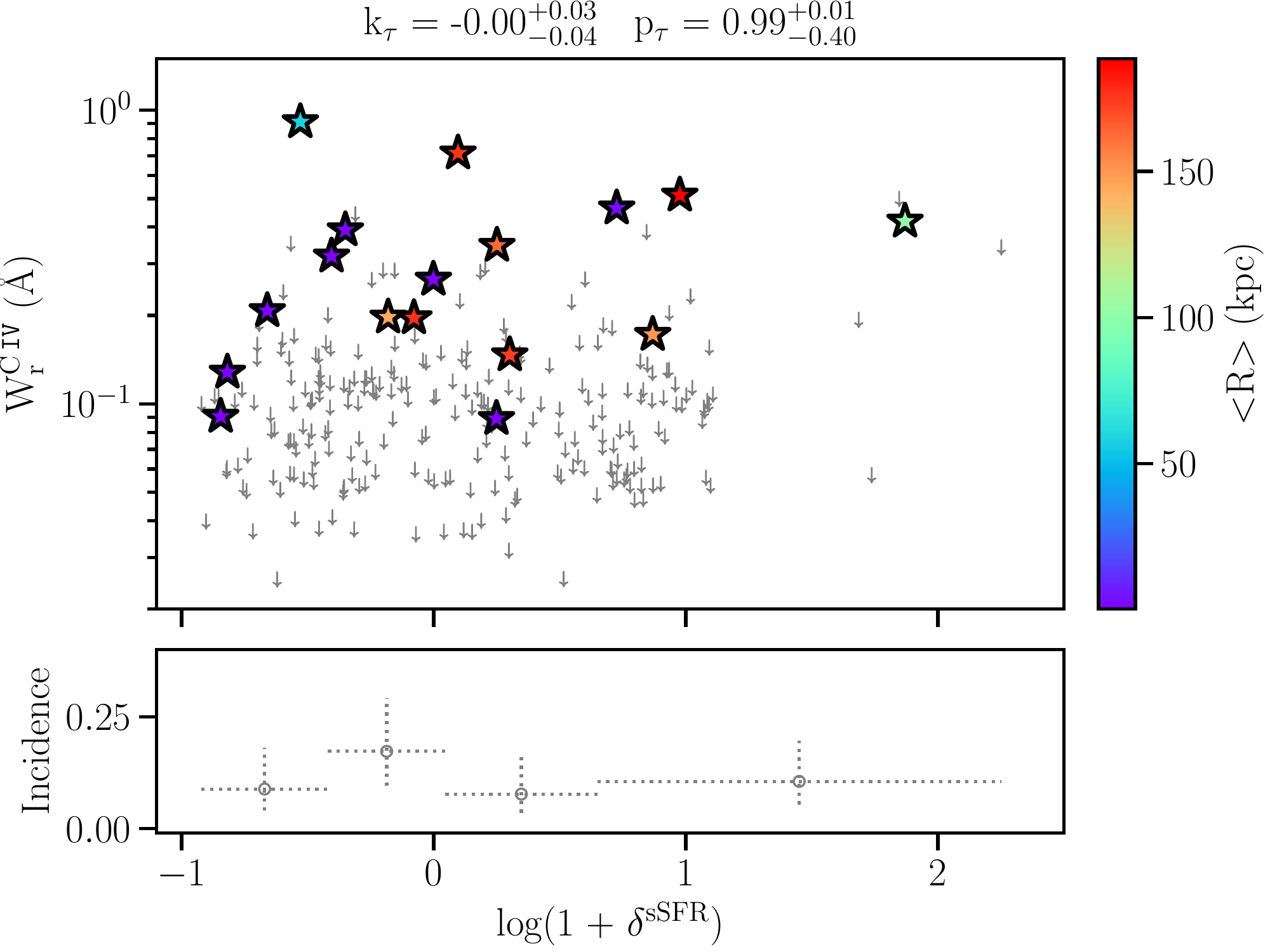}}
 \caption{Rest-frame equivalent width of \civ\ as a function of the galaxy number (a), stellar mass (b), SFR (c), and sSFR (d) overdensity over $0.1<z<1$ within a radius of 200\,kpc in the QSAGE sample. The symbols are the same as in Fig.~\ref{fig:mgii_ovden_combined}. The bottom panels show the incidence of \civ\ absorption as a function of the respective overdensity for a \wciv\ sensitivity limit of 0.1\,\AA. The uncertainties on the detection rates are 1$\sigma$ Wilson score confidence intervals.
 \wciv\ and incidence of \civ\ show increasing trends with overdensities in number and stellar mass of galaxies at $z<1$.
 }
 \label{fig:civ_ovden_zlt1}
\end{figure*}

Further, \wmgii\ and \mgii\ incidence are positively correlated with the stellar mass, SFR and sSFR overdensities. The dependence of \mgii\ absorption on sSFR overdensities is weaker than that on the mass and SFR overdensities. The incidence of \mgii\ absorption is $\approx$10\% for mass and SFR overdensities greater than 0.5 for \wmgii\ = 0.03\,\AA. We also find a strong dependence on the mass and SFR overdensities when we estimate the overdensities around galaxy redshifts as done above for $z>1$. Overall, the trends of \wmgii\ with the number, mass and SFR overdensities are similar at $z>1$ and at $z<1$. However, we note that the detection rates of \mgii\ absorption at log$(1 + \delta) \gtrsim 0.5-1$ tend to be $\approx$1.5-2 times higher at $z>1$. This is consistent with the higher covering fraction of \mgii\ gas found at $z>1$ (see Fig.~\ref{fig:mgii_gal_prop}).

\subsection{Dependence of \civ\ absorption on overdensity}
\label{sec_civ_overdensity}

Now we look at the connection of \civ\ absorption with galaxy overdensity using the QSAGE sample in two redshift ranges:  at $0.1<z<1$ within a radius of 200\,kpc, and at $1<z<1.6$ within a radius of 600\,kpc. First, in Fig.~\ref{fig:civ_ovden_zlt1} we show the dependence of \civ\ equivalent width and incidence on the galaxy number, mass, SFR and sSFR overdensities at $z<1$. We estimate the incidence for a sensitivity limit of \wciv\ = 0.1\,\AA. Note that we lack more stringent upper limits at $z<1$ to cut at a lower limit of 0.03\,\AA. On average, the equivalent width of \civ\ increases with the number overdensity. The detection rate of \civ\ absorption also increases with the overdensity, with incidence of $\approx$25\% at overdensities greater than 0.5. Note that the increasing trend of incidence with overdensity is not dependent on the choice of binning and holds if we estimate the incidence in fixed bins of overdensities instead of in equal quartiles. 

The highest overdensities lack detectable \civ\ absorption, although the \wciv\ upper limits in these cases are also relatively high ($\approx$0.2-0.3\,\AA). Note that the highest overdensity here is the same as the one discussed in Section~\ref{sec_mgii_overdensity}, arising in a large group of galaxies at $z\approx0.5$. Recall that in Section~\ref{sec_metals_groups} we had not found a significant difference in \civ\ absorption properties between group and single galaxies. Based on the analysis here, there appears to be a tendency for regions with overdensity $\approx$1 to give rise to stronger and more prevalent \civ\ absorption. However, we need to include more highly overdense [log$(1 + \delta) > 1$] regions in the sample to be able to say anything definitive about the \civ\ absorption at the highest overdensities. 

\begin{figure*}
 \subfloat[]{\includegraphics[width=0.48\textwidth]{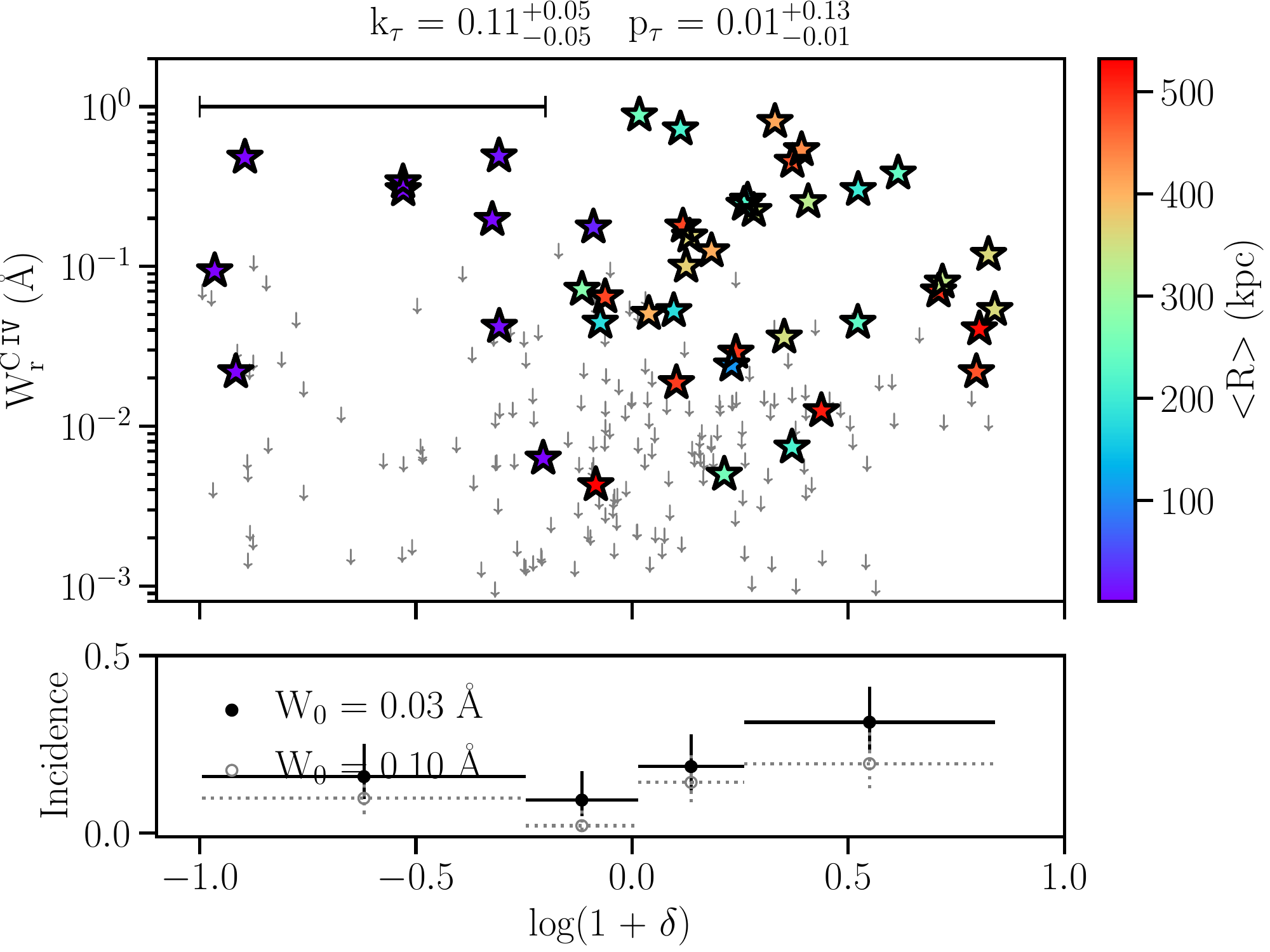}}
 \subfloat[]{\includegraphics[width=0.48\textwidth]{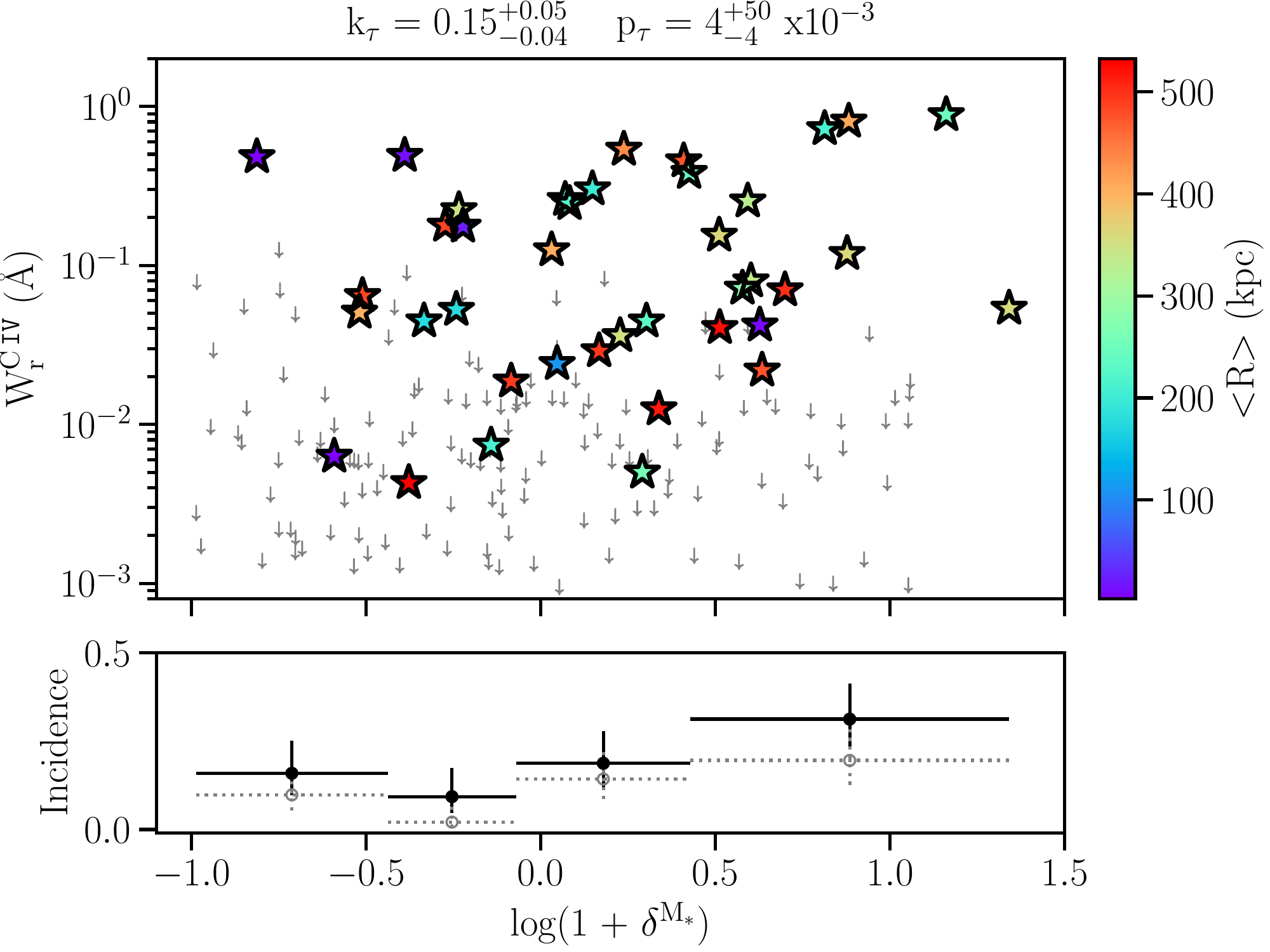}}
 \hspace{0.01cm}
 \subfloat[]{\includegraphics[width=0.48\textwidth]{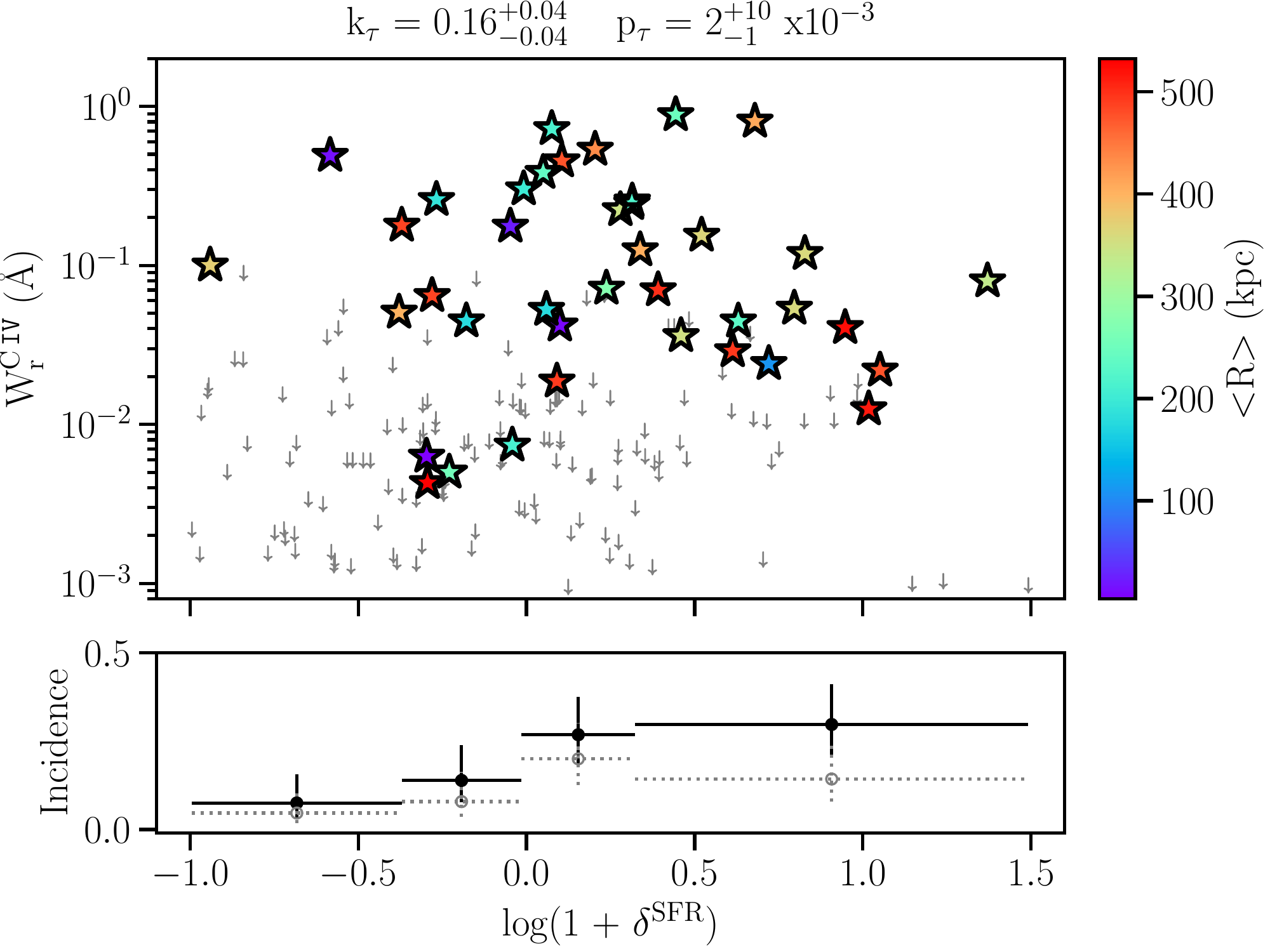}}
 \subfloat[]{\includegraphics[width=0.48\textwidth]{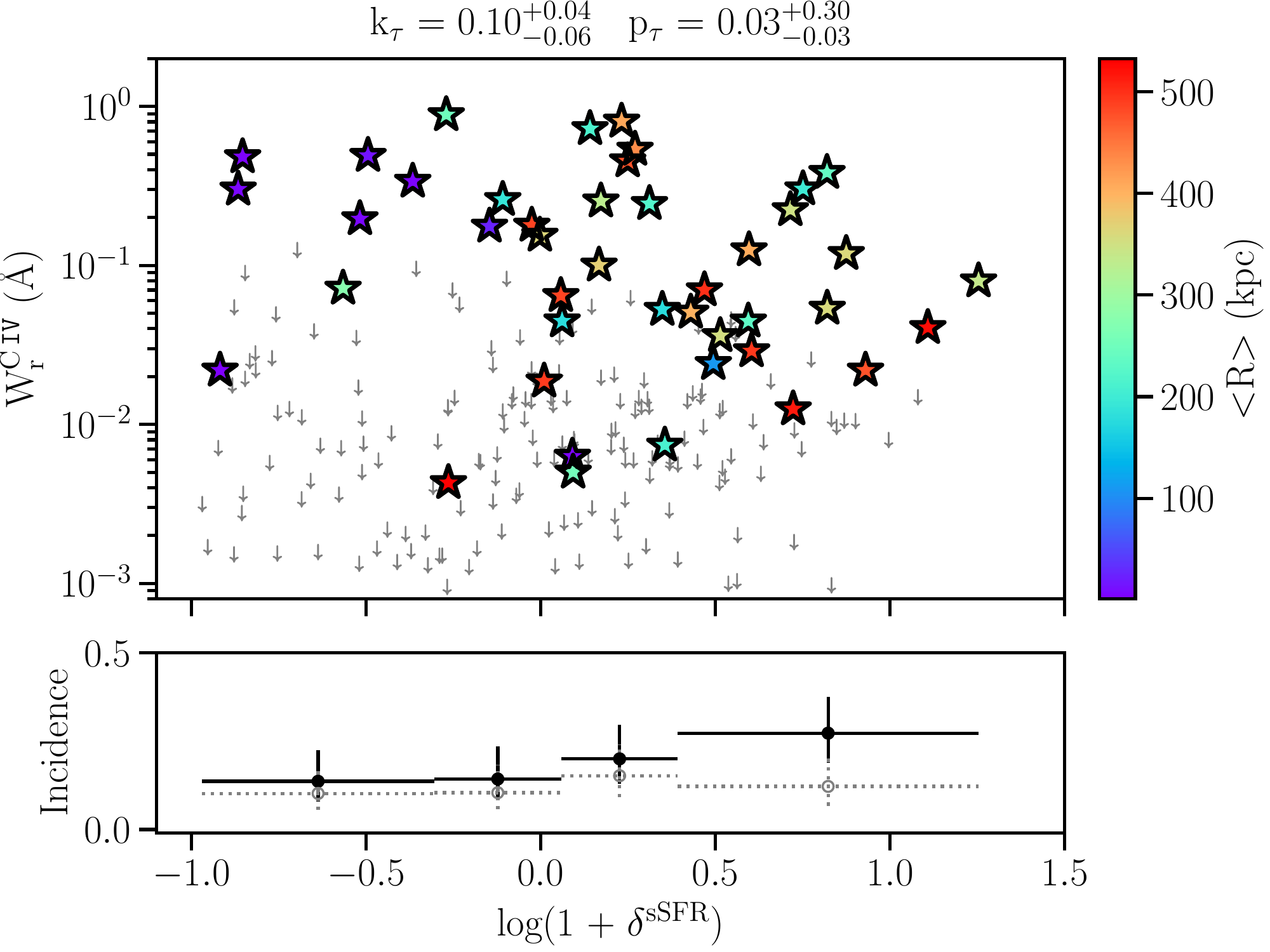}}
 \caption{Same as in Fig.~\ref{fig:civ_ovden_zlt1}, showing the dependence of \wciv\ and \civ\ incidence on galaxy overdensity over $1<z<1.6$ within a radius of 600\,kpc in the QSAGE sample. The bottom panels show the incidence of \civ\ absorption as a function of the respective overdensity for sensitivity limits of 0.03\,\AA\ (black solid lines) and 0.1\,\AA\ (grey dashed lines). 
 \wciv\ and incidence of \civ\ show increasing trends with overdensities in number, stellar mass and SFR of galaxies over $1<z<1.6$.
 }
 \label{fig:civ_ovden_zgt1}
\end{figure*}

In addition to the number overdensity, the \civ\ equivalent width also increases with the stellar mass overdensity. However, there is no strong dependence of \wciv\ on either the SFR or sSFR overdensities. The incidence of \civ\ absorption shows an increasing trend with mass and SFR overdensities, but no apparent trend with sSFR overdensities. The results based on the overdensity analysis here are in general consistent with what we found in Section~\ref{sec_metals_galaxies}, where the dependence of \civ\ absorption on stellar mass was found to be the strongest. 

Comparing with the results for \mgii\ in Section~\ref{sec_mgii_overdensity}, the dependencies of \mgii\ absorption on the different overdensities are found to be stronger than those of \civ\ absorption at $z<1$. On the other hand, at a fixed overdensity and equivalent width cut-off of 0.1\,\AA, the incidence of \civ\ absorption is found to be $\approx$2 times higher than that of \mgii\ absorption. As discussed in Section~\ref{sec_galaxies_groups}, this could imply a more extended distribution of the high-ionization gas as traced by \civ\ absorption than the low-ionization gas as traced by \mgii\ absorption.

Next, we study the dependence of \civ\ absorption on galaxy overdensities at $z>1$ in Fig.~\ref{fig:civ_ovden_zgt1}. For this, taking advantage of the full FoV of WFC3, we estimate overdensities for a radius of 600\,kpc. We are also able to estimate the incidence for two different sensitivity limits of \wciv\ = 0.03\,\AA\ and 0.1\,\AA. The average equivalent width of \civ\ increases with increasing stellar mass and SFR overdensities and to a lesser extent with the number and sSFR overdensities. The incidence of \civ\ absorption also shows an increasing trend with all the overdensities for the \wciv\ cut-off of 0.03\,\AA, with $\approx$30\% incidence above overdensities of 0.5, although the trend is less prominent for the \wciv\ cut-off of 0.1\,\AA. We find similar dependencies of \wciv\ and \civ\ incidence on the overdensities when we restrict to volumes around the galaxies. 

Note that due to the larger area considered here, the values of overdensities and incidence cannot be directly compared to the ones in Fig.~\ref{fig:civ_ovden_zlt1} at $z<1$. Therefore, it is not possible to definitely establish whether the differences found in the dependence of \civ\ absorption on overdensities at $z>1$ compared to $z<1$, like stronger dependence on SFR overdensity, are due to redshift evolution or due to the larger area considered. If we restrict to the same radius of 200\,kpc as done for the $z<1$ analysis, the sample size reduces by a factor of 10. We do not find any significant dependence of \wciv\ and incidence on the overdensities in this case. However, due to the small sample size, it is difficult to determine at present whether there is any significant redshift evolution in the dependence of \civ\ absorption on overdensities. 

%
\section{Discussion and Conclusions}
\label{sec_conclusion}

We first summarize the MAGG and QSAGE survey data used in this study and present the key results from this study in Section~\ref{sec_summary}. Then we discuss our results in the context of relevant results from the literature in Section~\ref{sec_literature}. Finally, we discuss one of the key results of this study, i.e. the dependence of cool gas on the galaxy environment, in Section~\ref{sec_discussion}.

\subsection{Summary of results}
\label{sec_summary}

We have presented a study of the correlation between low- and high-ionization metal-enriched gaseous haloes traced by \mgii\ and \civ\ absorption, respectively, and galaxy properties and environment characterized by overdensities at $z<2$. We have utilized two large galaxy surveys in quasar fields, MAGG \citep{lofthouse2020,dutta2020,fossati2021} and QSAGE \citep{bielby2019,stott2020}, for this study. MAGG, a MUSE optical IFU survey with FoV of $\approx1\times1$\,arcmin$^2$ in 28 \zqso\ $\approx$3-4 quasar fields, and QSAGE, an HST NIR grism survey with FoV of $\approx1.5\times1.5$\,arcmin$^2$ in 12 \zqso\ $\approx$1-2 quasar fields (8 of which have MUSE observations), are both "blind" (i.e. without any pre-selection), flux-limited galaxy surveys. The 90\% completeness limit of the MAGG $r$-band data and the QSAGE F140W band data is $\approx26$\,mag. 

Using these deep and complete galaxy surveys, we are able to study the gas-galaxy connection over a wide parameter space -- over stellar masses of 10$^{6-12}$\,\msun\ and up to $z\approx2$ and impact parameters of 750\,kpc (see Fig.~\ref{fig:qsage_magg_galprop}). Thanks to the wide survey areas and large spectroscopic galaxy samples ($\approx$750), we are additionally able to identify denser structures like groups using a Friends-of-Friends algorithm (see Fig.~\ref{fig:groups_single_galprop}) as well as define the local environment in a continuous and statistical manner using galaxy overdensities in fixed apertures. To study the multi-phase gaseous haloes, we make use of \mgii\ absorption lines identified blindly over $0.8<z<1.5$ in high-resolution optical spectra of quasars in the MAGG survey, and \mgii\ and \civ\ absorption lines identified blindly over $0.1<z<2$ in high-resolution optical and medium-resolution UV spectra of quasars in the QSAGE survey. Combining the properties of the gaseous haloes (equivalent width, covering fraction/incidence) and galaxies (redshift, stellar mass, SFR, sSFR), we have investigated the connection between metal-enriched gas and galaxy environment using different approaches -- by directly correlating absorption with galaxy and group properties, and by correlating absorption with galaxy overdensities. We summarize here the main results of this paper.

\begin{enumerate}
    \item \textit{Dependence of \mgii\ absorption on galaxy properties:}
    Using the combined MAGG and QSAGE sample, we find that \mgii\ equivalent width and covering fraction are negatively correlated with distance from galaxies (Fig.~\ref{fig:mgii_ew_imp}) and positively correlated with stellar mass and SFR of galaxies (Fig.~\ref{fig:mgii_gal_prop}). These results are corroborated by the correlation of \wmgii\ and \mgii\ incidence with overdensities of stellar mass and SFR (Figs.~\ref{fig:mgii_ovden_combined}, \ref{fig:mgii_ovden_galaxy}, \ref{fig:mgii_ovden_qsage}). However, we find weaker dependence of \mgii\ absorption on the specific SFR. Further, \wmgii\ and \mgii\ covering fraction do not show significant dependence on either SFR or sSFR in the sub-samples at low and high stellar mass. 
    \item \textit{Dependence of \civ\ absorption on galaxy properties:}
    In the QSAGE sample, the \civ\ covering fraction shows an increasing trend with stellar mass and SFR (Fig.~\ref{fig:civ_gal_prop}). The incidence of \civ\ gas also tends to be higher in regions of higher mass and SFR overdensities (Figs.~\ref{fig:civ_ovden_zlt1}, \ref{fig:civ_ovden_zgt1}). The dependence of \civ\ equivalent width on stellar mass and SFR is weaker than in the case of \mgii. \civ\ equivalent width and incidence do not show strong dependence on the sSFR, similar to what we find for \mgii\ gas.
    \item \textit{Comparison of low- and high-ionization gas distribution:}
    The relation of equivalent width of \civ\ with impact parameter is shallower than that of \mgii\ gas (Fig.~\ref{fig:civ_ew_imp}). Further, the covering fraction of \civ\ is $\approx$2 times higher on average than that of \mgii, at a given distance and equivalent width sensitivity (Fig.~\ref{fig:cf_grp}). The correlation length of \civ\ is higher than that of \mgii, based on cross-correlation analysis with galaxies, although consistent within the large uncertainties (Fig.~\ref{fig:cross_corr}). On the basis of all the above results, the distribution of high-ionization \civ\ gas is likely to be more extended around galaxies compared to the low-ionization \mgii\ gas.
    \item \textit{Dependence of distribution of metals on redshift:}
    We do not find any dependence of the average \mgii\ and \civ\ equivalent widths (including upper limits) on redshift. However, the covering fraction of \mgii\ and \civ\ gas tend to be higher at $z>1$ compared to at $z\le1$, by a factor of $\approx$1.5 and $\approx$4, respectively, within the virial radius (Figs.~\ref{fig:mgii_gal_prop}, \ref{fig:civ_gal_prop}). The incidence of \mgii\ gas at a given overdensity also tends to be higher at $z>1$, while for \civ\ gas it is difficult to directly compare the dependence on overdensity at $z<1$ and at $z>1$ in the current sample. 
    \item \textit{Dependence of \mgii\ absorption on galaxy environment:}
    The \mgii\ gas distribution around galaxies in groups (associations of two or more galaxies identified by FoF) is found to be different than that around single galaxies. Group galaxies are found to exhibit $\approx$2-3 times higher \mgii\ covering fraction at a given radius compared to single galaxies at all redshifts (Figs.~\ref{fig:cf_grp}, \ref{fig:cf_grp_zgal}). Furthermore, the covering fraction and equivalent width of \mgii\ gas are found to be similarly enhanced around group galaxies compared to a control sample of isolated galaxies that is matched in impact parameter, stellar mass and redshift (Fig.~\ref{fig:cf_grp_control}). This is supported by the correlation of \wmgii\ with number overdensity of galaxies. \mgii\ absorption is more prevalent at the highest overdensities, with comparatively fewer non-detections at overdensities greater than 1 (Figs.~\ref{fig:mgii_ovden_combined}, \ref{fig:mgii_ovden_qsage}).
    \item \textit{Dependence of \civ\ absorption on galaxy environment:}
    The covering fraction of \civ\ gas within impact parameter of 100\,kpc of group galaxies is two times higher compared to that of single galaxies, albeit only at 1$\sigma$ significance (Fig.~\ref{fig:cf_grp}). There is no significant difference in the average \civ\ covering fraction within larger radii between group and single galaxies. Furthermore, the \civ\ equivalent width and covering fraction around group galaxies are consistent with those around a control sample of single galaxies (Fig.~\ref{fig:cf_grp_control}). The \wciv\ and \civ\ incidence show an increasing trend with the number overdensity of galaxies up to overdensities of $\approx$1 (Figs.~\ref{fig:civ_ovden_zlt1}, \ref{fig:civ_ovden_zgt1}). 
\end{enumerate}
\subsection{Comparison with the literature}
\label{sec_literature}

As summarized in the previous section, we have found various connections between the metal-enriched halo gas as traced by \mgii\ and \civ\ absorption, and the galaxy properties, redshift and environment. Here we compare our results with those obtained by similar studies in the literature.

\begin{enumerate}
    \item \textit{Connection between metal-enriched halo gas and galaxy properties:}
    Similar dependencies of the cool, low-ionization gas, traced by \mgii\ absorption, on galaxy distance, mass and SFR as found in this study have been reported in the literature \citep[e.g.][]{chen2010,bordoloi2011,nielsen2013,lan2014,prochaska2014,rubin2018,dutta2020}. However, unlike some studies in the literature \citep[e.g.][]{bordoloi2011,lan2014}, we find weaker dependence of \mgii\ absorption on the specific SFR. Based on this and the lack of significant dependence on either SFR or sSFR in sub-samples at low and high stellar mass, we hypothesize that the dependence of \mgii\ absorption on SFR is mainly driven by the correlation between SFR and stellar mass, and that the stellar or halo mass is more likely to determine the distribution of \mgii\ absorbing gas, although this needs to be verified by including larger number of passive galaxies in the sample. Our result that the covering fraction of \civ\ increases with stellar mass and SFR is consistent with the results of \civ-galaxy studies at $z\le0.1$, which find more prevalent \civ\ absorption around more massive and star-forming galaxies \citep{bordoloi2014,burchett2016}. On the other hand, the lack of strong dependence of \civ\ equivalent width and incidence on the sSFR is contrary to what has been found in a few studies \citep{borthakur2013,bordoloi2014}, but is similar to what we find for \mgii\ gas. Hence, the stellar or halo mass could be the driving factor behind the distribution of \civ\ gas as well.
    \item \textit{Distribution of multi-phase halo gas:}
    Based on the radial profiles of equivalent widths and covering fractions, and angular cross-correlation functions, the distribution of the high-ionization \civ\ gas is inferred to be more extended around galaxies compared to the low-ionization \mgii\ gas. Studies using both observations and simulations have found that the distribution of the low-ionization gas tends to drop off more quickly with impact parameter than the high-ionization gas, and that the gaseous haloes become more highly ionized with increasing distance from galaxies \citep[e.g.][]{ford2013,liang2014,werk2014,oppenheimer2008}. Our results are qualitatively consistent with this. This could point towards a simple picture of the low-ionization, denser gas being embedded in the more extended, diffuse high-ionization gas around galaxies. Alternatively, the \civ\ absorption could be arising in interfaces between a hotter, more volume-filling phase and the low-ionization, cool gas phase.
    \item \textit{Redshift evolution of metals around galaxies:}
    We are able to trace the evolution of metal-enriched halo gas from intermediate to low redshifts, over a period where the cosmic star formation rate density declines dramatically ($z\approx0-2$). \citet{lan2020} reported a significant redshift evolution of the covering fraction of strong \mgii\ absorbers (\wmgii\ $>$1\,\AA) up to $z\approx1.3$ \citep[see also][]{schroetter2021}, similar to the SFR evolution of galaxies, but have not detected any evolution for the weaker absorbers (\wmgii\ $<$1\,\AA). The tentative increase in covering fractions of \mgii\ and \civ\ at $z>1$ found in this work, though not statistically significant at present, could be due to various factors such as galaxies having higher SFR on average at high redshift and populating gaseous haloes with larger amount of metals or redshift evolution in the ionizing background or gas density.
    \item \textit{Connection between metal-enriched halo gas and galaxy environment:}
    The higher equivalent width and covering fraction of \mgii\ gas around group galaxies, and the positive correlation between equivalent width and incidence of \mgii\ and galaxy overdensity, point towards significant influence of the local environment on the distribution of cool, low-ionization gas in haloes. This is in agreement with the results of other recent studies that have looked at \mgii\ absorption in group environments \citep{nielsen2018,fossati2019b,dutta2020,lundgren2021}. When it comes to \civ\ gas, \citet{burchett2016} report a dearth of \civ\ absorption in the inner halo of galaxies with \mhalo\ $>5\times10^{12}$\,\msun\ at $z<0.1$ \citep[see also][]{manuwal2021}. We do not detect \civ\ absorption in the most massive haloes in our sample, but find an increasing trend of equivalent width and incidence of \civ\ with overdensity. However, we lack a sufficient number of overdense structures to conclusively check the dependence of \civ\ absorption on the highest overdensities. We discuss the implications of the environmental dependence of the multi-phase halo gas in the next section.
\end{enumerate}

\subsection{Environmental dependence of metals in gaseous haloes}
\label{sec_discussion}

We have shown that the galaxy environment significantly influences the distribution of cool, metal-enriched, low-ionization gas around galaxies. It has been proposed that gravitational and hydrodynamic interactions could be the reason behind the observed enhancement in \mgii\ equivalent width and cross-section in denser environments \citep{fossati2019b,dutta2020}. The enhanced absorption could also be related to the higher gas density in these rich environments, which in turn could be related to the higher halo mass or galaxy-galaxy dynamics. Using cosmological magnetohydrodynamical simulations, \citet{nelson2021} found that the extent of the \mgii\ emission around galaxies with \mstar\ $\ge10^{10}$\,\msun increases with the local galaxy overdensities, due to the contributions of additional cool gas within and stripped from nearby satellites. It is likely that several processes acting in tandem affect the distribution of the cool gas around galaxies residing in overdense regions. The relative contributions of these processes to the boosting of \mgii\ strength and cross-section are also likely to evolve with redshift. Recall that the difference in cool gas properties with galaxy environment is seen across all redshifts at $z<2$, with slightly more prominent difference arising at $z<1$. The difference is also driven by the more massive (\mstar\ $>10^9$\,\msun) galaxies, which are more likely to be part of rich groups. By controlling for the redshift, stellar mass and impact parameter, we find that the \mgii\ cross-section is boosted around group galaxies compared to single galaxies. This implies that environmental processes, acting over and above secular processes, are likely to impact the distribution of the cool gas phase.

One of the most common environmental mechanisms affecting the neutral gas morphology at $z\approx0$ is ram pressure stripping, which is predicted to become more important at low redshifts and can act at large distances as well \citep{marasco2016,zinger2018}. The cool gas in the overdense regions in our sample can be stripped from the individual galaxy haloes by the intra-group medium. There are several examples of such stripping at $z\lesssim0.5$ \citep[e.g.][]{fumagalli2014,fossati2016,poggianti2017,johnson2018}, and of observations which suggest such stripping even at higher redshifts \citep[e.g.][]{elmegreen2010,rauch2014}. However, it needs to be determined whether the intra-group medium we probe is dense enough for ram pressure stripping to be effective. On the other hand, tidal interactions are believed to be more effective in the halo mass range probed here (\mhalo\ $<10^{14}$\,\msun), especially at smaller separations between galaxies \citep{boselli2006}. In the local Universe, there are several examples of gravitational interactions perturbing the neutral and ionized gas around galaxies in groups \citep[e.g.][]{yun1994,mihos2012,fossati2019a}. 

At $z\le0.2$, about 40-60\% of galaxies are identified to be in groups of two or more galaxies based on FoF and Voronoi-Delaunay techniques \citep{eke2004,tempel2012}, while this fraction is $\approx10-15$\% at $z\approx1$ \citep{knobel2009,knobel2012,gerke2012}. Note, however, that the completeness limit of the galaxy surveys varies with redshift. In the sample studied here, we find that on an average $\approx$43\% of the galaxies reside in "groups" of two or more galaxies. Compared with clusters, the group environment is more conducive to galaxy mergers due to the small velocity dispersion in groups. Mergers are found to occur more frequently at higher redshifts, with a merger fraction of 25\% at $z\approx1.4$ for galaxies with \mstar\ $\sim10^{9-10}$\,\msun\ \citep{conselice2008,conselice2014}. Hence, mergers and interactions could be playing a dominant role in disturbing the cool gas around galaxies in the overdense regions, especially at $z>1$. Indeed, in the local Universe, tidal interactions between pairs of dwarf galaxies (\mstar\ $<10^{10}$\,\msun) have been found to move a large amount of neutral gas to their outskirts, with $>50$\% of the total gas mass found beyond the extent of their stellar discs \citep{pearson2016}.

When it comes to the more highly ionized gas around galaxies, the influence of the environment is found to be less pronounced. We find a trend for the \civ\ equivalent width and cross-section to increase up to overdensities of 1 and halo masses of 10$^{13}$\,\msun, beyond which we do not detect \civ\ absorption. The \civ\ gas could be getting shock-heated to C\,{\sc v} in the more massive haloes. It needs to be confirmed with samples containing more massive structures whether the most overdense regions lack \civ\ absorbing gas. 
At $z=0$, the mass fraction of baryons in the CGM and the covering fractions of \civ\ and \ovi\ gas around galaxies with \mstar\ $\approx10^{10.2-10.7}$\,\msun\ are predicted to decline with increasing mass of their central super-massive black holes in the {\sc eagle} simulations \citep{davies2020,oppenheimer2020}. The galaxies with the more massive black holes at a fixed halo mass are further found to be in higher-density environments in these studies.
On the other hand, \citet{muzahid2021} have found a strong environmental dependence of \civ\ gas, with significantly stronger \civ\ absorption found around groups of multiple \lya\ emitters at $z\approx3$. Thus the effect of environment on the \civ\ gas could be evolving with redshift, and this will be investigated further using a sample of \civ\ absorption systems and \lya\ emitters at $z\approx3-4$ in the MAGG survey (Galbiati et al., in prep.). 

At $z<1$, \citet{burchett2018} report a dearth of warm-hot gas (10$^{5-6}$\,K) as traced by \ovi\ and broad \lya\ absorption in galaxy clusters, which could be attributed to environmental quenching. \citet{pointon2017} find smaller equivalent width and slightly smaller covering fraction of \ovi\ absorption in groups compared to isolated galaxies, suggesting that the warm gas is getting more highly ionized in group environments \citep[see also][]{bielby2019}. The above results are based on a handful of systems ($<$10) and need to be confirmed with larger samples spanning wider properties. However, combined with the results presented in this work, the emerging picture is that not only do properties of the gaseous haloes depend on the galaxy environment, different phases of the gas get transformed differently by environmental processes. Estimating and modeling the physical properties of gas across galaxy overdensities will provide further insights into the various environmental processes at play. 
%
%
\section*{Acknowledgements}
We thank the anonymous referee for their constructive comments that
improved the quality and presentation of the paper.
This project has received funding from the European Research Council (ERC) under the European Union’s Horizon 2020 research and innovation programme (grant agreement No 757535) and by Fondazione Cariplo (grant No 2018-2329).
SC gratefully acknowledges support from the Swiss National Science Foundation grant PP00P2\_190092 and from the ERC under the European Union’s Horizon 2020 research and innovation programme grant agreement No 864361.
RAC is a Royal Society University Research Fellow.
TMT was partially supported for this work by NASA grant HST-GO-13846.001-A from the Space Telescope Science Institute.
MTM thanks the Australian Research Council for \textsl{Discovery Projects} grants DP130100568 and DP170103470, which supported this work.
This work is based on observations collected at the European Organisation for Astronomical Research in the Southern Hemisphere under ESO programme IDs 197.A-0384, 065.O-0299, 067.A-0022, 068.A-0461, 068.A-0492, 068.A-0600, 068.B-0115, 069.A-0613, 071.A-0067, 071.A-0114, 073.A-0071, 073.A-0653, 073.B-0787, 074.A-0306, 075.A-0464, 077.A-0166, 080.A-0482, 083.A-0042, 091.A-0833, 092.A-0011, 093.A-0575, 094.A-0280, 094.A-0131, 094.A-0585, 095.A-0200, 096.A-0937, 097.A-0089, 099.A-0159, 166.A-0106, 189.A-0424, 094.B-0304, 096.A-0222, 096.A-0303, 103.A-0389, and 1100.A-0528.
Based on observations made with the NASA/ESA Hubble Space Telescope, obtained from the data archive at the Space Telescope Science Institute. STScI is operated by the Association of Universities for Research in Astronomy, Inc. under NASA contract NAS 5-26555.
Some of the data presented in this work were obtained from the Keck Observatory Database of Ionized Absorbers toward QSOs (KODIAQ), which was funded through NASA ADAP grants NNX10AE84G and NNX16AF52G along with NSF award number 1516777. 
This research has made use of the Keck Observatory Archive (KOA), which is operated by the W. M. Keck Observatory and the NASA Exoplanet Science Institute (NExScI), under contract with the National Aeronautics and Space Administration.
The Liverpool Telescope is operated on the island of La Palma by Liverpool John Moores University in the Spanish Observatorio del Roque de los Muchachos of the Instituto de Astrofisica de Canarias with financial support from the UK Science and Technology Facilities Council.
This work used the DiRAC Data Centric system at Durham University, operated by the Institute for Computational Cosmology on behalf of the STFC DiRAC HPC Facility (\url{www.dirac.ac.uk}). This equipment was funded by BIS National E-infrastructure capital grant ST/K00042X/1, STFC capital grants ST/H008519/1 and ST/K00087X/1, STFC DiRAC Operations grant ST/K003267/1 and Durham University. DiRAC is part of the National E-Infrastructure. 
This research has made use of the following {\sc python} packages: {\sc numpy} \citep{2020NumPy-Array}, {\sc scipy} \citep{2020SciPy-NMeth}, {\sc matplotlib} \citep{Matplotlib2007}, {\sc astropy} \citep{Astropy2013}, {\sc kde}py (\url{https://github.com/tommyod/KDEpy}).
%
%
\section*{Data Availability}
The data used in this work are available from the Mikulski Archive for Space Telescopes (\url{https://archive.stsci.edu}) and the European Southern Observatory archive (\url{https://archive.eso.org/}). The MUSE ESO P3 level data of the MAGG survey are available at \url{http://archive.eso.org/wdb/wdb/adp/phase3_main/form?collection_name=197.A-0384&release_name=DR1BOWER}. The list of galaxies and associated absorption properties used in this work is provided as supplementary material.
%

\bibliographystyle{mnras}
\bibliography{mybib} 


\appendix
\section{Star formation rate from emission lines}
\label{appendix_sfr}
\begin{figure*}
 \includegraphics[width=0.3\textwidth]{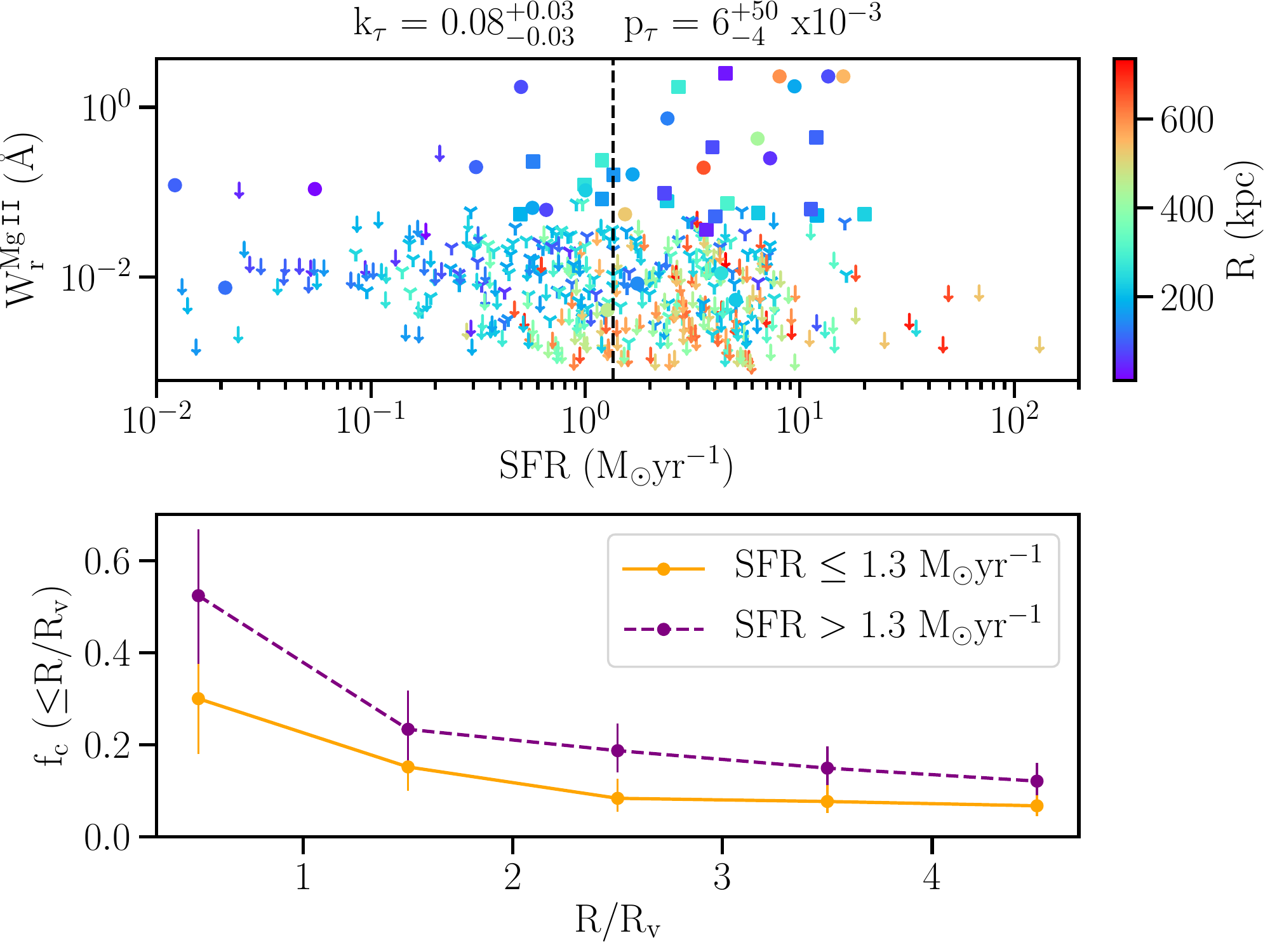}
 \includegraphics[width=0.3\textwidth]{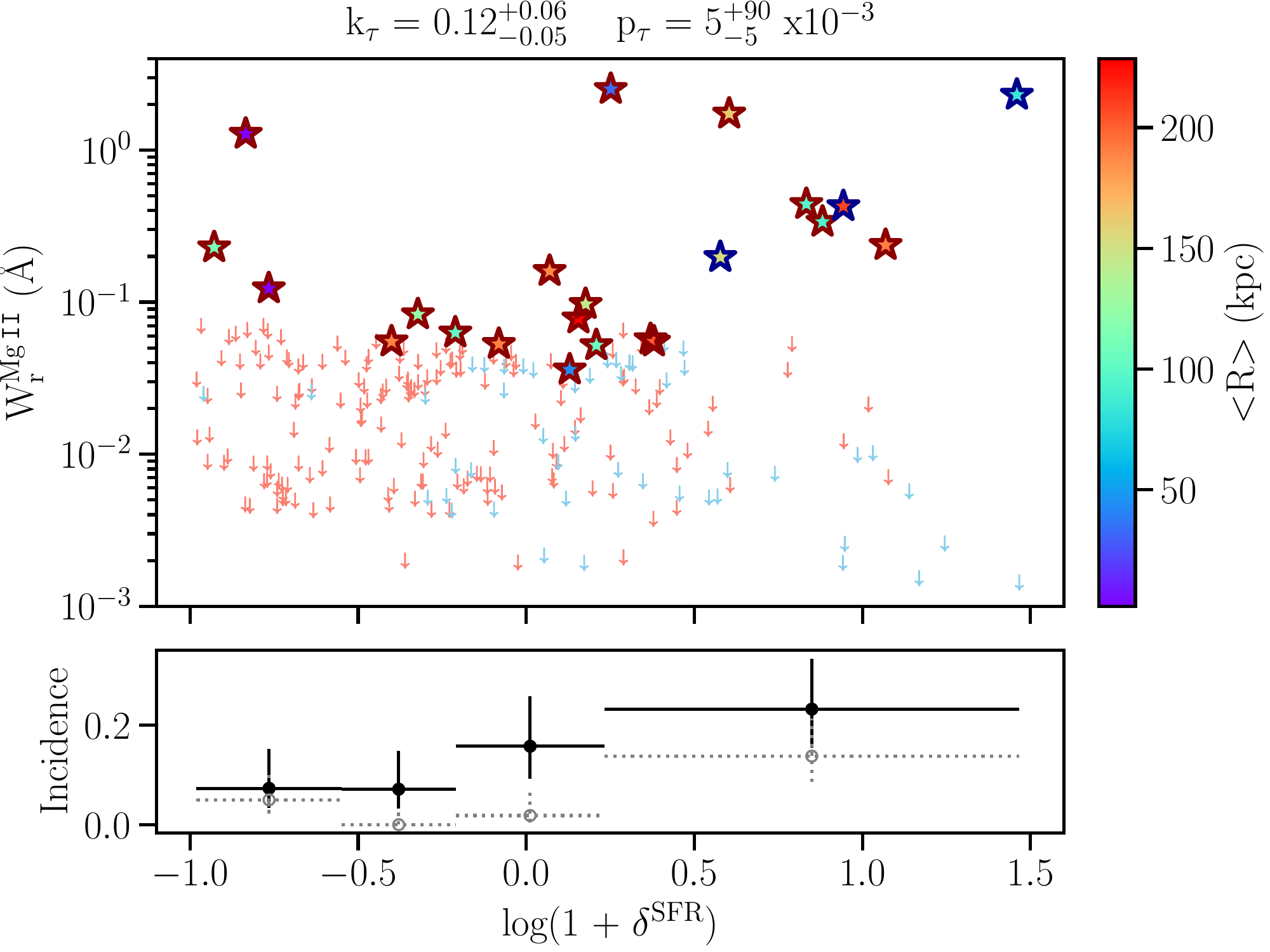}
 \includegraphics[width=0.3\textwidth]{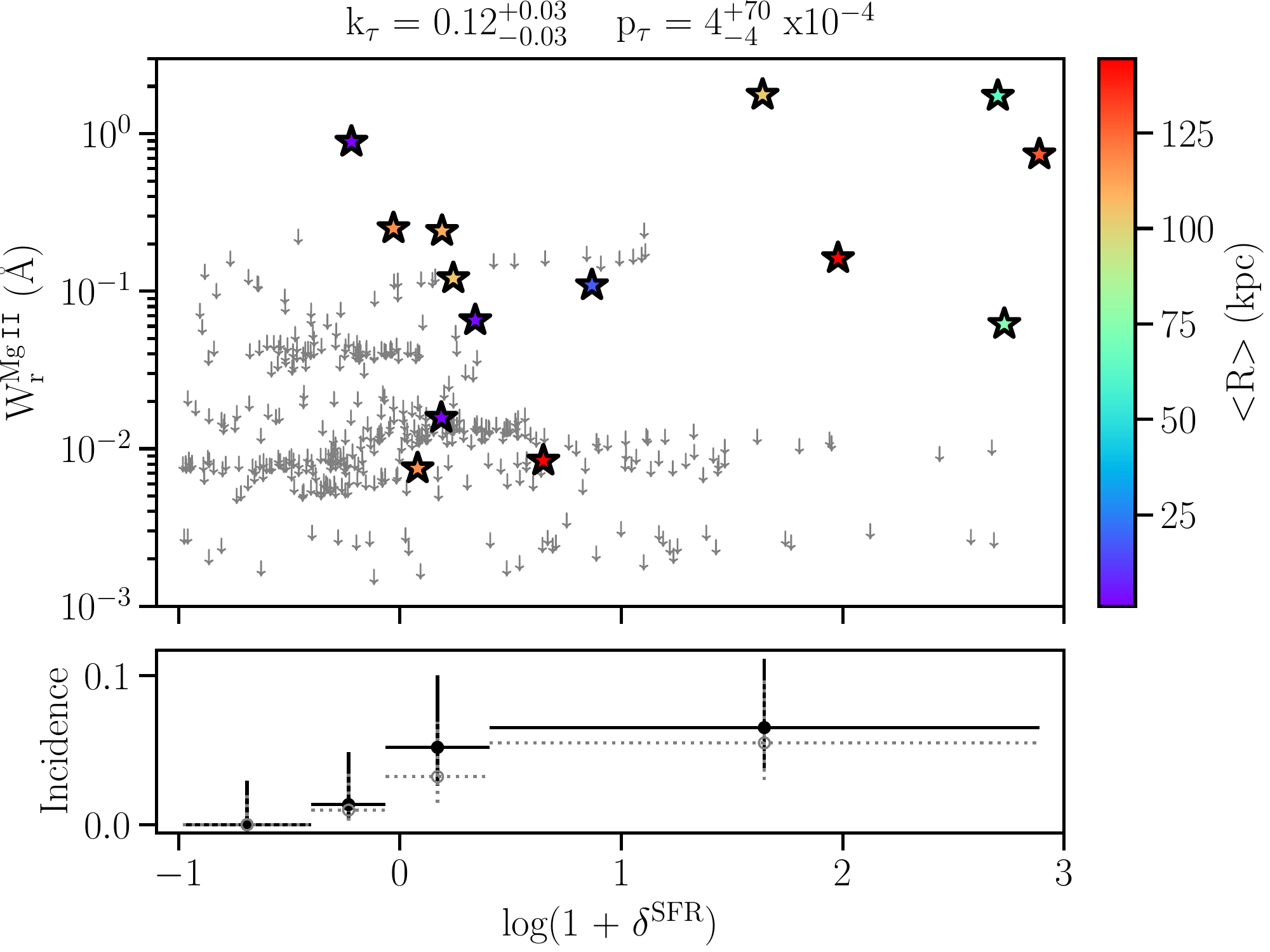}
 \caption{Dependence of \mgii\ absorption on emission line-based SFR. 
 {\it Left:} Dependence of \wmgii\ (top) and radial profile of \mgii\ covering fraction (bottom) on the SFR in the combined MAGG and QSAGE sample. The symbols are the same as in Fig.~\ref{fig:mgii_gal_prop}.
 {\it Centre:} Dependence of \wmgii\ (top) and \mgii\ incidence (bottom) on the SFR overdensity over $0.9<z<1.6$ within a radius of 240\,kpc in the combined MAGG and QSAGE sample. 
 {\it Right:} Dependence of \wmgii\ (top) and \mgii\ incidence (bottom) on the SFR overdensity over $0.1<z<1$ within a radius of 200\,kpc in the QSAGE sample.
 Symbols in the centre and right plots are the same as in Fig.~\ref{fig:mgii_ovden_combined}.
 The trends of absorption properties with emission line-based SFR are similar to the ones reported with SPS-based SFR in the main paper.
 }
 \label{fig:appendix_sfr}
\end{figure*}

The results presented in this paper are based on SFRs derived from fitting stellar population synthesis (SPS) models to the optical/NIR imaging and spectroscopy (see Section~\ref{sec_qsage_galaxy}). Here we estimate the effect of using SFRs estimated from the nebular emission lines on the results. To estimate the emission line-based SFR, we first correct the \ha\ and \oii\ luminosities for dust attenuation using the visual extinction (A$_{\rm v}$) estimated by {\sc mc-spf} and the extinction law of \citet{calzetti2000}, which includes a 2.27 attenuation factor to account for the extra extinction in young ($<$10 Myr) stars. The SFR is then estimated from the dust-corrected \ha\ luminosity following \citet{kennicutt1998}, after accounting for a factor of 1.7 \citep{zahid2012} to convert from the \citet{salpeter1955} IMF to the \citet{chabrier2003} IMF. If \ha\ is not covered, we use \oii\ luminosity and the relation from \citet{kewley2004}. Note that we correct the \ha\ fluxes, measured using WFC3 grism spectra, for contamination by the \nii\ line using a stellar mass and redshift dependent correction factor following the parameterization given in \citet{faisst2018}. The SPS-based SFRs and emission line-based SFRs follow each other in general, with an average scatter of $\approx$0.4\,dex. We find that the trends regarding SFR and sSFR reported in the paper hold when using the emission line-based SFRs as well. We show for example some of the plots which look at the dependence of \mgii\ absorption on SFR in Fig.~\ref{fig:appendix_sfr}, using now the emission line-based SFRs. The results are similar to the ones reported in the main paper using SPS-based SFRs. The results are similar in the case of \civ\ absorption as well. Thus, the exact method of estimating the SFR does not appear to have a significant effect on the results.

\section{Tests on overdensity estimates}
\label{appendix_density}
\begin{figure*}
 \subfloat[Box kernel]{\includegraphics[width=0.3\textwidth]{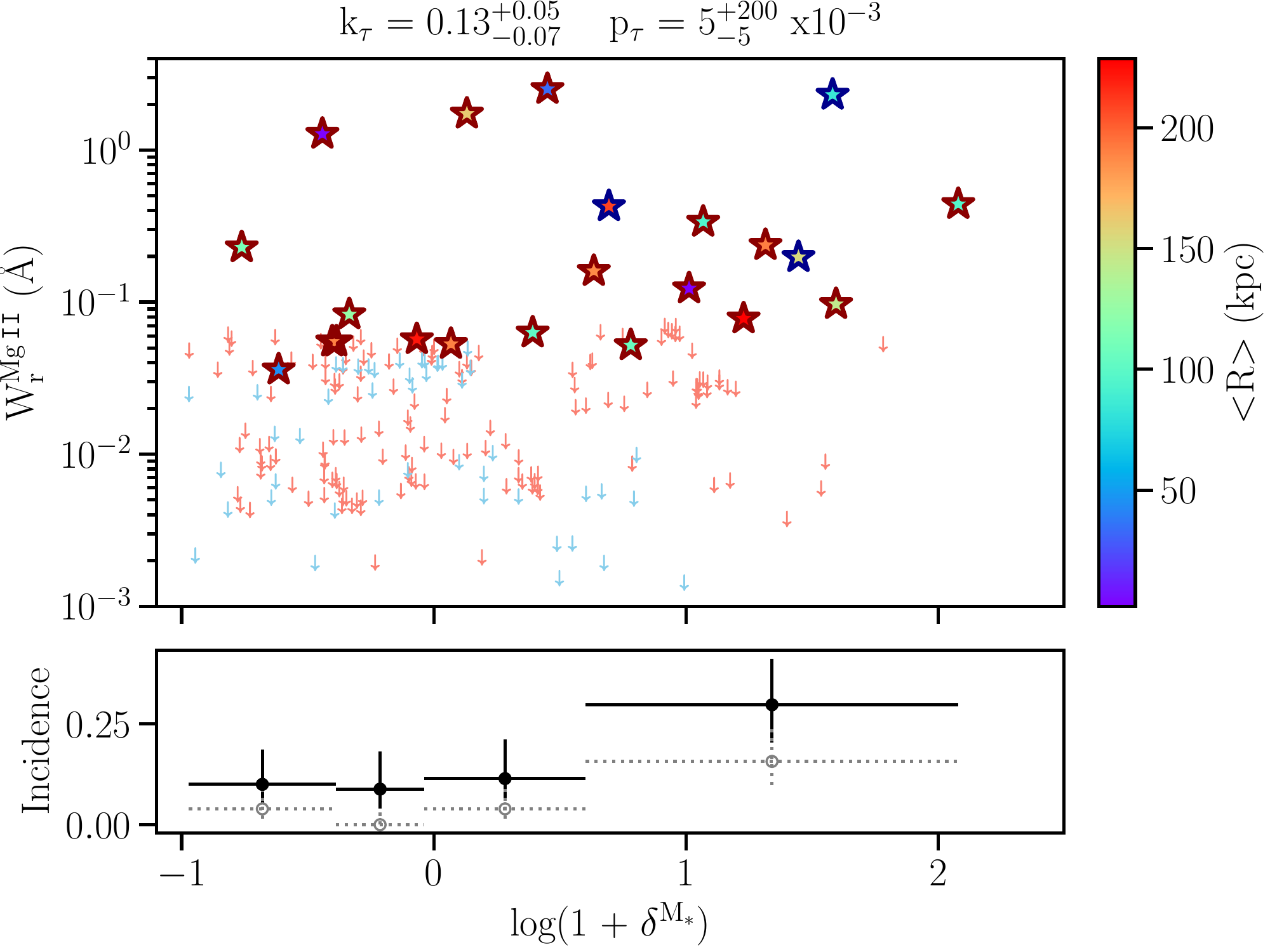}}
 \subfloat[Silverman's rule for bandwidth selection]{\includegraphics[width=0.3\textwidth]{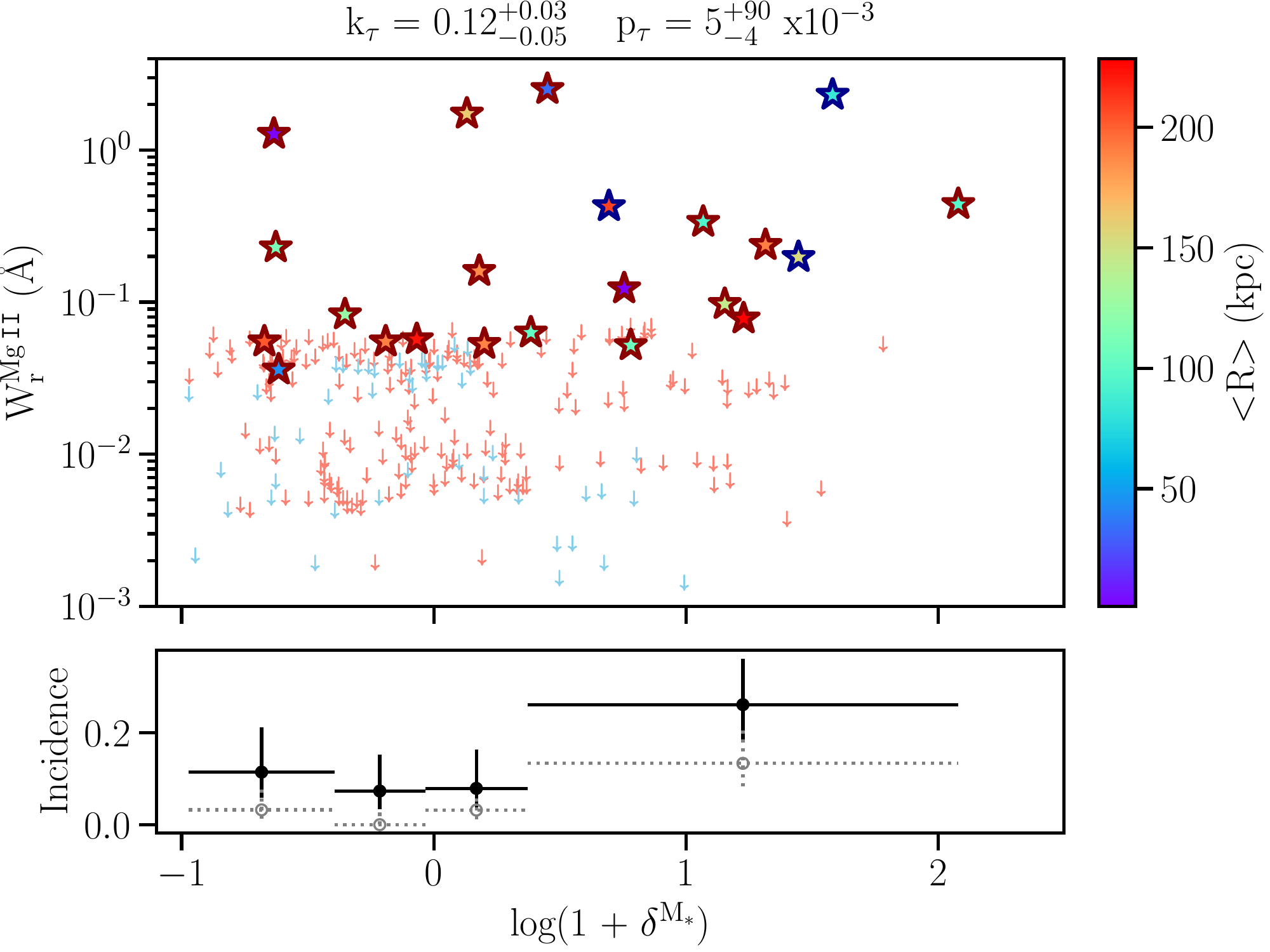}}
 \subfloat[Directly counting galaxies]{\includegraphics[width=0.3\textwidth]{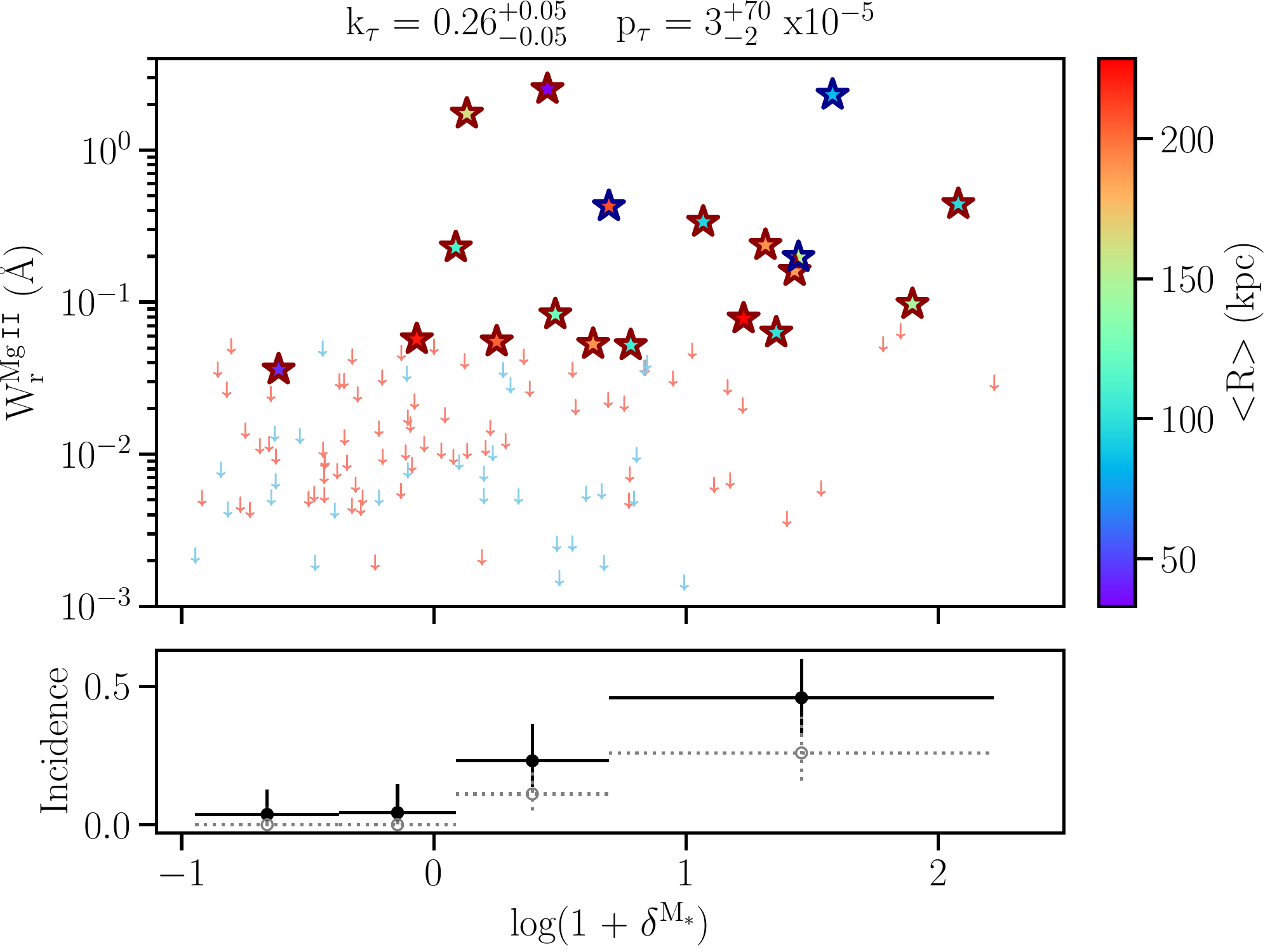}}
 \hspace{0.01cm}
 \subfloat[LoS start offset by 250\,\kms]{\includegraphics[width=0.3\textwidth]{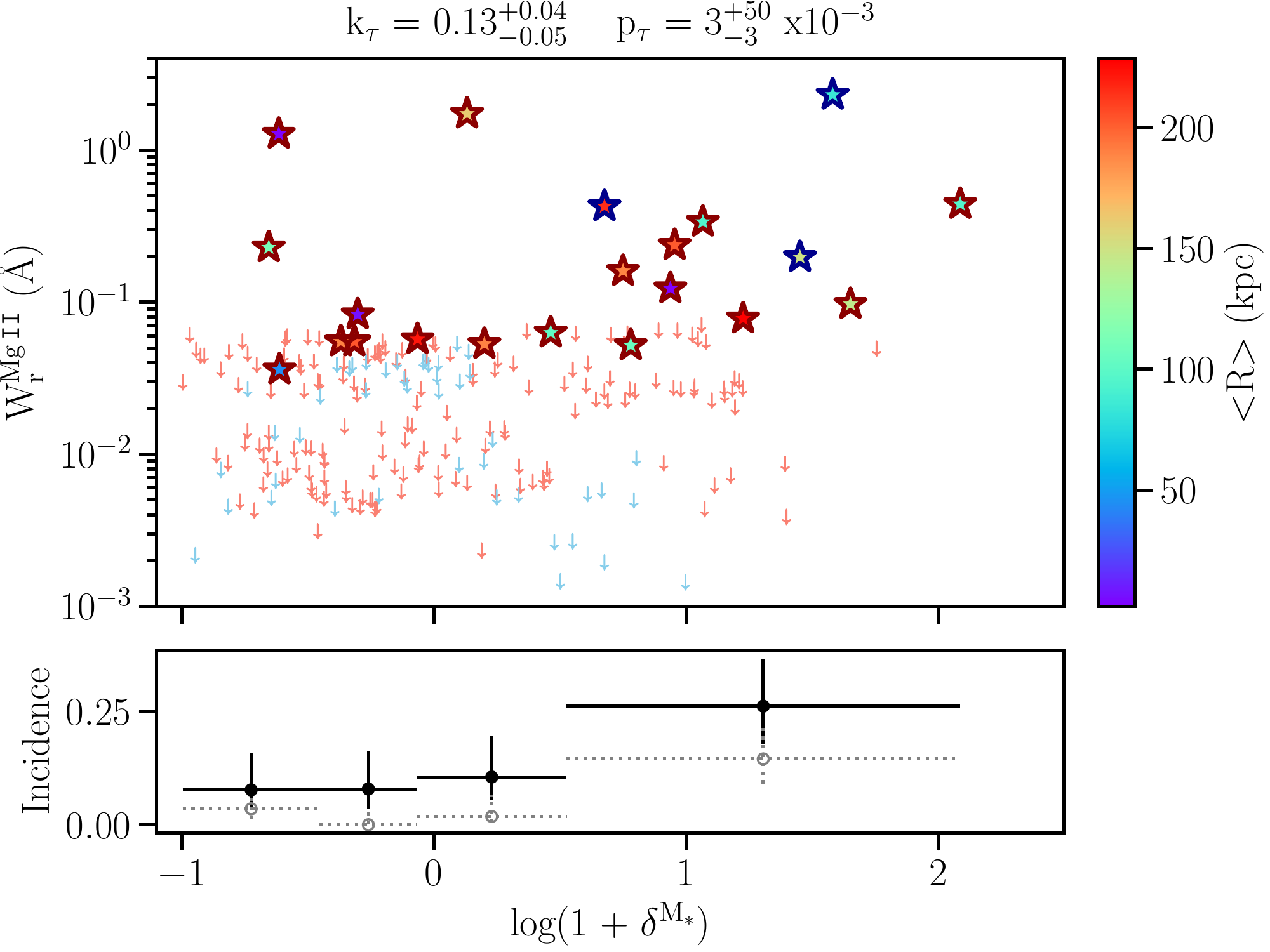}}
 \subfloat[LoS start offset by 500\,\kms]{\includegraphics[width=0.3\textwidth]{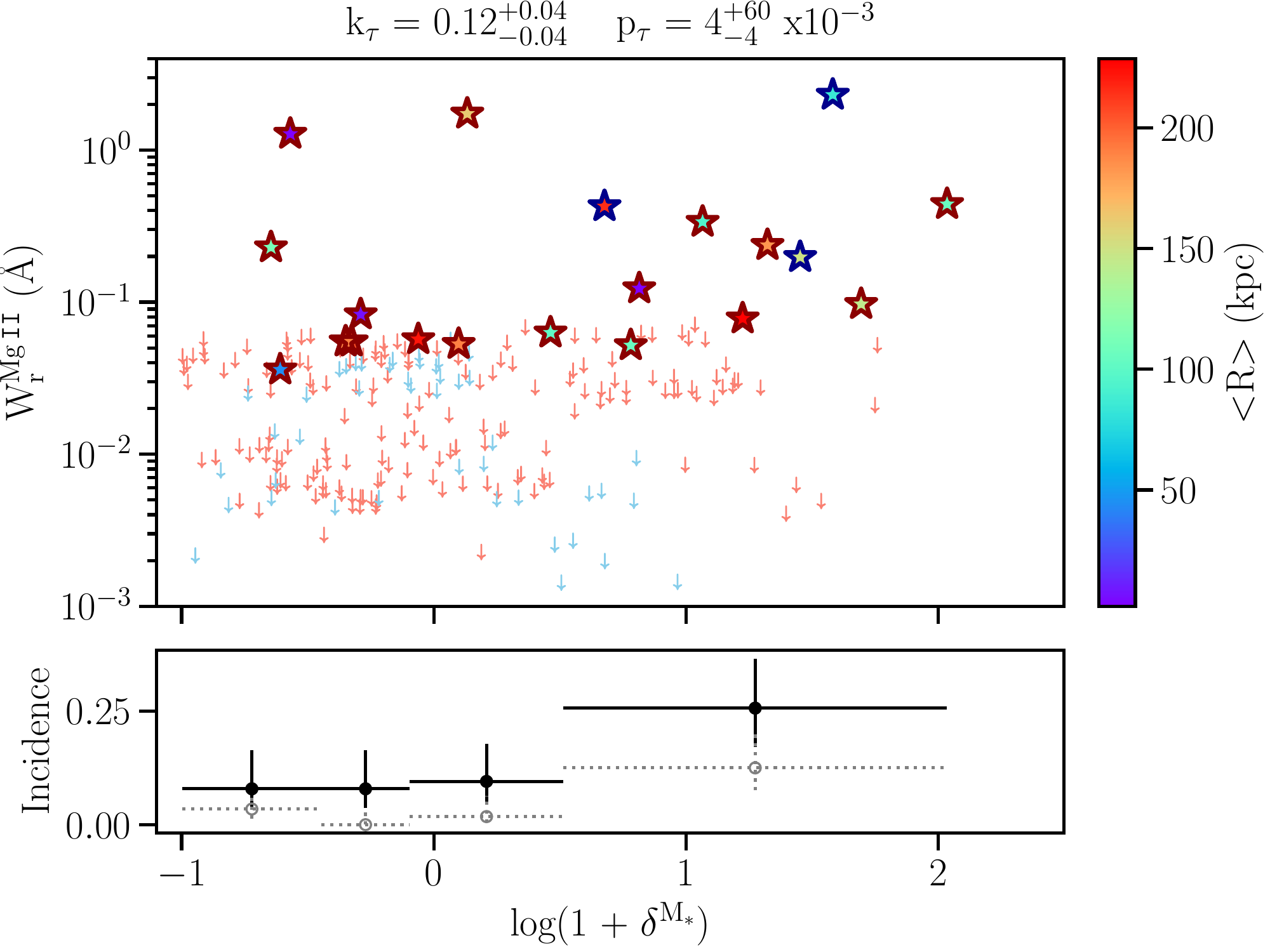}}
 \subfloat[LoS velocity window of $\pm$500\,\kms]{\includegraphics[width=0.3\textwidth]{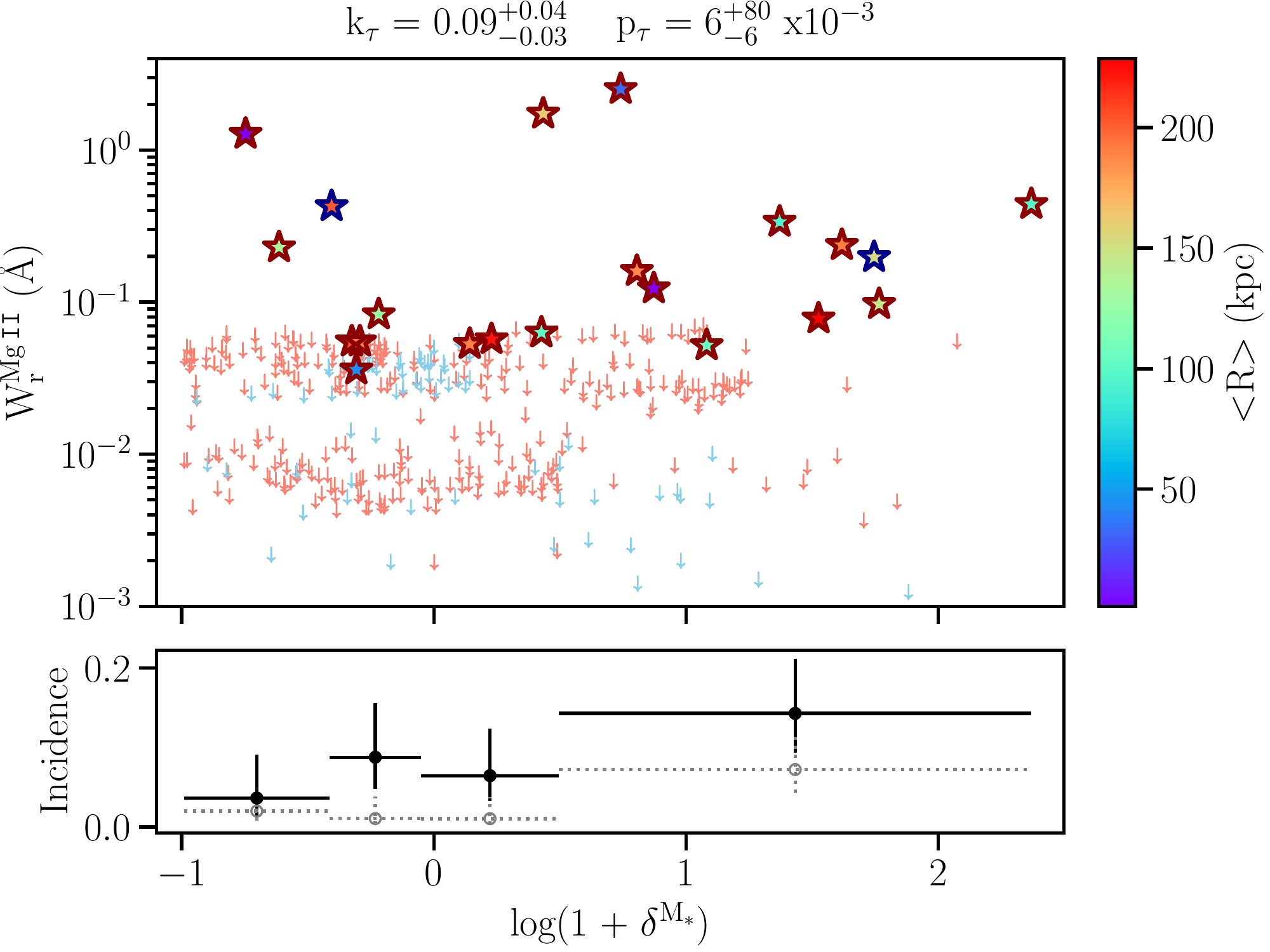}}
 \caption{Illustration of the effect of different changes in the method of estimating overdensities on the results. Dependence of \wmgii\ (top) and \mgii\ incidence (bottom) on the stellar mass overdensities in the combined MAGG and QSAGE sample over $0.9<z<1.6$. The symbols and color-coding are the same as in panel (b) of Fig.~\ref{fig:mgii_ovden_combined}. We make the following changes in estimating overdensities from panels (a) to (f): use box kernel in KDE, Silverman's rule for bandwidth selection in KDE, directly count the galaxies in volumes, offset the splitting up of the LoS into volumes by 250\,\kms\ and 500\,\kms, and use a LoS velocity window of $\delta v =\pm$500\,\kms\ to define the volumes and associate metal absorption lines. 
 The overall trends of equivalent width and incidence with overdensities are similar for different changes in the method of estimating overdensities.
 }
 \label{fig:appendix_density}
\end{figure*}

The results presented in the paper on connecting metals with galaxy overdensities are based on estimating the galaxy densities using 1D KDE with a Gaussian kernel and bandwidth selected by the Improved Sheather-Jones algorithm, and splitting the quasar LoS into equal volumes with a velocity window of $\delta v =\pm$1000\,\kms. We experimented with different kernels such as Boxcar and Epanechnikov, and different bandwidth selection algorithms such as Silverman's rule \citep{silverman1986} and Scott's rule \citep{scott1992}. These were not found to have a significant effect on the final results. We also estimated the galaxy densities by directly counting galaxies in the volumes instead of using the KDE. The dependence of equivalent width and incidence on the overdensities becomes slightly stronger in this case as we do not include the volumes where there are no galaxies present. Further, to check the robustness of the way we define the volumes and that the densities do not depend on the starting point along the quasar LoS, we offset the start of the LoS by 250\,\kms, 500\,\kms\ and 750\,\kms. We additionally use different velocity windows of $\delta v =\pm$500\,\kms\ and $\pm$700\,\kms\ to define the volumes and associate metal absorption lines. These changes in the way the volumes are defined do not have a significant effect on the results. We illustrate the effect of some of the above changes on the dependence of \mgii\ equivalent width and incidence on the stellar mass overdensities in the combined MAGG and QSAGE sample over $0.9<z<1.6$ in Fig.~\ref{fig:appendix_density}. Overall, we find that the results presented in the paper for both \mgii\ and \civ\ absorption are valid for different changes made to the method of estimating overdensities.


\bsp	
\label{lastpage}
\end{document}

%% file: overview_table.tex
\begin{table*}
\caption{Overview of the QSAGE and MAGG survey data used in this work.}
\centering
\begin{tabular}{ccc}
\hline
\hline
 & QSAGE & MAGG \\
\hline
No. of quasar fields & 12 & 28 \\
Galaxy data & HST WFC3, MUSE$^{\rm a}$ & MUSE \\
Quasar spectra & COS, STIS, UVES, HIRES & UVES, HIRES, MIKE, ESI, X-SHOOTER \\
Redshift range studied in this work & $0.1-2.2$ & $0.8-1.5$ \\
Maximum impact parameter at $z=1$ & $\approx750$\,kpc & $\approx350$\,kpc \\
90\% continuum completeness limit & F140W $\approx26$\,mag & $r\approx26.3$\,mag \\
90\% emission flux completeness limit & $\approx2\times10^{-17}$\,\ergscm\ $^{\rm b}$ & $\approx3\times10^{-18}$\,\ergscm\ $^{\rm c}$ \\
90\% stellar mass limit at $z=1$ & $\approx5\times10^8$\,\msun\ & $\approx4\times10^9$\,\msun\ \\
90\% (unobscured) SFR limit at $z=1$ & $\approx0.5$\,\msunyr\ $^{\rm d}$ & $\approx0.1$\,\msunyr\ $^{\rm e}$ \\
90\% \wmgii\ sensitivity limit & $\approx0.03$\,\AA\ & $\approx0.03$\,\AA\ \\
90\% \wciv\ sensitivity limit & $\approx0.1$\,\AA\ ($z\le1$), $\approx0.03$\,\AA\ ($z>1$) & --- \\
Description of data & \citet{bielby2019}, \citet{stott2020} & \citet{lofthouse2020}, \citet{dutta2020} \\
\hline
\hline
\end{tabular}
\label{tab:overview_table}
\begin{flushleft}
$^{\rm a}$ MUSE data are available for 8 of the 12 fields in QSAGE
$^{\rm b}$ based on WFC3 grism spectra
$^{\rm c}$ based on MUSE spectra
$^{\rm d}$ based on \ha\ 
$^{\rm e}$ based on \oii\
\end{flushleft}
\end{table*}

%% file: spectra_table.tex
\begin{table*}
\caption{Details of the QSAGE archival quasar spectra used in this work: (1) quasar name, (2) quasar redshift from WFC3 data \citep[see for details][]{stott2020}, (3) instrument name, (4) proposal ID of the HST/ESO programme or the database from which the spectrum is obtained, (5) wavelength range of the spectrum, (6) velocity dispersion per pixel, (7) typical S/N per pixel of the spectrum, measured around 1400$-$1500\,\AA, 1900$-$2000\,\AA, 2600$-$2700\,\AA, and 4000$-$5000\,\AA, in case of COS-FUV, COS-NUV, STIS, and UVES/HIRES, respectively.}
\centering
\begin{tabular}{ccccccr}
\hline
\hline
Quasar & \zqso\ & Instrument & Programme/ & Wavelength & Velocity   & S/N \\
       &        &            & Database   & Range      & Dispersion &     \\
       &        &            &            & (\AA)      & (\kms)     &     \\
(1)    & (2)    & (3)        & (4)        & (5)        & (6)        & (7) \\
\hline
J012017$+$213346 & 1.5041 & STIS    &       8673 & 2278$-$3072  & 4.9 &   9 \\
                 &        & UVES    & 185.A-0745 & 3060$-$6681  & 0.9 &  41 \\
J023507$-$040205 & 1.4428 & COS-FUV &      11741 & 1152$-$1800  & 7.4 &  10 \\
                 &        & COS-NUV &      13846 & 1800$-$2404  & 5.6 &   3 \\
                 &        & STIS    &       8673 & 2278$-$3072  & 4.9 &   6 \\
                 &        & UVES    &      SQUAD & 3301$-$9466  & 2.5 &  10 \\
J051707$-$441055 & 1.7332 & STIS    &       8288 & 2303$-$3072  & 4.9 &  11 \\
                 &        & UVES    &      SQUAD & 3051$-$10428 & 1.3 & 215 \\
J075054$+$425219 & 1.9151 & STIS    &       9040 & 2135$-$2941  & 4.9 &   6 \\
                 &        & HIRES   &     KODIAQ & 3043$-$4876  & 2.5 &  48 \\
J081331$+$254503 & 1.5103 & STIS    &       9040 & 2135$-$2941  & 4.9 &   4 \\ 
                 &        & UVES    &      SQUAD & 3048$-$6652  & 2.5 & 136 \\
J101956$+$274401 & 1.9298 & STIS    &       9186 & 2277$-$3071  & 4.9 &   2 \\
J112442$-$170517 & 2.4181 & STIS    &       9885 & 2200$-$3072  & 4.9 &   2 \\
                 &        & UVES    &      SQUAD & 3060$-$10429 & 1.5 & 160 \\
J113007$-$144927 & 1.1896 & STIS    &       9173 & 2278$-$3072  & 4.9 &   4 \\
                 &        & UVES    &      SQUAD & 3047$-$6808  & 2.5 &  77 \\
J143748$-$014710 & 1.3061 & COS-FUV &      11741 & 1152$-$1800  & 7.4 &  20 \\
                 &        & COS-NUV &      13846 & 1800$-$2112  & 5.6 &   4 \\
                 &        & STIS    &      13846 & 1985$-$2781  & 4.4 &   7 \\
                 &        & HIRES   &     KODIAQ & 3034$-$5881  & 1.3 &  35 \\
J152424$+$095829 & 1.3266 & COS-FUV &      11741 & 1152$-$1800  & 7.2 &  12 \\
                 &        & COS-NUV &      13846 & 1800$-$2112  & 5.7 &   4 \\
                 &        & STIS    &      13846 & 1985$-$2782  & 4.5 &   7 \\
                 &        & HIRES   &     KODIAQ & 3053$-$5878  & 1.3 &  19 \\
J163201$+$373750 & 1.4787 & COS-FUV &      11741 & 1152$-$1800  & 7.4 &  14 \\
                 &        & COS-NUV &      13846 & 1800$-$2404  & 5.3 &   3 \\
                 &        & STIS    &       8673 & 2278$-$3072  & 4.9 &   7 \\
                 &        & HIRES   &     KODIAQ & 3023$-$5878  & 1.3 &  27 \\
J163429$+$703132 & 1.3319 & STIS    &       7292 & 2277$-$3071  & 4.9 &  17 \\
\hline
\hline
\end{tabular}
\label{tab:spectra_table}
\end{table*}